\documentclass[article,12pt]{article} 
\usepackage{amscd,amsmath,amssymb,amsfonts,latexsym,mathrsfs,amsthm,mathtools}
\usepackage{graphicx} 
\graphicspath{{figures/}} %Setting the graphicspath
\usepackage{setspace} 
\usepackage{graphics,epsfig}
\usepackage{sectsty}
\usepackage[numbers]{natbib}
\usepackage[english]{babel}
\usepackage{bm}
\usepackage{subfigure}
\usepackage[usenames]{color}
\usepackage{rotating}
\usepackage{url}
\usepackage{algorithm}
\usepackage{algpseudocode}
\usepackage{flushend}
\usepackage{verbatim}
\usepackage{tabularx}
\usepackage{multirow} 
\usepackage{arydshln}
\usepackage{bm}

\newcommand{\x}{{\bf x}}

\newcommand{\y}{{\bf y}}
\newcommand{\z}{{\bf z}}
\newcommand{\post}{\bar{\pi}}

\newcommand{\norm}[1]{\left\lVert#1\right\rVert}

\newtheorem{thm}{Theorem}
\newtheorem{Rem}{Remark}

\newtheorem{propo}{Proposition}
\newtheorem{conj}{Conjecture}

\title{Adaptive quadrature schemes for Bayesian inference via active learning} 

\author{F. Llorente$^*$, L. Martino$^{\star}$, V. Elvira$^\dagger$, D. Delgado$^*$, J. L{\'o}pez-Santiago$^*$ \\
	{\small $^\dagger$ The University of Edinburgh, Edinburgh, United Kingdom.}\\
	{\small $^*$ Universidad Carlos III de Madrid, L{\'e}ganes, Madrid, Spain.} \\
	{\small $^{\star}$ Universidad Rey Juan Carlos I, M{\'o}stoles, Madrid, Spain.} 
}

\begin{document}

\maketitle
\thispagestyle{empty}

\begin{abstract}
We propose novel adaptive quadrature schemes based on an active learning procedure.  
We consider an interpolative approach for building a surrogate posterior density, combining it  with Monte Carlo sampling methods and other quadrature rules.
The nodes of the quadrature are sequentially chosen by maximizing a suitable acquisition function, which takes into account the current approximation of the posterior and the  positions of the nodes. This maximization  does not require additional evaluations of the true posterior.  We introduce two specific schemes based on Gaussian and Nearest Neighbors bases. For the Gaussian case, we also provide a novel procedure for fitting the bandwidth parameter, in order to build a suitable emulator of a density function. With both techniques, we always obtain a positive estimation of the marginal likelihood (a.k.a., Bayesian evidence).
An equivalent importance sampling interpretation is also described, which allows the design of extended schemes.
Several theoretical results are provided and discussed.  Numerical results show the advantage of the proposed approach, including a challenging inference problem in an astronomic dynamical model, with the goal of revealing the number of planets orbiting a star.
\newline
\newline 
{ \bf Keywords:} 
Numerical integration;  emulation;  Monte Carlo methods; Bayesian quadrature; experimental design; active learning.
\end{abstract}

%%%%%%%%%%%%%%%%%%%%%%
%%%%%%%%%%%%%%%%%%%%%%
\section{Introduction and brief overview}
\label{intro}
%%%%%%%%%%%%%%%%%%%%%
%%%%%%%%%%%%%%%%%%%%%
In this work, we consider the approximation of intractable integrals of type
\begin{align*}
I =  \int_\mathcal{X}f(\x)\post(\x)d\x,
\end{align*}
where $f(\x)$ is a generic integrable function and $\post(\x)$ is a probability density function (pdf). These integrals usually appear in Bayesian inference problems where $\post(\x)$ represents the posterior distribution of the variable of interest given the observed data. In the next subsections, we briefly review several approaches presented in the literature, which are related to the methodology presented this work.

%{\bf Numerical integration.} %\newline
%%%%%%%%%%%%%%%%%%%%%%%%%%%%
\subsection{Main families of quadrature methods}
%%%%%%%%%%%%%%%%%%%%%%%%%%%%
%random, 

With the term {\it numerical integration}, we refer to a broad family of algorithms for calculating definite integrals, and by extension, the term is also used to describe the numerical solution of differential equations. Although exact analytical solutions to integrals are always desirable, such ``unicorns'' are rarely available, specially in real-world systems. 
Indeed, many applications in signal processing, statistics, and machine learning  inevitably require the approximation of intractable integrals \cite{Burden00,Robert04,Niederreiter92}.
In particular, Bayesian methods need the computation of posterior expectations which, generally, are analytically intractable \cite{Robert04,martino2018independent}. %Consequently, numerical integration has become an indispensable tool in real-world applications, and it is far more common  to obtain solutions numerically rather than through exact mathematical expressions.  %%%See the next subsection for a brief classification of the existing so-called 
%%%%
The term numerical quadrature (or simply quadrature) is employed as a synonym for numerical integration  \cite{Burden00}.  More specifically, a quadrature formula is often stated as a weighted sum of integrand evaluations at specified points (a.k.a., nodes or knots) within the domain of integration.
% Some authors refer to numerical integration over more than one dimension as cubature; [1] others take quadrature to include higher-dimensional integration.
\newline
{\bf Deterministic quadratures.} A first family of numerical integration methods are the deterministic quadrature rules. A subclass within this family is  the Newton-Cotes quadrature rules \cite{Burden00}. The Newton-Cotes formulas are based on evaluating the integrand at equally spaced nodes and  are obtained by substituting the integrand function with a corresponding polynomial interpolation.  Smaller approximation errors can often be obtained by using the  Gaussian quadratures, where the nodes are optimally placed \cite{Burden00,liu1994note,jackel2005note}. However, their applicability is restricted to certain particular cases.
\newline
%\newline
{\bf Monte Carlo (MC) methods.} A second family is formed by stochastic quadrature rules based on MC sampling methods  \cite{Robert04,martino2018independent}, such as Markov chain Monte Carlo (MCMC) and importance sampling algorithms. In this framework, the nodes of the quadrature rules are randomly chosen. However, the resulting estimators often have a high variance, specially when the dimension of the problem grows. %generating (possibly correlated) samples distrubuted as $\post(\x)$ or  weighted samples... .
%\newline
\newline
{\bf Variance Reduction.} A third family, formed by the variance reduction techniques \cite{mcbookOwen,Robert04},  combines elements of the first two classes. In order to reduce the variance of the corresponding Monte Carlo estimators,
deterministic procedures are included within the sampling algorithms, e.g., conditioning, stratification, antithetic sampling, and
control variates \cite{mcbookOwen}. Other interesting  examples are the Riemann-based approximations which are combinations of a Riemann quadrature and random sampling  \cite[Chapter 4.3]{Robert04}.
%In the so-called variance reduction techniques
The Quasi-Monte Carlo (QMC) algorithms can be also included in this family.  In QMC,
deterministic sequences of points are generated (based on the concept of low-discrepancy) and then used as nodes of the corresponding quadrature \cite{Niederreiter92}.  Several other combinations of the previous classes above,  mixing determinism with random sampling schemes, can be found in the literature \cite{IGQ_2020,Deniz20,Liu2017BlackboxIS}. 
%\newline
\newline
{\bf Bayesian quadrature (BQ).} The BQ  framework represents a fourth approach which employs Gaussian Process (GP) regression algorithms for approximating the integrand function (and, as a consequence, the resulting integral as well) \cite{o1991Bayes,kennedy1998bayesian,rasmussen2003bayesian}. In the last years, this approach has raised the interest of several authors.  One problem with this approach is that, in some cases, a negative estimation of the marginal likelihood can be obtained. Some possible solutions have been proposed, although they are quite complex based on successive approximations \cite{osborne2012active,gunter2014sampling}. In this work, we provide two novel and much simpler alternatives for solving this issue.   
%Let us denote the integrand function as product of a generic function $f(\x)$ and the posterior density $\post(\x)$, i.e.,  $f(\x)\post(\x)$.
%considering the GP approximation either over $f(\x)$ or $\post(\x)$ (or jointly over both functions), 
Moreover, unlike this work, most contributions in BQ literature focus on the GP approximation of the function $f(\x)$ \cite{osborne2012active,gunter2014sampling,briol2019probabilistic}, although other papers on BQ describe quite general frameworks where $f(\x)$ can contain the likelihood or $\pi(\x)$  \cite{o1991Bayes,kennedy1998bayesian,rasmussen2003bayesian}. A connection between classical quadratures and BQ can be found in \cite{karvonen2017classical}. Finally, theoretical guarantees for adaptive BQ schemes can be found in the insightful work of  \cite{kanagawa2019convergence}.

%\vspace{-0.2cm}

%piecewise approx
%\newline
%{\bf Complex models and  emulation.}
%%%%%%%%%%%%%%%%%%%%%%%
\subsection{Emulation of complex models}
%%%%%%%%%%%%%%%%%%%%%%%
Many Bayesian inference problems involve the evaluation of computationally intensive models, because of {\bf (i)}  the use of particularly complex systems or {\bf (ii)}  a large number of available data (or both).
%When the model is particularly expensive (or its pointwise evaluation
%is impossible), generally two approaches are employed. 
%The first approach is the so-called approximate Bayesian computation (ABC) \cite{Beaumont10}. In the standard ABC scheme, model evaluation is substituted by evaluating a distance between the observed data and some
%artificial data generated according to the model. Therefore, ABC does not need to evaluate the model but to simulate  from it.
To overcome this issue, one possible strategy consists in replacing the true model  by a surrogate model
(a.k.a. an {\it emulator}), that could be also adaptively improved \cite{busby2009hierarchical,pratola2018optimal,Daniel2020}. Then, Bayesian inference is carried out on this approximate, cheaper model. 
\newline
%\newline
{\bf Use of the emulator.} The emulator can be applied  mainly in three different ways. {\bf (a)} One possibility is to apply  MC sampling methods considering the surrogate model as the target pdf \cite{ying2020moving,rasmussen2003gaussian}. This is used to speed up the MC algorithms. {\bf (b)} In order to improve the efficiency of MC estimators, a second option is to use the emulator as a proposal density within an MC technique, as we discuss in Section \ref{SectInterpProp} \cite{Gilks92,Gilks95,martino2018adaptive}. 
{\bf (c)}  A third possibility is to replace the true posterior with the  emulator in the integrals of interest, and computing them \cite{o1991Bayes,kennedy1998bayesian,rasmussen2003bayesian}.
Here, we mainly focus on the last approach, also combining it with MC methods (and other quadrature rules). 
%\newline
%\newline
% {\bf Remark.} Here, we focus mainly on the latter approach, but also we combine it with MC methods (and other quadrature rules). It is possible to show  (see Section \ref{ISinterp2}) that, in a particular case, the proposed approach can be also interpreted as a MC scheme where the surrogate posterior is considered as a target pdf, i.e., like the first strategy previously mentioned. %Namely, 
%%%%%%%%%%%%%%%%%%%%%%
%\newline
\newline
{\bf Construction of the emulator.} In the literature, the surrogate model is often built by using a regression algorithm, like a GP model or similar techniques \cite{butler2017measure,wang2018adaptive}. This probabilistic approach provides also uncertainty quantification that is used for estimating the approximation error and adapting the emulator \cite{stuart2018posterior}. Sometimes, the approximation regards only  some part of the model or is applied in a different domain (as the log-domain) \cite{bliznyuk2008bayesian,conrad2016accelerating,kandasamy2017query,jarvenpaa2020parallel}. Other
authors employ density estimation techniques for building the surrogate model, and then using it as a proposal density within MC algorithms \cite{chauveau2002improving,hanson2011polya,martino2017layered} or for replacing the true posterior (again within MC methods) \cite{delyon2016integral}. 
\newline
%\newline
%{\bf Remark.} We focus on an interpolative approximations of the posterior. This permits to relax the conditions over the kernel (basis) functions. For this reason, in some cases, we lose the probabilistic GP interpretation of our emulator but, at the same time, this allows us to employ more general kernel functions.

%{\bf Adaptation.} In order to build an emulator, we need a set of nodes where the true costly model is evaluated. The problem of choosing these points is treated in different parts of the statistics and machine learning literature, such as active learning  \cite{osborne2012active,wang2018adaptive,kandasamy2017query,Daniel2020}, and optimal experimental design strategies  \cite{auffray2012maximin,pronzato2012design,Pronzato17,pratola2018optimal}. Some authors consider the use of MCMC runs \cite{ying2020moving} or more sophisticated procedures combining  sampling and deterministic quadrature schemes for selecting the nodes \cite{van2020adaptive}. Other statistical tests, based on the distance between surrogate and true posterior, are also employed when the emulator is used as proposal density, as discussed below  \cite{martino2018adaptive}.   

\vspace{-0.15cm}

%%%%%%%%%%%%%%%%%%%%%%%%%%%%%%%%%%%%%%%%%%%%%%%%%
\subsection{Interpolative proposal densities within Monte Carlo schemes} \label{SectInterpProp}
%%%%%%%%%%%%%%%%%%%%%%%%%%%%%%%%%%%%%%%%%%%%%%%%%
The first use of an interpolative procedure for building a proposal density  is ascribable to the adaptive rejection sampling schemes \cite{Gilks92,Hoermann95,Gorur08rev,MartinoStatCo10}. The proposal is formed by polynomial pieces (constant, linear, etc.). Several works have proposed the use of interpolative  proposal densities within MCMC algorithms \cite{Gilks95,Meyer08,IA2RMS15,FUSS}. For more details, see also \cite[Chapters 4 and 7]{martino2018independent}.
Their use within an importance sampling scheme is considered in \cite{felip2019tree}. The adaptation is carried out considering different statistical tests, by measuring the discrepancy between the emulator  and the posterior  \cite{martino2018adaptive}.
\newline
The conditions needed for applying an emulator as an proposal density are discussed in  \cite{martino2018adaptive}. For this purpose,  we need to be able to: {\bf (a)} update the construction of the emulator, {\bf (b)} evaluate the emulator,  {\bf (c)} normalize the function defined by the emulator, and {\bf (d)}  draw samples from the emulator. It is not straightforward to find an interpolative construction which satisfies all those conditions jointly, for an arbitrary dimension of the problem. However,  the resulting algorithms (when they can be applied) provide good performance, confirming that the interpolative approach deserves more attention.

\vspace{-0.1cm}
%%%%%%%%%%%%%%%%%%%%%%
\subsection{Contributions}
%%%%%%%%%%%%%%%%%%%%%
%{\bf }
In this work, we leverage the advances in different fields of numerical integration and emulation, in order to design algorithms which build {\bf (a)} better emulators and {\bf (b)} more efficient quadrature rules.  %  combine the emulation with the numerical integration problem 
%Thus, the proposed schemes provide  two outcomes: {\bf (a)} an emulator of the posterior distribution and  {\bf (b)} an approximation of integrals involving this posterior. 
The novel algorithms are adaptive schemes which automatically select the nodes of the quadrature and of the resulting emulator. Namely,  the set of nodes used by the emulator is sequentially updated by maximizing a suitable acquisition function. Below, we list the main contributions of the work.
\newline
%$\checkmark$ 
$\bullet$ We propose a novel design of a suitable acquisition function defined as product of the posterior and a diversity term, taking into account the current positions of the nodes. Note that, unlike several  works in the literature, e.g., \cite{busby2009hierarchical,auffray2012maximin,pronzato2012design,pratola2018optimal}, we consider jointly both:  the information regarding the posterior and the distances among the current nodes. 
For the selection of the nodes, some authors also consider the use of MCMC runs \cite{ying2020moving} or more sophisticated procedures combining  sampling and deterministic quadrature schemes for selecting the nodes \cite{van2020adaptive}. Unlike \cite{ying2020moving,van2020adaptive}, our adaptive approach is based on an active learning procedure.
We also provide cheap versions of the acquisition function.
The cheap acquisition functions do not require the evaluation of the posterior but only the evaluation of the emulator.
The overall schemes are then {\it parsimonious} techniques which require the evaluation of the posterior density only at the nodes, sequentially selected by optimizing a cheap acquisition function. The proposed active learning strategy is also connected to the idea of obtaining a finite set of weighted {\it representative points}  which can summarize, in some sense, a distribution. This  topic  has gained attention in the last years \cite{Chen10,Lacoste15,SuppPoints,martino2018compressed}. 
\newline
$\bullet$  We consider an interpolative approximation of the posterior density $\post(\x)$, where  the interpolant is expressed as a linear combination of generic kernel-basis functions. Unlike several BQ techniques in \cite{osborne2012active,gunter2014sampling,briol2019probabilistic}, we approximate $\post(\x)$ instead of the function $f(\x)$ in the integral $I$. For this purpose, %differently from other works  \cite{o1991Bayes,kennedy1998bayesian,rasmussen2003bayesian}, 
we also propose the combination of the interpolant approach with  MC and other quadrature schemes. 
\newline  %chosen in advanced by the user. 
$\bullet$   With respect to other schemes in the literature \cite{kennedy1998bayesian,rasmussen2003bayesian}, our assumptions regarding the kernel-basis functions are less restrictive, e.g., they do not need to be symmetric. We could also employ different type of bases jointly, e.g., one different basis for each node.  For instance, our framework allows the use of nearest neighbors (NN) basis functions, which presents several advantages:  it does not require any matrix inversion and the coefficients of the linear combination (which defines the interpolator) are always positive \cite{karvonen2019positivity}, obtaining always a positive estimation of the marginal likelihood.  These benefits are very appealing as shown in \cite{osborne2012active,gunter2014sampling,karvonen2018fully,karvonen2019positivity}.   
%Furthermore, we provide a detailed description of two special cases: one of them considers Gaussian kernel functions and the other one considers nearest neighbor (NN) kernel functions.....
%Unlike in the Gaussian case, the NN kernels are not symmetric functions and, as a consequence, the probabilistic GP vision is not allowed.   
%In the first scenario in Section \ref{sec:GaussKernels}, we discuss the solution corresponding to $f(\x)$ being a polynomial function and, for a generic $f(\x)$, we suggest the combination with the Gauss-Hermite quadrature \cite{liu1994note,jackel2005note,IGQ_2020}. Moreover,  in this case,  the interpolant can be interpreted as a Gaussian Process (GP).  An heuristic procedure for tuning the hyper-parameters of the kernel functions is also introduced, as alternative to the marginal likelihood maximization (see Appendix \ref{App:MagicalHeuristicLuca}). 
\newline  
$\bullet$ Section \ref{ISinterp2} presents an importance sampling (IS) interpretation of the proposed schemes, where the weights involve the interpolant  instead of the true posterior density. This again shows that we can improve the Monte Carlo approximations without requiring additional evaluations of $\post(\x)$. Moreover, the alternative IS interpretation allows to design different techniques. One possible example is given in the final part of  Section \ref{ISinterp2}.  
\newline
$\bullet$ We also introduce a novel procedure for fitting the bandwidth parameter of the Gaussian kernel in order to build an {\it emulator of a density function}. In this scenario, the proposed strategy performs better than the standard maximization of the marginal likelihood of the corresponding GP.  Using this tuning procedure, we always obtain positive estimation of the marginal likelihood, even with Gaussian kernels (this is an important point; see   \cite{osborne2012active,gunter2014sampling}). 
\newline
We provide the theoretical support for the proposed methods in Section \ref{TeoSupp}. Most of the convergence results are mainly known in the scattered data approximation literature \cite{schaback1997reconstruction,wendland2004scattered,Pronzato17}.  
The efficiency of the proposed schemes is also confirmed by several numerical experiments (in Section \ref{NumEx}) with different target pdfs and dimensions of the problem. One of them is also a challenging astronomical application, where the goal is to  detect the number of exoplanets orbiting a star, and infer their orbital parameters.

%%%%%%%%%%%%%%%%%%%%%%%%%%%%%%%%%%%
%%%%%%%%%%%%%%%%%%%%%%%%%%%%%%%%%%%
\section{Interpolative quadratures for Bayesian inference}
%%%%%%%%%%%%%%%%%%%%%%%%%%%%%%%%%%%
%%%%%%%%%%%%%%%%%%%%%%%%%%%%%%%%%%%
%\subsection{Problem statement}
In many signal processing applications, the goal is to infer a variable of interest given a set of observations or measurements.
Let us denote the variable of interest by ${\bf x}\in \mathcal{X}\subseteq \mathbb{R}^{d_x}$, and let ${\bf y}\in \mathbb{R}^{d_y}$ be the observed data.
The posterior pdf is then
\begin{align*}
\bar{\pi}({\bf x})= p({\bf x}| {\bf y})= \frac{\ell({\bf y}|{\bf x}) g({\bf x})}{Z(\y)},
\label{eq:posterior}
\end{align*}
where $\ell({\bf y}|{\bf x})$ is the likelihood function, $g({\bf x})$ is the prior pdf, and $Z(\y)$ is the model evidence (a.k.a. marginal likelihood).
Generally, $Z(\y)$ is unknown, so we are able to evaluate the unnormalized target function,
\begin{equation*}
\pi({\bf x})=\ell({\bf y}|{\bf x}) g({\bf x}).
\label{eq:target}
\end{equation*}
Usually, the analytical computation of the posterior density $\bar{\pi}({\bf x}) \propto \pi({\bf x})$ is unfeasible, hence numerical approximations are required. Our goal is to approximate integrals of the form 
\begin{align}\label{eq:IntegralOfInter}
I =  \int_\mathcal{X}f(\x)\post(\x)d\x =\frac{1}{Z} \int_\mathcal{X}f(\x)\pi(\x)d\x,
\end{align}
where $f(\x)$ is some integrable function, and 
\begin{align}\label{eq:MargLike}
Z = \int_\mathcal{X}\pi(\x)d\x.
\end{align}
In the literature, random sampling or deterministic quadratures are often used \cite{martino2018independent,caflisch1998monte,Robert04}. In this work, we consider alternative quadrature rules based on an adaptive interpolative procedure.  The adaptation is obtained by applying an active learning scheme.
%
%%%%%%%%%%%%%%%%%%%%%%%%%%%%%%%%%%%
\subsection{Interpolative approach}
%%%%%%%%%%%%%%%%%%%%%%%%%%%%%%%%%%%
Let us consider a set of distinct nodes $\x_1,\dots,\x_N \in \mathcal{X}$  and some non-negative kernel or basis function,
$k(\x,\x'): \mathcal{X} \times \mathcal{X}\to \mathbb{R}^+\cup \{0\}$ (i.e., $k(\x,\x')\geq 0$). From now on, we use the terms basis or kernel as synonyms.  
The interpolant of $\pi(\x)$ is as follows
\begin{align}\label{eq:Interpolant of pi(x)}
	\widehat{\pi}(\x)=\sum_{i=1}^N\beta_ik(\x,\x_i),
\end{align}
where the coefficients $\beta_i$ must be such that $\widehat{\pi}(\x)$ interpolates the points $\pi(\x_1),\dots,\pi(\x_N)$, that is, 
$\widehat{\pi}(\x_i) = \pi(\x_i)$ for $i=1,\dots,N$.
%$\widehat{\pi}(\x_i)= \pi(\x_i)$ for $i=1,\dots,N$.
Hence, the $\beta_i$ are the solutions to the following linear system
\begin{gather}\label{Computation of coeff beta}
\left\{
\begin{split}
&\beta_1 k(\x_1,\x_1)+....+\beta_N k(\x_1,\x_N) = \pi(\x_1), \\
&\beta_1 k(\x_2,\x_1)+....+\beta_N k(\x_2,\x_N) = \pi(\x_2),  \\
& \qquad \vdots \\
& \beta_1 k(\x_N,\x_1)+....+\beta_N k(\x_N,\x_N) = \pi(\x_N).
\end{split}
\right.
\end{gather}
Denoting  $({\bf K})_{i,j}=k(\x_i,\x_j)$ ($1\leq i,j \leq N$), $\bm{\beta} = [\beta_1,\dots,\beta_N]^\top$ and ${\bf d} = [\pi(\x_1),\dots,\pi(\x_N)]^\top$, Eq. \eqref{Computation of coeff beta} can be written in matrix form as ${\bf K}\bm{\beta} = {\bf d}$. Thus, the coefficients are given by
\begin{align}\label{eq:InterpCoeffs}
\bm{\beta} = {\bf K}^{-1}{\bf d}.
\end{align}
Note that, depending on the choice of kernel and its parameters, these coefficients can be negative.
%\newline
\begin{Rem}
	%{\bf Remark.} 
	The only requirement regarding the functions $k(\x,\x')$ is that the interpolation matrix ${\bf K}$ must be non-singular (i.e., invertible) for any set of distinct nodes. The symmetry of $k(\x,\x')$ is not required. Different type of bases can be employed, for instance, one for each node ${\bf x}_i$, i.e., $k_i(\x,\x_i)$.
\end{Rem}
\begin{Rem}
	For simplicity, in this first part of the paper, we consider a fixed number of nodes $N$. However, a key point of the work is the adaptation procedure  in Section \ref{sec:adaptiveProcedure}, where new nodes are sequentially added.  
\end{Rem}
\noindent
A detailed theoretical analysis is provided in Section \ref{TeoSupp}.

%%%%%%%%%%%%%%%%%%%%%%%%%%%%%%%%%%%
\subsection{Interpolative quadrature schemes}\label{IQschemes}
%%%%%%%%%%%%%%%%%%%%%%%%%%%%%%%%%%%

We can approximate both $Z$ and $I$ by substituting the true $\pi(\x)$ with its interpolant  $\widehat{\pi}(\x)$.
\newline
%\newline
{\bf Approximation of $Z$.}
Let $\int_\mathcal{X} k(\x,\x_i)d\x=C_i> 0$ be the measure of the $i$-th kernel. An approximation of $Z$ can be obtained, by substituting Eq. \eqref{eq:Interpolant of pi(x)} in \eqref{eq:MargLike},
\begin{align}\label{estimate of Z}
\widehat{Z}=\int_{\mathcal{X}}\widehat{\pi}(\x)d\x =\sum_{i=1}^{N}\beta_i\int_{\mathcal{X}}k(\x,\x_i)d\x = \sum_{i=1}^{N}\beta_iC_i.
\end{align}
If the kernels are normalized, i.e., $C_i=1$, note that $\widehat{Z}=\sum_{i=1}^{N}\beta_i$. 
\begin{Rem}
	Although $Z>0$, $\widehat{Z}$ can take negative values, since the coefficients $\beta_i$ can be negative. However, in this work, we suggest two schemes (with Gaussian bases and a suitable tuning procedure, and with NN bases) which ensure a positive estimation of $Z$.
\end{Rem}
%\newline
\noindent
{\bf Approximation of $I$.}
%Now the interpolant $\widehat{\pi}(\x)$ can be used to approximate any expected value w.r.t. $\post(\x)$. Let $f(\x)$ be some function integrable w.r.t. $\post(\x)$, and let $I=\E_{\post}[f(\x)]$ be its posterior expectation, 
By substituting \eqref{eq:Interpolant of pi(x)} and \eqref{estimate of Z} in \eqref{eq:IntegralOfInter}, we obtain an approximation of $I$ as
\begin{align}\label{I approx by substituting pi with interpolator}
I \approx \widehat{I} = \frac{1}{\widehat{Z}}\int_\mathcal{X}f(\x)\widehat{\pi}(\x)d\x. %= \frac{1}{\widehat{Z}} \widehat{J},
%=\frac{\sum_{i=1}^{N'}\beta_i\int_\mathcal{D}f(\x)k(\x_i,\x)d\x}{\sum_{i=1}^{N'}\beta_i\int_\mathcal{D}k(\x_i,\x)d\x}.
\end{align}
Note that, given $\widehat{\pi}(\x) = \sum_{i=1}^{N}\beta_ik(\x,\x_i)$, the approximation of $I$ in \eqref{I approx by substituting pi with interpolator} 
can be expressed as
\begin{align}\label{BGHQ Integral 1}
\widehat{I} = \frac{1}{\widehat{Z}} \sum_{i=1}^{N}\beta_i\int_\mathcal{X}f(\x)k(\x,\x_i)d\x&=\frac{1}{\widehat{Z}}\sum_{i=1}^{N} \beta_i J_i,   \\
&= \frac{1}{\widehat{Z}} \sum_{i=1}^{N}\nu_i \pi(\x_i), \nonumber 
\end{align}
where $J_i=\int_\mathcal{X}f(\x)k(\x,\x_i)d\x$, $\bm{\nu}= [\nu_1,...,\nu_N]^{\top} = {\bf K}^{-1}{\bm{\zeta}}$ with ${\bm{\zeta}}=[J_1,\dots,J_N]^\top$ being the vector of integrals. Clearly, the performance of  $\widehat{I} $ depends on the discrepancy between $\widehat{\pi}(\x)$ and $\pi(\x)$, as shown by Theorem \ref{Thm:errorBoundInftyNorm}. This discrepancy is reduced by properly adding new nodes, as suggested in Section \ref{sec:adaptiveProcedure}. 

%  Note that in last equation above, we have expressed the approximation $\widehat{I} $ as a linear combination of the target evaluations. 
%\newline
%{{\bf Remark.}  sobre adaptativo y llamar a la section de teoria... key point of the work is....} 

%In the following sections, we show two different choices of kernel that allow us to compute \eqref{BGHQ Integral 1}. 
%\begin{align}\label{BGHQ Integral 2}
%\int_\mathcal{D}\widehat{\pi}(\x)d\x = \sum_{i=1}^{N}\beta_i\int_\mathcal{D}k(\x,\x_i)d\x.
%\end{align}

%\newline
%\newline

%%%%%%%%%%%%%%%%%%%%%%%%%%%%%%%%%%%
\subsection{Monte-Carlo based interpolative quadrature schemes}\label{MC_INT_quad_sect}
%%%%%%%%%%%%%%%%%%%%%%%%%%%%%%%%%%%
In this work, we assume that the evaluation of the target function $\pi(\x)$ is the main computational bottleneck \cite{busby2009hierarchical,Daniel2020}. We consider that other operations, such as sampling and evaluating different proposal densities, are negligible with respect to the target evaluation.   
The techniques, presented in this section, do not require additional target evaluations with respect to Eq. \eqref{BGHQ Integral 1}. In some specific cases, we can compute the integrals $J_i$ and $C_i$ analytically (e.g., see next section). %and if the measures $C_i$ are known, we can directly use the approximation given in Eq. \eqref{BGHQ Integral 1}. 
Otherwise, we need to approximate $J_i$, and in some cases, also $C_i$.  Some general ideas are described below.
\newline
{\bf Normalized kernels ($C_i=1$).} If the values $C_i=1$ are known,\footnote{For the sake of simplicity and without loss of generality, we assume  $C_i=1$.} we can compute $\widehat{Z}=\frac{1}{N}\sum_{n=1}^N \beta_i$. Moreover, if we are able to draw samples from each $k(\x,\x_i)$,  we have 
\begin{align}\label{eq:KernelIntegral}
J_i=\int_{\mathcal{X}}f(\x)k(\x,\x_i) d\x\approx \widehat{J}_i = \frac{1}{M} \sum_{m=1}^{M} f(\z_{i,m}),  
\end{align}
with $\z_{i,m} \sim k(\x,\x_i)$, hence
\begin{align}
\widehat{I} \approx \frac{1}{\widehat{Z}M} \sum_{i=1}^N  \beta_i \sum_{m=1}^M f(\z_{i,m}).  \label{estOtraVez1}%= \frac{1}{M}\sum_{m=1}^M\sum_{i=1}^N \bar{\beta}_i f(\z_{i,m})
\end{align}
%{ podria ser el caso que $C_i = 1$ pero no s\'e muestrear de $k(\x,\x_i)$...}
%where $\bar{\beta}_i = \frac{\beta_i}{\widehat{Z}}$.
If we know $C_i$, another possible scenario is when we are not able to draw from $k(\x,\x_i)$. In this case,  we can employ  the  importance sampling (IS)  procedure described below to approximate the integrals $J_i$.
\newline
%%%%%%%%%%%%%%%%%%%%%%%%%%%%%%%%%%%
{\bf Kernels with unknown $C_i$.} In this case, we also have to approximate  $\int_{\mathcal{X}}k(\x,\x_i)d\x = C_i$. For this purpose, we can employ IS with proposal densities $q_i(\x)$, with $i=1,...,N$, obtaining
\begin{align} \label{EqAqui_Ci}
C_i\approx %\int_{\mathcal{X}} \frac{k(\x,\x_i)}{q_i(\x)} q_i(\x)d\x = \frac{1}{M} \sum_{i=1}^{M} \frac{k(\z_{i,m},\x_i)}{q_i(\z_{i,m})} 
\widehat{C}_i= \frac{1}{M} \sum_{m=1}^{M} w_{i,m}, % \qquad i=1,...,N,
\end{align}
where the weights are $w_{i,m}=\frac{k(\z_{i,m},\x_i)}{q_i(\z_{i,m})}$   and $\z_{i,m} \sim q_i(\x)$. Moreover, we also obtain
%\quad \mbox{(en el caso nuestro es 0 o 1 tipo rejection sampling)}$
\begin{align}\label{EqAqui2_Ii}
J_i %=\int_{\mathcal{X}}f(\x) \frac{k(\x,\x_i)}{q_i(\x)} q_i(\x) d\x
\approx \widehat{J}_i = \frac{1}{M} \sum_{m=1}^{M} w_{i,m} f(\z_{i,m}).  
\end{align} 
%where again $w_{i,m}=\frac{k(\z_{i,m},\x_i)}{q(\z_{i,m})}$ and $\z_{i,m} \sim q_i(\x)$.
Replacing \eqref{EqAqui_Ci}-\eqref{EqAqui2_Ii} into \eqref{BGHQ Integral 1}, the final estimator is given by
\begin{align}
\widehat{I} &\approx\frac{1}{\sum_{i=1}^N \beta_i \sum_{m=1}^{M} w_{i,m}} \sum_{i=1}^N\beta_i \sum_{m=1}^M w_{i,m} f(\z_{i,m}), \\
& = \sum_{m=1}^M\sum_{i=1}^N \bar{\rho}_{i,m} f(\z_{i,m}), \label{estOtraVez2}
\end{align}
where $\bar{\rho}_{i,m} =\frac{  \beta_i w_{i,m}}{\sum_{j=1}^N\sum_{k=1}^{M}   \beta_j w_{j,k}}$. %and $\sum_{m=1}^{M}\sum_{i=1}^{N}\bar{\rho}_{i,m}=1$. 
\begin{Rem}
	Note that, in any of the scenarios above, we do not need to evaluate the target $\pi(\x)$ at the samples $\z_{i,m}$. Namely, we do not require additional target evaluations with respect to Section \ref{IQschemes}.  Moreover, as $M \rightarrow \infty$, the estimators in Eqs. \eqref{estOtraVez1}-\eqref{estOtraVez2} converge  to the expression \eqref{BGHQ Integral 1}, under standard MC arguments \cite{Robert04}.  
\end{Rem}
\noindent 
For further details, see the theoretical results in Section \ref{NoiseSect} and Theorems \ref{Thm:ErrorBoundFillAndSeparDists} and  \ref{Teo7}. So far we have considered Monte Carlo approaches to estimate $J_i$ and $C_i$.
Other particular and more efficient approaches (such as deterministic quadratures) are possible if we consider specific kernel functions. In the next sections, we analyze two specific cases (with Gaussian and NN kernels).

%%%%%%%%%%%%%%%%%%%%%%%%%%%%%%%%%%%%%
%%%%%%%%%%%%%%%%%%%%%%%%%%%%%%%%%%%%%
\section{Interpolation with Gaussian kernels}\label{sec:GaussKernels}
%%%%%%%%%%%%%%%%%%%%%%%%%%%%%%%%%%%%%
%%%%%%%%%%%%%%%%%%%%%%%%%%%%%%%%%%%%%
Let us consider the case of Gaussian kernels (with an  unbounded support $\mathcal{X}=\mathbb{R}^{d_x}$), 
%{\scriptsize
\begin{align}\label{eq:GaussianKernel}
&k_G(\x,\x_i) = \nonumber\\
&\frac{1}{(2\pi)^{\frac{d_x}{2}}|{\bf \Sigma}|^{\frac{1}{2}}}
\exp\left(-\frac{1}{2}(\x-\x_i)^{\top}{\bf \Sigma}^{-1}(\x-\x_i) \right ),
%k_G(\x,\x_i) = \frac{1}{(2\pi)^{\frac{d_x}{2}}|{\bf \Sigma}|^{\frac{1}{2}}}
%\exp\left(-\frac{1}{2}(\x-\x_i)^{\top}{\bf \Sigma}^{-1}(\x-\x_i) \right ),
\end{align} 
where ${\bf \Sigma}$ is a  positive definite matrix. We take ${\bf \Sigma} = h^2{\bf I}$ where $h>0$ is the bandwidth hyperparameter that needs to be tuned (see Section \ref{Sec:FromIntToReg}). 
Alternatively, note that we can also use unnormalized Gaussian kernels 
$k_G(\x,\x_i) = A\exp\left(-\frac{1}{2}(\x-\x_i)^{\top}{\bf \Sigma}^{-1}(\x-\x_i) \right )$, where $A$ is another parameter to possibly tune,
and then consider $C_i=A(2\pi)^{\frac{d_x}{2}}|{\bf \Sigma}|^{\frac{1}{2}}$.
\newline 
%\newline 
{\bf Polynomial functions $f(\x)$.} The use of Gaussian kernel functions $k_G(\x,\x_i)$ with $f(\x)$ being polynomial, {ensures that} the integrals in \eqref{BGHQ Integral 1} {are} available in closed-form.
Let ${\bf f}(\x) = \x^r = [x_1^r,\dots,x_{d_x}^r]^\top$ be componentwise powers of $\x \in \mathbb{R}^{d_x}$ ($r = 1, 2, \dots$). Then, 
\begin{align*}
J_i=\int_{\mathbb{R}^{d_x}}{\bf f}(\x)k_G(\x,\x_i) d\x= \int_{\mathbb{R}^{d_x}} \x^r k_\text{G}(\x,\x_i)d\x,
\end{align*}
corresponds to the $r$-th marginal moments of a multivariate Gaussian centered at $\x_i$.  Note that the marginal moments of a Gaussian density are well-known. Some instances are
\begin{align*}
&\int_{\mathbb{R}^{d_x}} \x k_\text{G}(\x,\x_i)d\x = \x_i   \qquad (r=1), \\
&\int_{\mathbb{R}^{d_x}} \x^2 k_\text{G}(\x,\x_i)d\x = \x_i^2 + \text{diag}({\bf \Sigma}),\qquad (r=2),  
\end{align*}
where the power $\x_i^2$ is considered a componentwise operation. Then, in this case, we can directly replace the values of $J_i$ in Eq.  \eqref{BGHQ Integral 1}.
\newline 
%\newline
{\bf Generic functions $f(\x)$.}  Each of the $N$ integrals on the right hand of \eqref{BGHQ Integral 1} may be  also approximated efficiently with a {\it  Gauss-Hermite quadrature} (GH) \cite{liu1994note,jackel2005note}, i.e.,
\begin{eqnarray*}
	%\label{eq:MultivariateGaussKernelIntegralApproxbyGH}
	\int_{\mathbb{R}^{d_x}}f(\x)k_G(\x,\x_i) d\x\approx \widehat{J}_i= \sum_{m=1}^{M} \bar{w}^\text{GH}_{m}f(\z_{i,m}),
\end{eqnarray*}
where $\bar{w}^\text{GH}_{m}$ and $\z_{i,m}$ are the weights and nodes of the GH quadrature used for $i$-th integral. Note the quadrature weights are independent of $i$ and are normalized, i.e., $\sum_{m=1}^M\bar{w}^\text{GH}_{m}=1$. Moreover, we have $\z_{i,m} = \widetilde{\z}_m + \x_i$, that is, the only difference is a translation of a single set of GH nodes $\widetilde{\z}_m$ \cite{jackel2005note} (see also the \texttt{Suppl. Material}).   Again, we do not need extra evaluations of the target $\pi(\x)$. 
Note that, with enough number of points $\z_{i,m}$, Gauss-Hermite quadrature is also exact when $f(\x)$ are polynomial functions  \cite{elvira2020importance}. Theoretical results, valid for positive definite radial basis functions, can be found in Section \ref{RBFteo}.

%%%%%%%%%%%%%%%%%%%%%%%%%%%%%%%%
\subsection{Probabilistic interpretation }\label{Sec:FromIntToReg}
%%%%%%%%%%%%%%%%%%%%%%%%%%%%%%%%%
If $k(\x,\x')=k(\x',\x)$ (i.e., it is symmetric) and $k(\x,\x')$ is semi positive definite, as in the Gaussian case, we can interpret the construction of the interpolant  $\widehat{\pi}(\x)$ as a Gaussian process (GP) \cite{rasmussen2006gaussian}.
%%% In a GP regression, we aim to make inference about an unknown function of which some observations are available.
In our setting, ${\bf d} = [\pi(\x_1),\dots,\pi(\x_N)]^\top$ represents the observed vector. The process starts by placing a GP prior on $\pi(\x)$,
$\pi(\x) \sim {\mathcal{GP}}({\bf 0}, k(\x,\x'))$,
where the GP mean is ${\bf 0}$ and $k(\x,\x')$ is the covariance function. 
Conditioning on {\bf d}, it can be shown that the posterior of $\pi(\x)$ is given by
\begin{align*}
\pi(\x) | {\bf d} \sim \mathcal{GP}(\widehat{\pi}(\x), C(\x,\x')),
\end{align*}
where the mean function is the interpolant  $\widehat{\pi}(\x)$ given in \eqref{eq:Interpolant of pi(x)}, and the posterior covariance function is $C(\x,\x')= k(\x,\x') - {\bf k}({\x})^{\top}{\bf K}^{-1}{\bf k}({\x}')$,  with 
$$
{\bf k}({\x}) = [k(\x,\x_1),\dots,k(\x,\x_N)]^\top,
$$ 
and $({\bf K})_{i,j}  = k(\x_i,\x_j)$. The variance at $\x$ is
\begin{align}\label{eq:VarOfGP}
V(\x)=C(\x,\x)= k(\x,\x) - {\bf k}({\x})^{\top}{\bf K}^{-1}{\bf k}({\x}).
\end{align}
Observe that $V(\x_i)=0$ for all $i=1,\dots,N$. If we assume that  the vector of evaluations ${\bf d}$ is noisy, we can relax the exact fit requirement by introducing a regularization term, replacing ${\bf K}$ with the matrix ${\bf K} + \sigma^2{\bf I}$, where ${\bf I}$ is an $N\times N$ identity matrix. The noise term $\sigma^2$ also provides numerical stability.
The probabilistic interpretation of the integrals involving $\pi$ is given in  Appendix \ref{App:BayesQuadview}.

%%%%%%%%%%%%%%%%%%%%%%%%%%%%%%%%%%%%
\subsection{Tuning of hyperparameters}\label{sec:hyperparTuning}
%%%%%%%%%%%%%%%%%%%%%%%%%%%%%%%%%%%%
Let us denote as $\bm{\theta}$ the vector as hyperparameters of the kernel functions $k(\x,\x')$. 
A standard way of fitting the hyperparameters $\bm{\theta}$ is to maximize the marginal likelihood of the GP \cite{rasmussen2006gaussian}. In this case, the evaluations of $\pi(\x)$ play the role of data. Given the evaluations ${\bf d} = [\pi(\x_1),\dots,\pi(\x_N)]^\top$, the marginal likelihood is given by
$p({\bf d}|\bm{\theta})=\mathcal{N}({\bf d}|{\bf 0}, {\bf K})$, and its  log-version is
\begin{align*}
%\label{eq:LogMargLikeGP}
\log p({\bf d}|\bm{\theta}) =  -\frac{1}{2}{\bf d}^\top{\bf K}^{-1}{\bf d} - \frac{1}{2}\log |{\bf K} | +c, 
\end{align*}
where $c$ is a constant. Note that ${\bf K}$ depends on $\bm{\theta}$. 
However, for fitting the bandwidth parameter $h$ of the Gaussian kernels, we propose an alternative procedure described in Appendix \ref{App:MagicalHeuristicLuca}, specifically designed for building an emulator of a {\it density function}.  In this context, the proposed procedure performs better then the maximization of $p({\bf d}|\bm{\theta})$.
\begin{Rem}
	Using the  novel tuning procedure  in Appendix \ref{App:MagicalHeuristicLuca}, the corresponding estimator $\widehat{Z}$  takes always positive values.
\end{Rem}

%%%%%%%%%%%%%%%%%%%%%%%%%%%%%%%%%%%%%%%%%%%
%%%%%%%%%%%%%%%%%%%%%%%%%%%%%%%%%%%%%%%%%%%
\section{Constant kernels based on Nearest Neighbors}\label{sec:ConstKernels}
%%%%%%%%%%%%%%%%%%%%%%%%%%%%%%%%%%%%%%%%%%%
%%%%%%%%%%%%%%%%%%%%%%%%%%%%%%%%%%%%%%%%%%%
Given the set of nodes $\{\x_i\}_{i=1}^N$ in a bounded domain $\mathcal{X}$, consider now the use of constant kernels with finite support
\begin{align}\label{eq:constantKernels}
k(\x,\x_i) = \mathbb{I}_{\mathcal{R}_i}(\x),
\end{align}
where $\mathbb{I}_{\mathcal{R}_i}(\x)$ is the indicator function in $\mathcal{R}_i$, i.e., $\mathbb{I}_{\mathcal{R}_i}(\x)=1$ for all $x \in \mathcal{R}_i$ and zero otherwise. 
Each $\mathcal{R}_i$ consists of the points $\x \in \mathcal{X}$ that are closest to $\x_i$, i.e., 
\begin{align*}
\mathcal{R}_i = \{\x\in \mathcal{X}: \norm{\x - \x_i}_p \leq \min_{j\neq i}\norm{\x - \x_j}_p  \},
\end{align*}
where $\norm{\cdot}_p$ denotes the $p$-norm.
That is, $\mathcal{X} = \cup_{i=1}^N \mathcal{R}_i$ is
the Voronoi partition of $\mathcal{X}$ using $\{\x_i\}_{i=1}^N$ as support points. In this case, solving \eqref{eq:InterpCoeffs} for the coefficients $\bm{\beta}$ is straightforward since the matrix ${\bf K}$ is the identity matrix, and thus 
\begin{align*}
\beta_i = \pi(\x_i) \enskip \text{for} \enskip i=1,\dots,N.
\end{align*}
Note that all $\beta_i \geq 0$ with this kernel. Hence the interpolant is given by
\begin{align}\label{eq:NNInterpolator}
\widehat{\pi}(\x) = \sum_{i=1}^N\pi(\x_i)\mathbb{I}_{\mathcal{R}_i}(\x).
\end{align}
Note that to evaluate $\widehat{\pi}(\x)$ at any $\x$ we need to find just the closest node. We do not need to  know the borders of regions $\{\mathcal{R}_i\}_{i=1}^N$ for this purpose. This choice of kernels has three clear advantages:
\begin{itemize}
	\item[(i)] no need to solve the linear system in \eqref{eq:InterpCoeffs} since ${\bf K} = {\bf I}$ and hence $\bm{\beta} = {\bf d}$, 
	\item[(ii)] the coefficients $\bm{\beta} = {\bf d}$ are always non-negative (this ensures that $\widehat{Z}\geq0$),
	\item [(iii)] no need of tuning the bandwidth hyperparameter. 
\end{itemize}
The difficulty, however, is determining the Voronoi partition, as well as the measures $C_i=\int_\mathcal{X}k(\x,\x_i)d\x$. We show how to address these issues in Section \ref{Sec:ApproxVoroRegions}. In this case, 
$$
C_i =\int_{\mathcal{X}}\mathbb{I}_{\mathcal{R}_i}(\x)d\x=  |\mathcal{R}_i|,
$$
where $|\mathcal{R}_i|$ denotes the measure of the $i$-th Voronoi region. The approximation of $Z$ is given by
\begin{align}\label{eq:marglikeConsKern}
\widehat{Z} = \sum_{i=1}^N\pi(\x_i)C_i,
\end{align}
%where $C_i$ is the measure of $i$-th region $\mathcal{R}_i$
and Eq. \eqref{BGHQ Integral 1} is expressed as
\begin{align}\label{eq:postExpConsKern}
\widehat{I} &=\frac{1}{\widehat{Z}} \sum_{i=1}^N\pi(\x_i)\int_{\mathcal{R}_i}f(\x)d\x, \nonumber \\
&=\frac{1}{\sum_{k=1}^N\pi(\x_k)C_k} \sum_{i=1}^N\pi(\x_i)\int_{\mathcal{R}_i}f(\x)d\x.
%	\\	{\approx \sum_{i=1}^N\pi(\x_i)f(\x_i).}
\end{align}
The convergence of this scheme is guaranteed as $N$ grows, as shown by Theorems \ref{Thm:LipsTarget} and \ref{TeoRiemann}. Further theoretical analysis are provided in Section \ref{NNteo}.
Note that we need to estimate the measures $C_i$, as well as the integrals $\int_{\mathcal{R}_i}f(\x)d\x$ to compute $\widehat{Z}$ and $\widehat{I}$. The next section is devoted to this purpose.

%%%%%%%%%%%%%%%%%%%%%%%%%%%%%%%%%%%%%%%%%
\subsection{Approximating Voronoi regions and resulting estimators}\label{Sec:ApproxVoroRegions}
%%%%%%%%%%%%%%%%%%%%%%%%%%%%%%%%%%%%%%%%%%%%%%%
%and describe two options for dealing with the integrals in the right-hand side of \eqref{eq:postExpConsKern}.
%%Obtaining analytically the Voronoi regions $\mathcal{R}_i$ and  their measures $C_i$ is not an easy task. 
In order to approximate $C_i$,  we can generate $M$ uniform vectors $\{{\bf z}_m\}_{m=1}^M$ in $\mathcal{X}$ via Monte Carlo sampling or Quasi-Monte Carlo sequences (e.g. a Sobol sequence) \cite{caflisch1998monte}. 
Define the set $\mathcal{U}_i$ as
\begin{align*}
\mathcal{U}_i &= \{{\bf z}_m: \norm{{\bf z}_m - \x_i}_p \leq \min_{j\neq i} \norm{{\bf z}_m - \x_j}_p \} \\
&=\{{\bf z}_{\ell_i} \}_{\ell_i=1}^{|\mathcal{U}_i|},
\end{align*}
i.e., the $|\mathcal{U}_i|$ vectors closest to $\x_i$ in $p$-norm, which form a discrete approximation of $\mathcal{R}_i$. Note that $\sum_{i=1}^N|\mathcal{U}_i|=M$. Hence, the measure $C_i$ can be approximated by noting that $\frac{C_i}{|\mathcal{X}|} \approx \frac{|\mathcal{U}_i|}{M}$, hence
\begin{align}\label{eq:ApproxMeasuresRi}
C_i \approx \frac{|\mathcal{U}_i|}{M}|\mathcal{X}|,
\end{align}
where $|\mathcal{X}|$ is the measure of $\mathcal{X}$. Thus, the estimator in Eq. \eqref{eq:marglikeConsKern} can be rewritten as
\begin{align}\label{eq:margLikeConsKernel_Approx}
\widehat{Z} 
%	&= \sum_{i=1}^N\pi(\x_i)\frac{L_i}{L}V_\mathcal{X} \\
&\approx\frac{|\mathcal{X}|}{M}\sum_{i=1}^N\pi(\x_i)|\mathcal{U}_i|.
\end{align}
We can also obtain an approximation of the integral $J_i=\int_{\mathcal{R}_i}f(\x)d\x$  by leveraging a QMC or MC approximation of the Voronoi regions.
Specifically, the uniform vectors ${\bf z}_{\ell_i}$ in $\mathcal{U}_i$ can be used to approximate the integral in \eqref{eq:postExpConsKern} as follows
\begin{align}\label{eq:ApproxFintViaMonteCarlo}
J_i=\int_{\mathcal{R}_i}f(\x)d\x &\approx  \frac{C_i}{|\mathcal{U}_i|}\sum_{\ell_i=1}^{|\mathcal{U}_i|}f({\bf z}_{\ell_i}) \approx  \frac{|\mathcal{X}|}{M}\sum_{\ell_i=1}^{|\mathcal{U}_i|}f({\bf z}_{\ell_i}),
\end{align}
where we used \eqref{eq:ApproxMeasuresRi} again in \eqref{eq:ApproxFintViaMonteCarlo}. The procedure above can be seen as an accept-reject method, and the estimators are also unbiased \cite[Chapter 3 and Section  6.6]{martino2018independent}.
Note that a simpler possible  approximation with one point is $J_i=\int_{\mathcal{R}_i}f(\x)d\x\approx  f(\x_i) C_i$. Thus, %a refined version of Eq. \eqref{eq:NN_FirstEstimate_with_MC} is 
replacing the expressions  \eqref{eq:margLikeConsKernel_Approx}-\eqref{eq:ApproxFintViaMonteCarlo} in  \eqref{eq:postExpConsKern}, the final estimator becomes
\begin{align}\label{eq:consKer_final_approx_I}
\widehat{I} 
&\approx \frac{1}{\sum_{k=1}^N\pi(\x_k)|\mathcal{U}_k|}
\sum_{i=1}^N\pi(\x_i)\sum_{\ell_i=1}^{|\mathcal{U}_i|}f({\bf z}_{\ell_i}).
\end{align}
%\newline
%\newline
{\bf Connection with Section \ref{MC_INT_quad_sect}}. The estimators  above can be interpreted as the application of an importance sampling (IS) scheme as described in Section \ref{MC_INT_quad_sect}, for  kernel functions with unknown $C_i$. However, unlike in Section \ref{MC_INT_quad_sect}, here we consider a unique and uniform proposal density
$$
q_i(\x)=q(\x) = \frac{1}{|\mathcal{X}|}  \mathbb{I}_\mathcal{X}(\x), \qquad \forall i=1,...,N. 
$$ 
Then, we can also remove the subindex $i$ in the sample ${\bf z}_{i,m} \sim q(\x)$, i.e., we have only $M$ samples ${\bf z}_m \sim q(\x)$.
Hence, following Eqs. \eqref{EqAqui_Ci}-\eqref{EqAqui2_Ii},  we have
\begin{align}\label{CiapproxinNN}
C_i&\approx  \frac{1}{M} \sum_{m=1}^{M} w_{i,m}, \\
J_i=\int_{\mathcal{R}_i}f(\x)d\x  &\approx \frac{1}{M} \sum_{m=1}^{M} w_{i,m} f(\z_{m}), 
\end{align}
where $\z_{m} \sim q(\x)= \frac{1}{|\mathcal{X}|}  \mathbb{I}_\mathcal{X}(\x)$, and the weights are 
\begin{gather}
w_{i,m}= \frac{k(\z_{m},\x_i)}{q(\z_{m})}=\left\{
\begin{split}
& |\mathcal{X}|  \quad \mbox{ if } \quad   \z_{m} \in \mathcal{R}_i, \\  %||\z_{m}-\x_iÊ||<  ||\z_{m}-\x_k || \quad \forall k\neq i  \\ 
& 0   \qquad \mbox{ if } \quad \z_{m} \notin \mathcal{R}_i. \\ %\mbox{ otherwise.} \\
\end{split}
\right.
\end{gather}
Replacing the expression of the weights $w_{i,m}$ into the formulas above, we recover the estimators in \eqref{eq:margLikeConsKernel_Approx} and \eqref{eq:consKer_final_approx_I}.

%%%%%%%%%%%%%%%%%%%%%%%%%%%%%%%%%%
%%%%%%%%%%%%%%%%%%%%%%%%%%%%%%%%%%
\section{An alternative IS interpretation} \label{ISinterp2}
%%%%%%%%%%%%%%%%%%%%%%%%%%%%%%%%%%
%%%%%%%%%%%%%%%%%%%%%%%%%%%%%%%%%%
In this section, we discuss a special case of the IS scheme given in Section \ref{MC_INT_quad_sect},   
when a unique proposal $q_i(\x)=q(\x)$ is employed   and only $M$ samples $\z_m\sim q(\x)$ are drawn (as already considered in the previous section). In this scenario, the IS procedure  in Section \ref{MC_INT_quad_sect} has another relevant interpretation, which allows us to design other different schemes.
Considering a generic kernel $k(\x,\x_i)$ and Eq. \eqref{CiapproxinNN},  we can rearrange $\widehat{Z} $ as
\begin{align*}
\widehat{Z} = \sum_{i=1}^{N}\beta_iC_i &\approx\sum_{i=1}^{N}\beta_i\frac{1}{M}\sum_{m=1}^Mw_{i,m} \\
&= \sum_{i=1}^{N}\beta_i\frac{1}{M}\sum_{m=1}^M\frac{k(\z_m,\x_i)}{q(\z_m)} \\
&=\frac{1}{M}\sum_{m=1}^M\frac{\sum_{i=1}^{N}\beta_ik(\z_m,\x_i)}{q(\z_m)}.
\end{align*}
Then, recalling that $\widehat{\pi}(\x)=\sum_{i=1}^{N}\beta_ik(\x,\x_i)$ and replacing this expression above, we finally obtain
\begin{align}
\widehat{Z}  &\approx \frac{1}{M}\sum_{m=1}^M\frac{\widehat{\pi}(\z_m)}{q(\z_m)} = \frac{1}{M}\sum_{m=1}^M \gamma_{m},
\end{align}
where $\gamma_{m} =\gamma(\z_{m}) =\frac{\widehat{\pi}(\z_m)}{q(\z_m)}$ for $m=1,...,M$. Moreover,  with similar steps, we can obtain
\begin{align}
\widehat{I} \approx \frac{1}{M\widehat{Z}}\sum_{m=1}^M \gamma_{m}  f({\bf z}_m), 
\end{align}
\begin{Rem}
	The weights $\gamma_m$ have the form of the standard IS weights with the target function $\widehat{\pi}$ in the numerator, and the proposal density $q$ in the denominator. Hence, the entire sampling procedure can be interpreted as a standard IS scheme where the target function is $\widehat{\pi}$ instead of $\pi$. This shows again that we do not need extra target evaluations and, hence, we can employ an arbitrary large value of $M$. 
\end{Rem}
\begin{Rem}
	Note that this result is valid for any kernel $k(\x,\x_i)$, and we use a unique proposal $q(\x)$ in the procedure described in Section \ref{MC_INT_quad_sect}.
\end{Rem}
\noindent
Below, we consider the NN case with a uniform proposal $q(\x)$, deriving the same formulas in Section \ref{Sec:ApproxVoroRegions}. 
\newline 
{\bf Uniform proposal density and NN interpolator.} Let us consider $q(\x)=\frac{1}{|\mathcal{X}|} \mathbb{I}_{\mathcal{X}}(\x)$, i.e., a uniform density in $\mathcal{X}$, and the NN kernel function. For each sample $\z_m$, the corresponding weight $\gamma_{m}$ is  
\begin{align*}
\gamma_{m}=\gamma(\z_m)=\frac{\widehat{\pi}(\z_m)}{\frac{1}{|\mathcal{X}|}}=
\frac{\pi(\x_{k_m})}{\frac{1}{|\mathcal{X}|}}=|\mathcal{X}| \pi(\x_{k_m}),
\end{align*}
where $\x_{k_m}$ is the closest node to sample ${\bf z}_m$, i.e.,  $\x_{k_m}=\arg \min_{j} \norm{{\bf z}_m - \x_j}_p$.
%\begin{align}
%\norm{{\bf z}_m- \x_{k_m}}_p \leq \min_{j\neq {k_m}} \norm{{\bf z}_m - \x_j}_p. 
%\end{align}
Then, the IS approximation of $\widehat{Z}$ is 
\begin{align*}
\widehat{Z} \approx \frac{1}{M}\sum_{m=1}^M \gamma_{m} = 
\frac{|\mathcal{X}|}{M}\sum_{m=1}^M \pi(\x_{k_m}) = \frac{|\mathcal{X}|}{M}\sum_{k=1}^{N} \pi(\x_{k})|\mathcal{U}_k|,
\end{align*}
where $|\mathcal{U}_k|$ counts the number of ${\bf z}_m$ whose closest node is $\x_k$ ($k=1,\dots,N$). Note that this expression is the same as in \eqref{eq:margLikeConsKernel_Approx}. Similarly, the IS estimate of $\widehat{I}$ is given by
\begin{align*}
\widehat{I} \approx\frac{1}{M\widehat{Z}}\sum_{m=1}^M \gamma_m f({\bf z}_m)
&=\frac{|\mathcal{X}|}{M\widehat{Z}}  \sum_{m=1}^M \pi(\x_{k_m}) f({\bf z}_m) \nonumber\\
&=\frac{|\mathcal{X}|}{M\widehat{Z}}  \sum_{k=1}^{N}  \pi(\x_{k})  \sum_{\ell_k=1}^{|\mathcal{U}_k|}f({\bf z}_{\ell_k}),
\end{align*}
which is the same expression as in \eqref{eq:consKer_final_approx_I}. However, this alternative IS interpretation allows us to design different schemes using a different proposal density, as shown below.
\newline
\noindent{\bf Gaussian mixture proposal.} We consider now an alternative to the uniform proposal in $\mathcal{X}$.  More specifically, we propose drawing $\{{\bf z}_{\ell}\}_{m=1}^M$ from a Gaussian mixture proposal pdf built considering the set of nodes $\{\x_i\}_{i=1}^N$, i.e.,
\begin{align*}
{\bf z}_m \sim q(\x) = \sum_{i=1}^{N} \xi_i \mathcal{N}(\x|\x_i,{\bf C}_i),
\end{align*}
where the mixture weights $ \xi_i $ are 
\begin{align*}
\xi_i = \frac{\pi(\x_i)}{\sum_{n=1}^{N} \pi(\x_n)}, \qquad i=1,...,N,
\end{align*}
and the covariances ${\bf C}_i$ can be determined by the minimum distance of $\x_i$ to its closest node. In this case, the IS weights are given by
\begin{align*}
\gamma({\bf z}_m) = \frac{\widehat{\pi}({\bf z}_m)}{\sum_{i=1}^{N}\xi_i \mathcal{N}({\bf z}_m|\x_i,{\bf C}_i)}= \frac{\pi(\x_{k_m})}{\sum_{i=1}^{N}\xi_i \mathcal{N}({\bf z}_m|\x_i,{\bf C}_i)},
\end{align*}
where $\x_{k_m}$ is the closest node to ${\bf z}_m$, with $m=1,\dots,M$. %This scheme can have better performance than the method described in the previous section, specially in high dimensional spaces.

%%%%%%%%%%%%%%%%%%%%%%%%%%%%%%
%%%% FIGURA BANANAS NN_AQ %%%%
%%%%%%%%%%%%%%%%%%%%%%%%%%%%%%
\begin{figure*}[!t]
	\centering
	%	\centerline{
	\subfigure[True target pdf]{\includegraphics[width=0.23\textwidth]{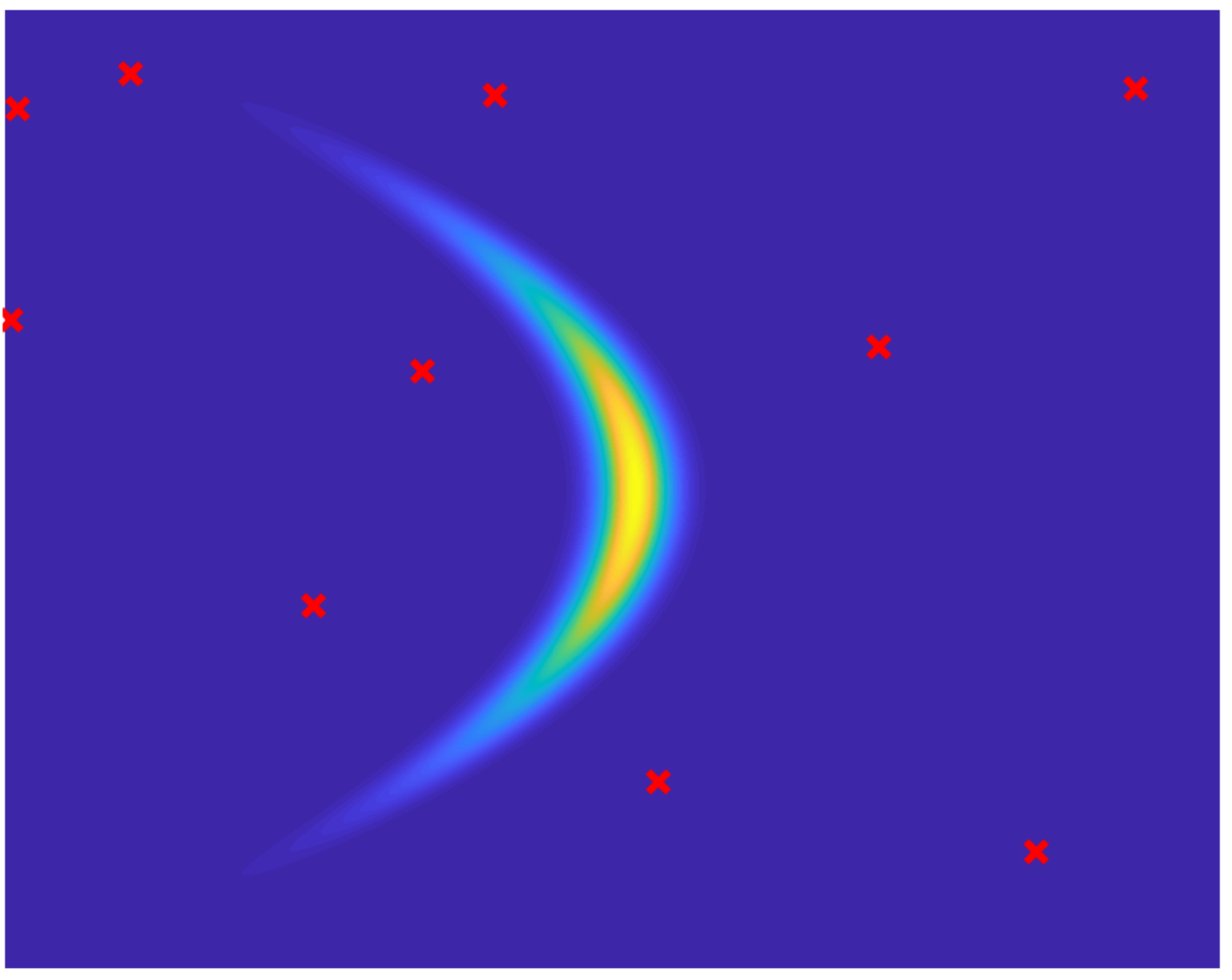}}
	\subfigure[$E=50$]{\includegraphics[width=0.23\textwidth]{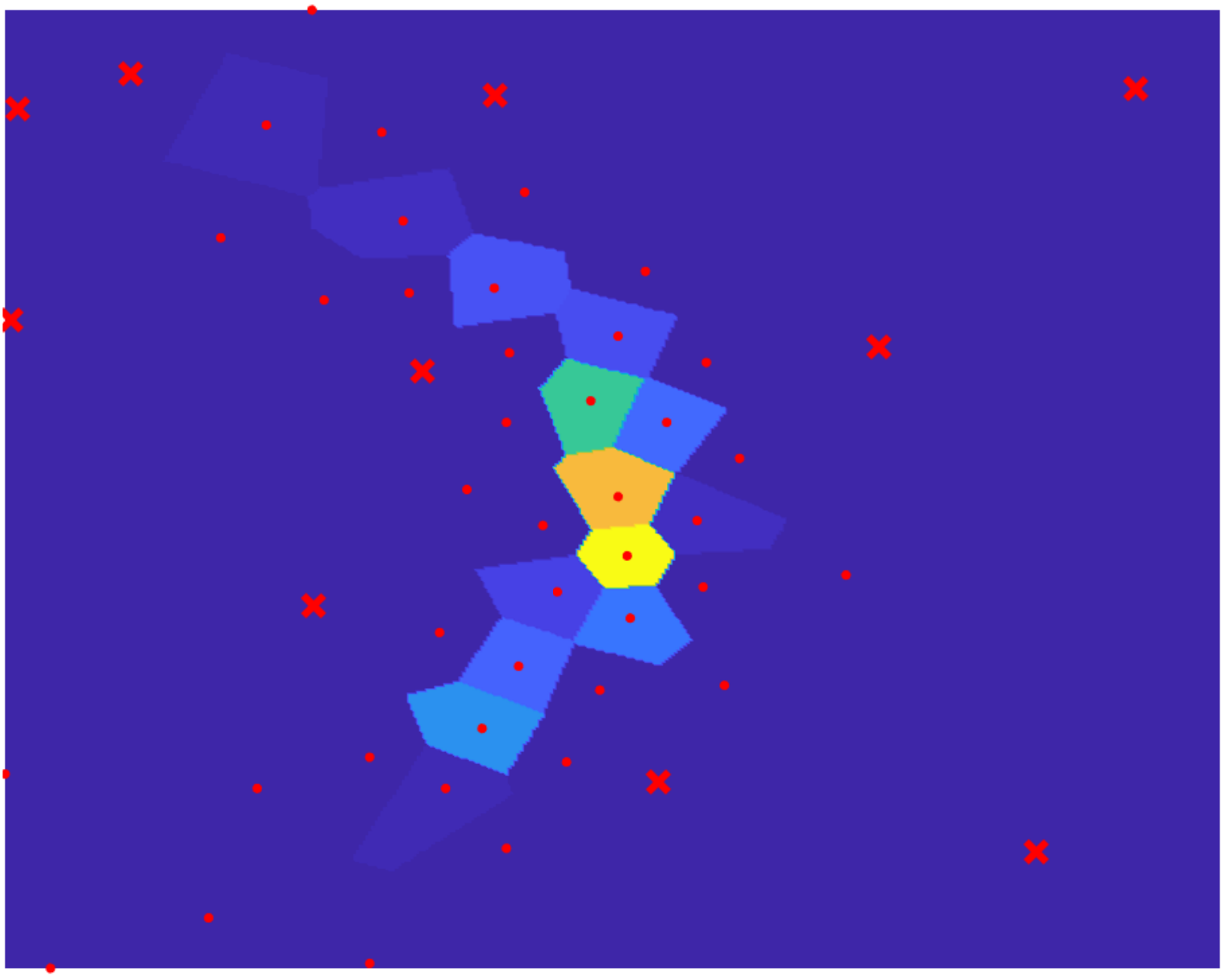}}
	\subfigure[$E=250$]{\includegraphics[width=0.23\textwidth]{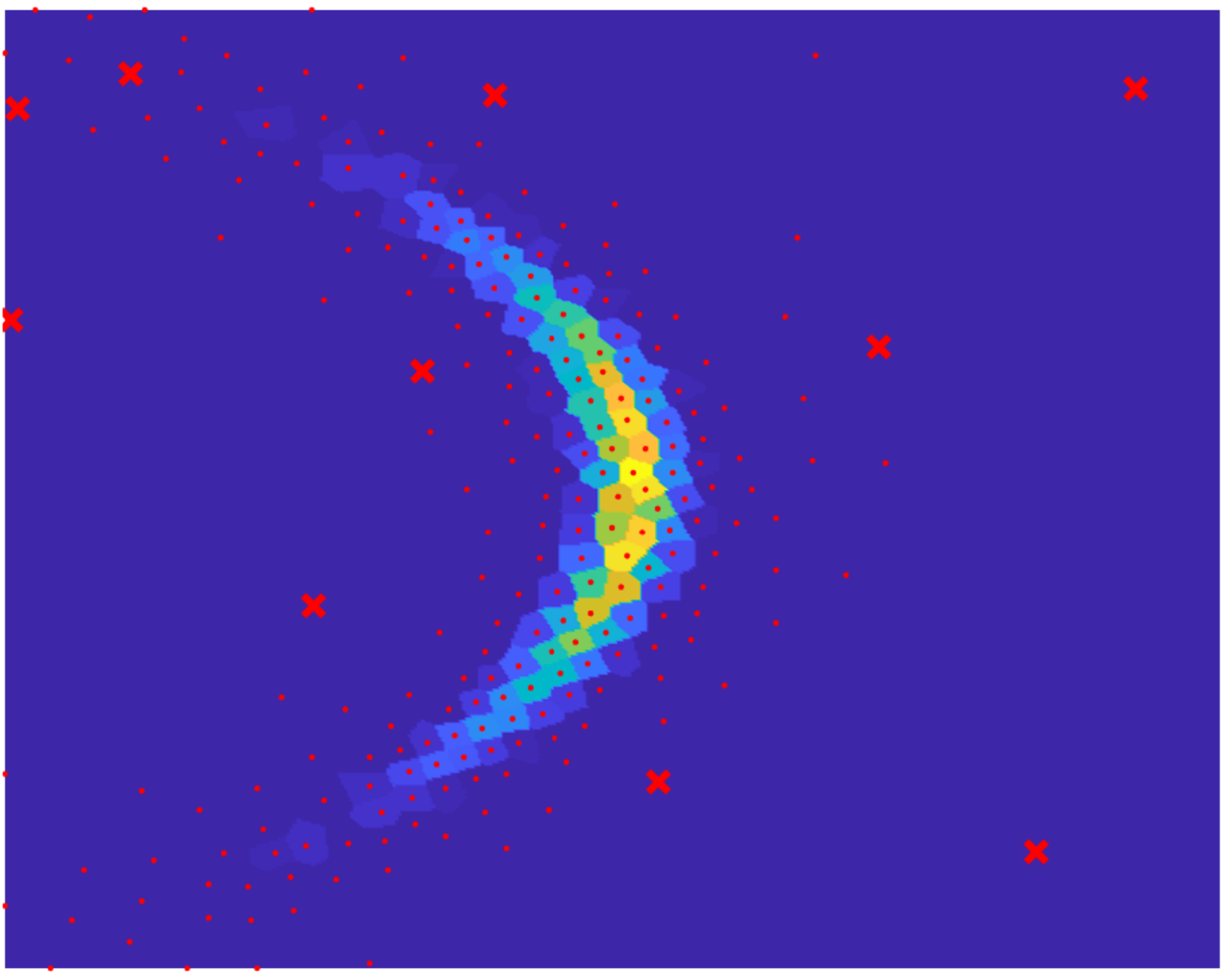}}
	\subfigure[$E=10^3$]{\includegraphics[width=0.23\textwidth]{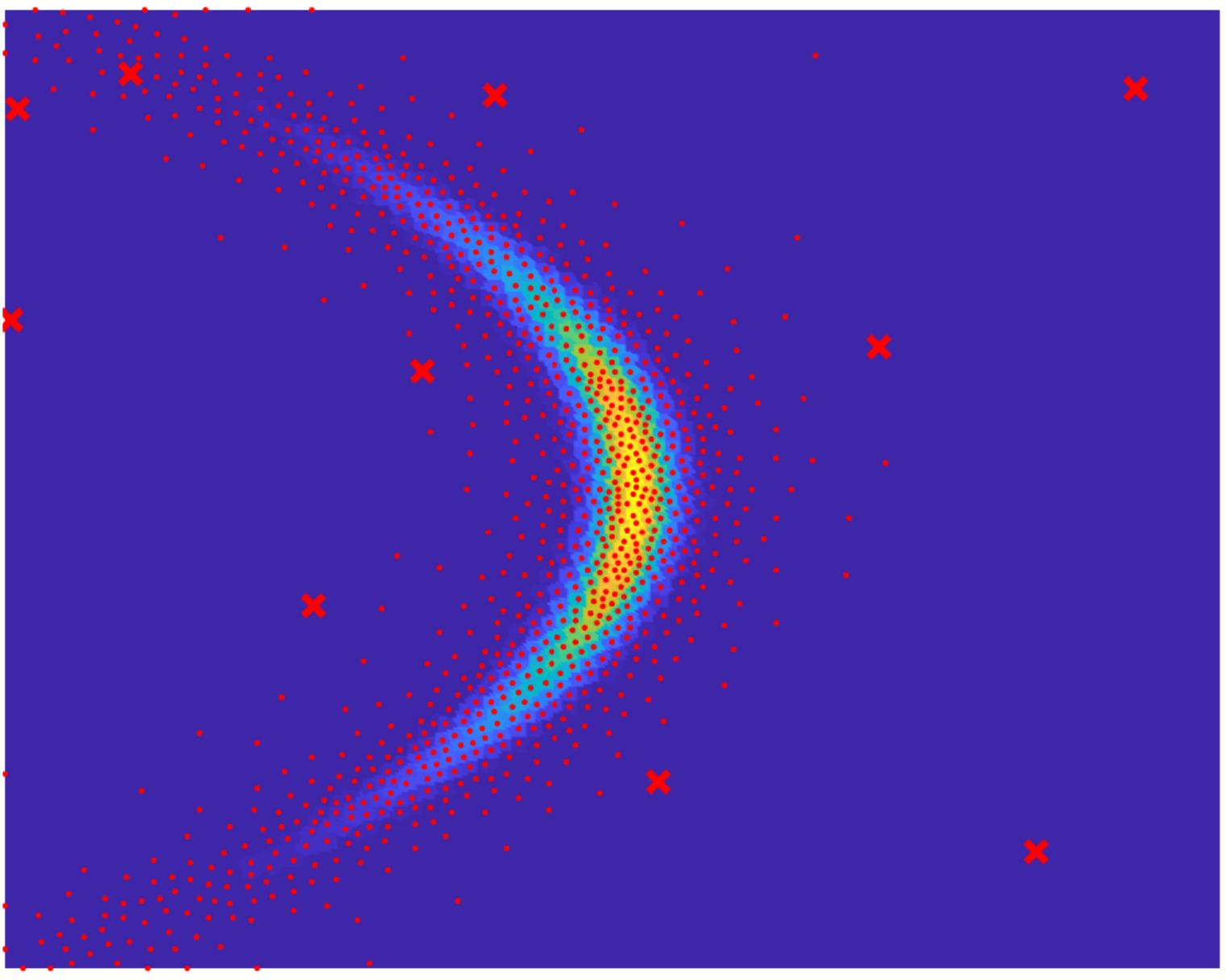}}
	%	}
	\vspace{-0.3cm}
	\caption{\footnotesize Example of application of NN-AQ. The cross-marks represent the starting nodes, while the points added adaptively by NN-AQ are shown with dots. {\bf (a)} The banana-shaped target  and the starting nodes. {\bf (b)}-{\bf (c)}-{\bf (d)} The NN-AQ emulator with $E=50,250,10^3$ number of target evaluations.  }
	\label{fig_bananas}
\end{figure*}

%%%%%%%%%%%%%%%%%%%%%%%%%
%%%%%%%%%%%%%%%%%%%%%%%%%
\section{Adaptive procedure}\label{sec:adaptiveProcedure}
%%%%%%%%%%%%%%%%%%%%%%%%%
%%%%%%%%%%%%%%%%%%%%%%%%%
In this section, we present an adaptive mechanism to add new nodes to the interpolant.
Our algorithm {adds} nodes sequentially with the aim to discover high-valued regions of $\pi(\x)$ while fostering the exploration of the state space. We employ an active learning procedure where a new point is obtained by maximizing a suitable acquisition function. 
%Let $t$ denote an iteration index. 
The resulting adaptive algorithm is shown in Table \ref{table_general_algorithm}. Note that the final number of nodes is $N_T=T+N_0$.
The adaptive quadrature scheme based on the Gaussian kernels is denoted as GK-AQ, whereas the other scheme based on the Nearest Neighbors (NN) kernels is denoted as NN-AQ. Figure \ref{fig_bananas} depicts an example of application of {the} NN-AQ.

%%%%%%%%%%%%%%%%%%%%%%%%%%%%%%%
\subsection{Building suitable acquisition functions} \label{sec:adaptiveProcedure2}
%%%%%%%%%%%%%%%%%%%%%%%%%%%%%%
Let us denote as $t \in \mathbb{N}$ the {$t$th} iteration of the algorithm.
In the update stage, we decide to add a new node where the acquisition function, $A_t: \mathcal{X} \to \{0\}\cup\mathbb{R}^+$, is maximum. The acquisition function takes into account the shape of ${\pi}(\x)$ and the spatial distribution of the current nodes. More specifically, it must fulfill 
\begin{align*}
A_t(\x_i) = 0 \enskip  \text{for all }  t \mbox{ and } \enskip i=1,\dots,N_t,
\end{align*}
and grow as we move apart from the nodes. We consider acquisition functions $A_t(\x)$ of the form
\begin{align}\label{eq:AcWithPi}
A_t(\x) =\pi(\x)D_t(\x),
\end{align}
where $D_t(\x)$ is a diversity term that  penalizes the proximity to the current nodes.  Note that the information of $f(\x)$ could be also included as $A_t(\x) =f(\x)\pi(\x)D_t(\x)$. In some settings, the function $A_t(\x)$ above could be directly used after choosing a diversity term $D_t(\x)$. However, in this work, we consider that evaluating $\pi(\x)$ is costly, so we propose cheaper versions of \eqref{eq:AcWithPi}.

%%%%%%%%%%%%%%%%%%%%%%%%%%%%%%%%%%%%%%%%%%%%%%%%%%%%%%%%%
%%%%%%%%%% TABLA %%%%%%%%%%%%%%%%%%%%%%%%%%%%%%%%%%%%%%%%
\begin{table}[!t]
	%	\centering
	%\small
	\caption{\textbf{Adaptive Quadrature algorithm.}}
	%\vspace{-0.2cm}
	\begin{tabular}{|p{0.95\columnwidth}|}
		\hline
		%\footnotesize
		%\newline
		\vspace{0.1cm}
		{\bf Initialization:} Set $N_0$ initial nodes and set ${\bf X}_0 = \{\x_1,\dots,\x_{N_0}\}$, ${\bf d}_0 = [\pi(\x_1),\dots,\pi(\x_{N_0})]^\top$.
		\newline
		\newline 
		{\bf For $t=0,\ldots,T$:}
		\begin{enumerate}
			\item {\it Build the interpolator}. Use the set ${\bf X}_t= \{\x_1,\dots,\x_{N_t}\}$  and corresponding evaluations ${\bf d}_t = [\pi(\x_1),\dots,\pi(\x_{N_t})]^\top$  to build $\widehat{\pi}_t(\x)$ using Gaussian kernels (see Section \ref{sec:GaussKernels}) or constant kernels (see Section \ref{sec:ConstKernels}).
			
			\item {\it Build the acquisition function}. Use $\widehat{\pi}_t(\x)$ and the set of current nodes ${\bf X}_t$ to build the acquisition function $A_t(\x)$, e.g., Eqs. \eqref{eq:AcforGauss}-\eqref{eq:AcforConst}.
			
			%		\item {\it Estimation stage}. Obtain estimates $\widehat{Z}_t$ and $\widehat{I}_t$.
			
			\item {\it Update stage}. Obtain new node $\x_{N_t + 1}$ by 
			\begin{align}\label{eq:MaximizationOfAc}
			\x_{N_t+1} = \arg \max_{\x\in\mathcal{X}} A_t(\x),
			\end{align}
			%		Sample a new node $\x'$: iterate some steps of a MCMC algorithm (e.g. Metropolis-Hastings algorithm) with target $V_t(\x)\pi(\x)$ and store the last state $\x'$. 
			append ${\bf X}_{t+1} = \{{\bf X}_t, \x_{N_t+1}\}$ and ${\bf d}_{t+1} = [{\bf d}_t,\pi(\x_{N_t+1})]^\top$.
			
			\hspace{-6mm}{\bf Outputs:} Build the final interpolant  $\widehat{\pi}_T(\x)$ and obtain the approximations $\widehat{I}$ and $\widehat{Z}$. 
			\vspace{0.15cm}	
		\end{enumerate} \\
		\hline 
	\end{tabular}
	\label{table_general_algorithm}
\end{table}
%%%%%%%%%% TABLA %%%%%%%%%%%%%%%%%%%%%%%%%%%%%%%%%%%%%%%%
%%%%%%%%%%%%%%%%%%%%%%%%%%%%%%%%%%%%%%%%%%%%%%%%%%%%%%%%%

%%%%%%%%%%%%%%%%%%%%%%%%%%%%%%
%%%% FIGURA AC's en 1D %%%%%%%
%%%%%%%%%%%%%%%%%%%%%%%%%%%%%%
\begin{figure*}[!t]
	\centering
	%	\centerline{
	\subfigure[Initial state]{\includegraphics[width=0.23\textwidth]{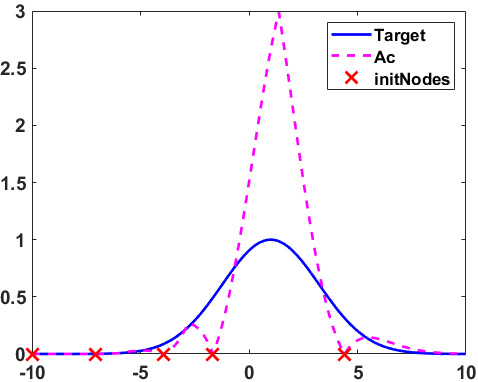}}
	\subfigure[Iteration 1]{\includegraphics[width=0.23\textwidth]{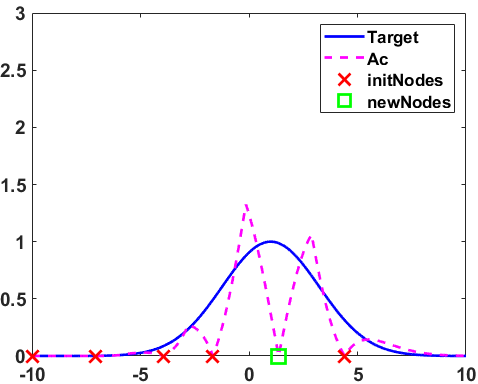}}
	\subfigure[Iteration 2]{\includegraphics[width=0.23\textwidth]{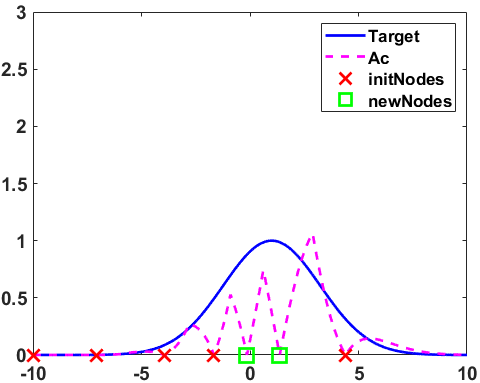}}
	\subfigure[Iteration 3]{\includegraphics[width=0.23\textwidth]{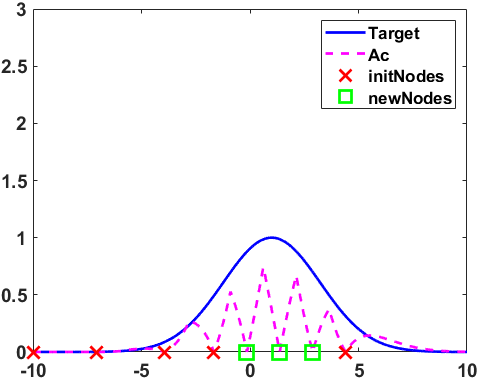}}
	%	}
	\vspace{-0.3cm}
	\caption{\footnotesize 1D example of application of $A_t(\x) = \pi(\x)D_t(\x)$ with the diversity term $D_t(\x)=\min\limits_{i=1,\dots,N_t}|x-x_i|$. At each iteration, the new node, shown with a green square, is added where $A_t(x)$ is maximum. }
	\label{fig:AcquFunDemo}
\end{figure*}
%%%%%%%%%%%%%%%%%%%%%%%%%%%%%%
%%%%%%%%%%%%%%%%%%%%%%%%%%%%%%

%%%%%%%%%%%%%%%%%%%%%%%%%%%%%%%%%%%%%%%%%%
\subsection{Cheap acquisition functions}\label{sec:adaptiveProcedure3}
%%%%%%%%%%%%%%%%%%%%%%%%%%%%%%%%%%%%%%%%
%{\bf Cheap acquisition functions.}  
We recall that the most costly step is the evaluation of the target function $\pi(\x)$. This is often due to the use of complex models and/or large amounts of data. For that reason, we propose a cheap type of $A_t(\x)$,
\begin{align}\label{eq:AcWithInter}
A_t(\x) =\widehat{\pi}_t(\x)D_t(\x),
\end{align}
so that no evaluations of the true $\pi(\x)$ are required. In this case, in terms of posterior evaluations $E$, the cost of the overall algorithm in Table \ref{table_general_algorithm},  is $E=N_0+T$. 
%The acquisition function in \eqref{eq:AcWithInter}
\begin{Rem}
	The particular case $A_t(\x)=D_t(\x)$ corresponds to the space-filling experimental designs (e.g., see \cite{pronzato2012design,Pronzato17,muller2001coffee} and Theorem \ref{teo3}). In the other particular case with $A_t(\x)=\widehat{\pi}_t(\x)$, the resulting schemes are similar to other approaches in literature which combine sampling and optimization (e.g., see \cite{Deniz20}).
\end{Rem}
\noindent
In the Gaussian kernel scenario, we may use the variance in \eqref{eq:VarOfGP} as diversity term
\begin{align}\label{eq:AcforGauss}
A_t(\x) = \widehat{\pi}_t(\x)V_t(\x),
\end{align}
where we have set $D_t(\x)=V_t(\x)$, that fulfills $V_t(\x_i)=0$ for $i=1,\dots,N_t$. 
This choice is motivated by the fact that the approximation error is bounded by the maximum value of $V_t(\x)$ (e.g., see Theorem \ref{Thm:ErrorBoundGPVar}).
Since the function $V_t(\x)$ is unfeasible with constant NN kernels, we suggest a diversity term of the form
\begin{align}\label{eq:AcforConst}
A_t(\x) = \widehat{\pi}_t(\x)\min_{i=1,\dots,N_t} \norm{\x - \x_i}_p.
\end{align}
Note that the term $D_t(\x)=\min\limits_{i=1,\dots,N_t} \norm{\x - \x_i}_p$ is zero when evaluated at any current node: for each $\x_j \in {\bf X}_t$ the minimum distance is w.r.t. itself, which is zero. 
This choice is motivated by Theorem \ref{teo3}, since  the approximation error is also bounded by the maximum value of $D_t(\x)$. %Indeed, the maximum of $D_t(\x)=\min\limits_{i=1,\dots,N_t} \norm{\x - \x_i}_p$ corresponds to the so-called ``fill distance'', which measures the overall performance of the interpolation and should be as low as possible.
Figure \ref{fig:AcquFunDemo} depicts an example with this choice of $D_t(\x)$.  Note that the choice $D_t(\x)=\min\limits_{i=1,\dots,N_t} \norm{\x - \x_i}_p$ can be also employed in the Gaussian kernel scenario.
\newline
\newline
Another alternative is to consider tempering versions of the acquisition function, 
\begin{align}\label{temperedA}
A_t(\x) = \left[\widehat{\pi}_t(\x)\right]^\alpha \left[D_t(\x)\right]^\beta,
\end{align}
where $\alpha \geq 0$ can be used to prioritize moving towards high-valued zones of $\widehat{\pi}_t(\x)$, while $\beta \geq 0$ to encourage exploration.  The values $\alpha$ and $\beta$ can also vary with the iteration $t$. The maximization of $A_t(\x)$ can be performed by simulated annealing or other optimization techniques. The performance of different acquisition functions have been compared in Figure \ref{fig_SIMU2} (see Section \ref{sec:bananaTar}). One can observe that maximizing the proposed acquisition functions provides much better results than adding uniformly random nodes.
\newline
{\bf Observations.} For the GK-AQ algorithm, the most costly step corresponds to the inversion of the  $N_t \times N_t$ matrix  ${\bf K}_t$, needed to be done in order to build the acquisition function in Eq. \eqref{eq:AcforGauss}. Note that the inverse ${\bf K}_t^{-1}$ is used for both 
%the coefficients, $\bm{\beta}^{(t)} = ({\bf K}^{(t)})^{-1}{\bf d}_t$, and thus for 
evaluating the interpolant  $\widehat{\pi}_t(\x)$ and computing the variance $V_t(\x)$. 
We can alleviate the cost of this step by building ${\bf K}_{t}^{-1}$ iteratively from ${\bf K}_{t-1}^{-1}$. The recursion formula is given in Appendix \ref{Inverse of bordered matrix}.
In the case of NN-AQ, evaluating the acquisition function in \eqref{eq:AcforConst} requires only to calculate the distances with respect to each node. This computation can be used for both evaluating the interpolant and the diversity term $D_t(\x) = \min%\limits_{i=1,\dots,N_t}
\norm{\x - \x_i}_p$. Note that the cost of searching for the nearest neighbor has only a weak dependence on the dimension of the space. 
%\newline

%%%%%%%%%%%%%%%%%%%%%%%%%%%%%%%%%%%%%%%%%%
%%%%%%%%%%%%%%%%%%%%%%%%%%%%%%%%%%%%%%%%%%
\section{Theoretical support} \label{TeoSupp}
%%%%%%%%%%%%%%%%%%%%%%%%%%%%%%%%%%%%%%%%%%
%%%%%%%%%%%%%%%%%%%%%%%%%%%%%%%%%%%%%%%%%%

In this section, we provide some theoretical results supporting the proposed schemes.
% First, we show that the approximate integral (the one based on the interpolant rather than the true target) converges to the integral of interest as the interpolant gets closer to the target. 
%Let , known up to a normalizing constant.
We consider  $\bar{\pi}(\x)=\frac{1}{Z} \pi(\x)$ a bounded target pdf and a bounded domain $\mathcal{X}\subset\mathbb{R}^{d_x}$.  Let also $f(\x): \mathcal{X} \to \mathbb{R}$ be an integrable function.
In this section, we consider $J = \int_\mathcal{X}f(\x)\pi(\x)d\x$ as the integral of interest. For a generic $f(\x)$,  $J$ corresponds to the numerator of the integral $I$ in Eq. \eqref{eq:IntegralOfInter}. For $f(\x)=1$, $J$ becomes the normalizing constant of $\pi(\x)$, i.e., $J=Z$, which is the denominator of $I$. Thus, working with $J$ is equivalent to working with $I$.  Let also $ \widetilde{J} = \int_\mathcal{X}f(\x)\widehat{\pi}(\x)d\x$, 
be the approximation of $J$ given by substituting the interpolant  $\widehat{\pi}(\x)$. A first general result valid for any interpolation procedure is given below. 
\begin{thm}\label{Thm:errorBoundInftyNorm}
	The error incurred by substituting $\pi(\x)$ with $\widehat{\pi}(\x)$ in $J$ is bounded, 
	\begin{align*}
	|J - \widetilde{J}| 
	&\leq \norm{f(\pi - \widehat{\pi})}_1 \\
	&\leq \norm{f}_2 \norm{\pi -  \widehat{\pi}}_2 \\
	& \leq |\mathcal{X}| \norm{f}_\infty\norm{\pi - \widehat{\pi}}_\infty,		
	\end{align*}
	where $ \norm{\cdot}_1$, $\norm{\cdot}_2$ and $\norm{\cdot}_\infty$ denote the $L^1$, $L^2$ and $L^\infty$ norms respectively. 
	\begin{proof}
		See Appendix \ref{Teo1}.
	\end{proof}
\end{thm}
\noindent  Therefore, if are able to build an interpolant  $\widehat{\pi}$ in a way such  $\norm{\pi -\widehat{\pi}}_\infty$ vanishes to zero, then the approximation $\widetilde{J}$ will converge to $J$. Note that, in this section, we ensure the convergence of numerator $J$ and denominator $Z$ of $I = \frac{J}{Z}$, independently.  A complete treatment (yet more complicated) should consider the convergence of the two quantities at the same time.
%{ 
%\begin{Rem}
%Note that, in the general case,  the proposed schemes contain two sources of error: one due to the substitution of $\post$ with $\widehat{\pi}$ and, the second one is due to the Monte Carlo (or other) approximations. The first source is controlled by the adaptive process (with a suitable selection of additional nodes), and last one is reduced by increasing $M$ (which does not require extra evaluation of the true posterior $\post$). 
%\end{Rem}}
%\noindent
For the rest of results, we need to distinguish between the case of Gaussian kernel and constant kernel interpolators. To establish convergence of both schemes we need to make some preliminary definitions  and considerations.

\vspace{-0.15cm}

%%%%%%%%%%%%%%%%%%%%%%%%%%%%%%%%%%%%%
\subsection{Space-filling measures and related results}
%%%%%%%%%%%%%%%%%%%%%%%%%%%%%%%%%%%%%
We introduce two well-known measures of dispersion widely employed in the function approximation literature. In this section, we always consider a bounded support $\mathcal{X}$. 
\newline
{\bf  Fill distance.} Given the set of nodes $\{\x_i\}_{i=1}^N\subset \mathcal{X}$, let us define the following quantity
\begin{align}\label{eq:fillDist}
r = \max_{\x \in \mathcal{X}} \min_{1\leq i \leq N}\norm{\x - \x_i}_2,
\end{align}
which is the fill distance. 
\newline
{\bf Separation distance.}  The separation distance is defined as
\begin{align}\label{eq:SeparDist}
s = \min_{i\neq j} \norm{\x_i-\x_j}_2,
\end{align}
i.e., the minimal distance between two nodes.  Note that $s \leq 2r$. Having a small $s$ increases the numerical instability and can have a detrimental effect in the error bounds. 
%\newline
The adaptive procedure described in Sect. \ref{sec:adaptiveProcedure} produces a sequence of nodes that sequentially minimizes $r$.
\begin{propo}\label{Fillprop}
	Consider the acquisition function given in Eq. \eqref{temperedA} with $\alpha=0$ and $\beta=1$, and the choice 
	$A_t(\x) = \min\limits_{i=1,\dots,N_t} \norm{\x - \x_i}_2$,
	%	\end{align}
	where $\{\x_i\}_{i=1}^{N_t}$ are the current nodes of the interpolator. The maximum of this function is the fill distance $r_t$ in Eq. \eqref{eq:fillDist}, at iteration $t$.
	Adding the point $\x_{N_t + 1}$ corresponding to $r_t$ to the set of current nodes ensures that 
	\begin{align*}
	r_{t+1} = \max \min_{i=1,\dots,N_{t+1}} \norm{\x - \x_i}_2 \leq r_t,
	\end{align*} 
	and that $r_t \to 0$ when $t\to \infty$. 
	\vspace{-0.2cm}
	\begin{proof} 
		See  Sect. 4.1 in \cite{Pronzato17} and \cite{auffray2012maximin}. This procedure is related to the ``coffee house design'' in \cite{muller2001coffee}. 
	\end{proof}
\end{propo}
\begin{propo}\label{Thm:loDelFinlandes_Cojonen}
	For isotropic kernels,  the variance function  $V(\x)$ given in Eq. \eqref{eq:VarOfGP} satisfies that $\max\limits_{\x \in \mathcal{X}} [V(\x)]^{\frac{1}{2}} \leq \Phi(r)$, where $\Phi(r)$ is an increasing function of $r$, depending on the kernel function.  In the case of Gaussian kernels, $\Phi(r)$ is an exponential function.
	\vspace{-0.2cm}
	\begin{proof}
		See Sect. 2.1 in \cite{Pronzato17} and Sect. 2 in \cite{auffray2012maximin}. 
		%		See also \cite{Schaback1995}.
	\end{proof}
\end{propo}
\begin{propo}\label{PropV}
	Consider the acquisition function given in Eq. \eqref{temperedA} with $\alpha=0$ and $\beta=1$, i.e.,  and the choice 
	$A_t(\x) = V_t(\x)$.  Let us set also $\varphi_t=\max\limits_{\x \in \mathcal{X}}V_t(\x)$. By adding new nodes according to the rule 
	$$
	\x_{N_t+1} = \arg \max A_t(\x),
	$$ 
	we  are minimizing $\varphi_t$ over the iterations $t$, i.e., $\varphi_t$ is a non-increasing function of $t$ and $\varphi_t \rightarrow 0$ as $t \rightarrow \infty$. %For isotropic kernels, such as the Gaussian kernel, we have $\max\limits_{\x \in \mathcal{X}}[V(\x)]^{\frac{1}{2}} \leq \Phi(r)$, where $\Phi$ is an increasing function of $r$ { citar \cite{Pronzato2017}.}	
	%	 \begin{proof}
	%	  From Propositions \ref{Fillprop}  and \ref{Thm:loDelFinlandes_Cojonen},  $r_t\rightarrow 0$ as $t \rightarrow \infty$, then $\Phi(r_t)\rightarrow 0$ and $\varphi_t\rightarrow 0$ as $t \rightarrow \infty$.  
	%	 \end{proof}
	%\vspace{-0.2cm}
	\begin{proof}
		This algorithm is known as $p$-greedy algorithm in \cite{santin2017convergence}. See the behavior of the variance of a GP interpolant  \cite{rasmussen2006gaussian}. This acquisition function is commonly used in the kriging literature. For instance, see \cite{mackay1992information} and \cite{pronzato2012design}. 
		%paper ESPECTACULAR
	\end{proof}
\end{propo}
\begin{propo}\label{PropNN}
	Consider the acquisition function given in Eq. \eqref{eq:AcforConst} with $\alpha=0$ and $\beta=1$, and the choice
	$A_t(\x) = \min\limits_{i=1,\dots,N_t} \norm{\x - \x_i}_2$,
	where $\{\x_i\}_{i=1}^{N_t}$ are the current nodes of the interpolator. The sequence of nodes obtained as 
	%	$\{\x_i\}_{i=1}^{N_t}$, where $\x_{N_t}$ is obtained by taking $\arg \max A_{t-1}(\x)$ ($t=1,2,\dots$), 
	$\x_{N_t+1} = \arg \max A_t(\x)$, for $t\in \mathbb{N}^+$,  is a  uniform low-discrepancy sequence in a bounded $\mathcal{X}$  \cite{niederreiter1992random}. 
%	\vspace{-0.5cm}
	\begin{proof}
		This  procedure can be  interpreted as deterministic and sequential version of the well-known latin hypercube sampling (LHS) \cite{niederreiter1992random}.  %If 
		%{ frase o reference de Cantor sets... quiza decir que escogiendo una init en grid entonces el algoritmo procede completandolo y luego haciendolo mas fino?} 
	\end{proof}
\end{propo}

\begin{Rem}
	Note that the proposed schemes do not need that the space is covered uniformly. The only requirement, for decreasing the fill distance $r$, is to be able to reach any subset of the domain $\mathcal{X}$ with a non-null probability (strictly positive). 
\end{Rem}
%%%%%%%%%%%%%%%%%%%%%%%%%%%%%%%%%%%%%%%%%%
\subsection{Results for interpolators based on radial basis functions (RBFs)} \label{RBFteo}
%%%%%%%%%%%%%%%%%%%%%%%%%%%%%%%%%%%%%%%%%%%%
In this section, we consider that $k(\x,\x')$ is the Gaussian kernel considered in Sect. \ref{sec:GaussKernels}. More generally, the results from this section are valid for any $k(\x,\x')$ that is a (positive definite) radial basis function (RBF).

%%%--------------------------------%%%
\subsubsection{Exact computation of $J_i$}
%%%--------------------------------%%%
Recall $\widehat{\pi}(\x) = \sum_{i=1}^N\beta_i k(\x,\x_i)$, where the weights are $\bm{\beta} = [\beta_1,\dots, \beta_N] = {\bf K}^{-1}{\bf d}$ using the interpolation matrix ${\bf K}$ and the vector of target evaluations ${\bf d}$.
% = [\pi(\x_1),\dots, \pi(\x_N)]^\top$. 
The approximation $\widetilde{J}$ can be written as  
\begin{align*}
\widetilde{J} &= \int_\mathcal{X}f(\x)\widehat{\pi}(\x)d\x 
%	&=\sum_{i=1}^N\beta_i\int_{\mathcal{X}}f(\x)k(\x,\x_i)d\x \\
=\sum_{i=1}^{N}\beta_i J_i = \sum_{i=1}^{N}\nu_i \pi(\x_i),
\end{align*}
where $J_i = \int_\mathcal{X}f(\x)k(\x,\x_i)d\x$, and the weights $\bm{\nu} =[\nu_1,\dots,\nu_N]^\top$ are given by $\bm{\nu} = {\bf K}^{-1}{\bm{\zeta} }$ with ${\bm{\zeta}}$
%=[J_1,\dots,J_N]^\top$ 
being the vector of $J_i$'s. In this form, $\widetilde{J}$ is expressed as a combination of evaluations of $\pi(\x)$, i.e., a quadrature. 
The following theorem establishes that the weights $\bm{\nu} ={\bf K}^{-1}{\bm{\zeta} }$ are optimal for a quadrature of this kind. Note that the Gaussian kernels are symmetric positive definite functions, and are special cases of radial basis functions (RBF).

\begin{thm}\label{Thm:OptimalWeights}
	Let us consider a symmetric kernel function $k(\x_i,\x_j)=k(\x_j,\x_i)$ which always defines a positive definite matrix ${\bf K}$.  The native space related to $k(\x,\x')$ is a reproducing kernel Hilbert space (RKHS) \cite{aronszajn1950theory,schaback1999native}. Given the points $\{\x_i\}_{i=1}^N$ and $\bm{\nu} = {\bf K}^{-1}\bm{\zeta}$, the quadrature $\widetilde{J} = \sum_{i=1}^N \nu_i\pi(\x_i)$ is optimal in the sense of Golomb-Weinberg \cite{golomb1958optimal}, i.e., the weights $\nu_i$ minimizes the norm of the integration error functional in the dual space \cite{aronszajn1950theory,schaback1999native}.
	
	\begin{proof}
		A sketch of the proof is in App. \ref{Teo2}.
		See also \cite{sommariva2006numerical} and \cite{briol2019probabilistic} and references therein. 
	\end{proof}
\end{thm}

\begin{thm}\label{Thm:ErrorBoundGPVar}
	Suppose that $\pi(\x)$ belongs to the RKHS generated by the kernel function $k(\x,\x')$. The interpolant  $\widehat{\pi}(\x) = \sum_{i=1}^N\beta_i k(\x,\x_i)$ satisfies $|\pi(\x)-\widehat{\pi}(\x)|\leq \norm{\pi}_\mathcal{H}[V(\x)]^{\frac{1}{2}}$ for all $\x\in\mathcal{X}$ and hence
	%	\begin{align}
	$
	\norm{\pi - \widehat{\pi}}_\infty \leq \norm{\pi}_\mathcal{H}\max_{\x \in \mathcal{X}}[V(\x)]^{\frac{1}{2}},
	$
	%	\end{align}
	where $\norm{\cdot}_\mathcal{H}$ denotes the norm in the RKHS, and $V(\x)$ is the variance function given in Eq. \eqref{eq:VarOfGP}. Hence, from Theorem \ref{Thm:errorBoundInftyNorm}, we have 
	\begin{align*}
	| J - \widetilde{J} | \leq|\mathcal{X}| \norm{f}_\infty \norm{\pi}_\mathcal{H}\max_{\x \in \mathcal{X}}[V(\x)]^{\frac{1}{2}}.
	\end{align*}
	%	Moreover, for isotropic kernels $V(\x)$ satisfies that $\max_{\x \in \mathcal{X}} [V(\x)]^{\frac{1}{2}} \leq \Phi(r)$, where $\Phi(r)$ is an increasing function of $r$, depending on the kernel function. 
	\begin{proof}
		See Sect. 2.1 in \cite{Pronzato17} and Sect. 2 in \cite{auffray2012maximin}. 
	\end{proof}
\end{thm}

\noindent
The theorem above, jointly with Proposition \ref{PropV}, justify the choice of the diversity term $D_t(\x)=V_t(\x)$ in Section \ref{sec:adaptiveProcedure3}.
The next theorem, based on results from the literature on approximating functions with RBFs, establishes that the approximation error tends to zero when $r \to 0$, and that the rate of convergence can be exponentially fast in the case of infinitely smooth RBFs, such as the Gaussian kernels.
\begin{thm}\label{teo3}
	The error of the quadrature $\widetilde{J}$ is
	\begin{align*}
	| J - \widetilde{J} | \leq|\mathcal{X}| \norm{f}_\infty\norm{\pi - \widehat{\pi}}_\infty = \mathcal{O}(\lambda(r)),
	\end{align*}
	where $\lambda(r) \to 0$ as $r \to 0$, with $r$ being the fill distance given in Eq. \eqref{eq:fillDist}.  The convergence rate depends on the regularity degree of $\pi(\x)$. For $\pi(\x)$ sufficiently regular (technically, belonging to the RKHS induced by the RBF kernel), and Gaussian RBF the bound $\lambda(r)$ decreases exponentially
	\begin{align*}
	\lambda(r) = e^{-c_h|\log r|/r},
	\end{align*}  	
	with a certain constant $c_h>0$, which generally depends  on the bandwidth $h$.		
	
	\begin{proof}
		See Sect. 11.3 and table in page 188 of \cite{wendland2004scattered}.
	\end{proof}
\end{thm}
% \newline
\noindent  
Recall that the diversity term in \eqref{eq:AcforConst} produces a monotonically decreasing sequence of fill distances that converges to zero in the limit of $t \to \infty$, as stated in Proposition \ref{Fillprop}. The next theorem states that the approximation error tends to zero as $N \to  \infty$, and provides a  quite pessimistic upper bound.
\begin{thm} Given a sequence of nodes $\{\x_i\}_{i=1}^N$ generated as in Proposition \ref{PropNN}, it can be shown that 
	$r \leq C_{d_x,\mathcal{X}} {N^{-1/d_x}}{\log N}$,
	where $C_{d_x,\mathcal{X}}$ is a constant that probably depends on the dimension $d_x$ and the measure of $\mathcal{X}$. Then, the following (pessimistic) upper bound can be provided
	\begin{align*}
	|J - \widetilde{J}| = \mathcal{O}\left( e^{-c_1\frac{1}{N^{-1/d_x}\log N}  
		-c_2\frac{\left|\log\left({N^{-1/d_x}}{\log N}\right)\right|}{{N^{-1/d_x}}{\log N}}}
	\right),
	%	|J - \widetilde{J}| = \mathcal{O}\left( e^{-c_h\frac{|\log C_{d_x,\mathcal{X}}|}{C_{d_x,\mathcal{X}} {N^{-1/d_x}}{\log N}}  
	%	-c_h\frac{\left|\log\left({N^{-1/d_x}}{\log N}\right)\right|}{C_{d_x,\mathcal{X}} {N^{-1/d_x}}{\log N}}}
	%	\right).
	\end{align*}
	where $c_1>0$ and $c_2>0$ are constants depending on $h$, $d_x$ and the measure of $\mathcal{X}$.
	\begin{proof}
		See Sect. 2.5.1 in \cite{Pronzato17} and \cite{niederreiter1992random}.
	\end{proof}
\end{thm}

%%%--------------------------------%%%
\subsubsection{Noisy computation of $J_i$} \label{NoiseSect}
%%%--------------------------------%%%
Theorem \ref{teo3} above states that the convergence of $\widetilde{J}$ is achieved when the fill distance $r$ goes to zero.  Recall that in $\widetilde{J}=\sum_{i=1}^{N}\beta_i J_i$ we consider the exact computation of $J_i= \int_\mathcal{X}f(\x)k(\x,\x_i)d\x$. In this section, we consider of approximating $J_i$ by the estimator $\widehat{J}_i$, so that we finally have a noisy version of $\widetilde{J}$, i.e.,  
$\widehat{J} = \sum_{i=1}^{N}\beta_i \widehat{J}_i$. %\approx  \widehat{J} % = \sum_{i=1}^{N}\nu_i \pi(\x_i),
Below, we show some results related to $\widehat{J}$, but
we need some previous definitions.
\newline
%\newline
{\bf Stability.}  The numerical stability of the solution depends on the inversion of the interpolation matrix ${\bf K}$ and it is connected to the separation distance $s$.
Clearly, if two nodes are very close, then the corresponding two rows of the interpolation matrix are almost identical and the matrix  becomes ill-conditioned \cite{schaback1995error,wendland2004scattered}. 
\newline
%\newline
{\bf Reproduction quality.} Roughly speaking, an interpolant built with more nodes (i.e., $N$ grows) filling the space, generally yields a better approximation. This concept is connected to the fill distance $r$ in Eq. \eqref{eq:fillDist}. Recall that the fill distance is a measure of how well the data fills the space \cite{wendland2004scattered}. 
\newline
%\newline
\noindent{\bf Uncertainty principle.} A typical problem when reconstructing functions is the trade-off between reproduction quality and numerical stability. Let us consider RBF kernels with a {\it fixed bandwidth}, as $N$ grows. Generally, when one aims at a very good approximation of the function of interest, the numerical stability gets compromised, and conversely, if one aims to have good numerical stability, the approximation will be poor. This is known in the literature as  uncertainty principle \cite{schaback1995error}. 
\newline
\newline
Let us denote as $h$ the parameter which controls the bandwidth of the RBFs, as ${\bm \Sigma}=h^2 {\bf I}$ in the Gaussian kernel.   
The next theorem illustrates the case where the numerical instability combined with the error in computing the vector of integrals $\bm{\zeta}=[J_1,\dots,J_N]^\top$ deteriorates the error bound of Theorem \ref{teo3} (for a fixed $h$). Let us denote the vector of approximated integrals by $\widehat{\bm{\zeta}}=[\widehat{J}_1,\dots,\widehat{J}_N]^\top$ and recall ${\bf d}= [\pi(\x_1),\dots, \pi(\x_N)]^\top$ is the vector of evaluations of $\pi$.

\begin{thm}\label{Thm:ErrorBoundFillAndSeparDists} {\bf (for a fixed bandwidth $h$)}
	Let us consider a bounded support $\mathcal{X}$. If we take into account the error in the evaluation of the integrals $\bm{\zeta}=[J_1,\dots,J_N]^\top$, denoted by $\widehat{\bm{\zeta}} = [\widehat{J}_1,\dots,\widehat{J}_N]^\top$, the corresponding approximation $\widehat{J} = \sum_{i=1}^N\beta_i \widehat{J}_i$ has an error of
	\begin{align*}
	|J - \widehat{J}| &\leq |\mathcal{X}|\norm{f}_\infty\norm{\pi - \widehat{\pi}}_\infty + ||{\bf K}^{-1}||_2 ||{\bf d}||_2 ||\bm{\zeta}-\widehat{\bm{\zeta}}||_2 \\
	&= \mathcal{O}(\lambda(r)) + \mathcal{O}(\upsilon(s,h)) 
	||\bm{\zeta} - \widehat{\bm{\zeta}}||_2,
	\end{align*}
	where $\lambda(r) \to 0$ as $r \to 0$,  $\upsilon(s,h) \to \infty$ as $s \to 0$, with $r$ and $s$ being, respectively, the fill distance and separation distance given in Eqs. \eqref{eq:fillDist} and \eqref{eq:SeparDist}. The parameter $h$, which determines the bandwidth of the radial kernel, is considered fixed. 
	The function $\upsilon(s,h)$ is an upper bound for $\norm{{\bf K}^{-1}}_2$, which is a measure of stability (note that $||{\bf K}^{-1}||_2$ corresponds to the inverse of the lowest eigenvalue of ${\bf K}$). 	
	\begin{proof}
		See Appendix \ref{Sect_ErrorBoundFillAndSeparDists}. For the bound $\upsilon(s,h)$ see Corollary 12.4 in \cite{wendland2004scattered}.
	\end{proof}
\end{thm}
\noindent 
The bound in Theorem \ref{Thm:ErrorBoundFillAndSeparDists} expresses the uncertainty relation.
Indeed, we see that making $s \to 0$ poses a problem if we use a fixed bandwidth $h$. Indeed, the interpolation matrix ${\bf K}$ becomes  ill-conditioned  as two nodes are too close, and the error $||\bm{\zeta} - \widehat{\bm{\zeta}}||_2$ is amplified. 
The growing rate of $\upsilon(s,h)$, as $\lambda(r)$, depends on the smoothness of the RBF. For Gaussian kernels, the rates of $\upsilon(s,h)$ and $\lambda(r)$ are both exponential.  However, with a Monte Carlo approximation, we can always improve the approximation $\widehat{\bm{\zeta}}$ by increasing the number of samples $M$, so that $||\bm{\zeta} - \widehat{\bm{\zeta}}||_2 \to 0$. Recall that the increase of the number of Monte Carlo samples $M$ does not require additional evaluations of the target $\pi$ in the proposed schemes. Furthermore, even with a fixed $M$, we can control the value $||{\bf K}^{-1}||_2$ by decreasing the bandwidth $h$ of the kernel function. The following results consider these two cases.

\begin{thm} {\bf (for a fixed bandwidth $h$ and $M \to \infty$)}\label{Teo7}
	Given a bounded support $\mathcal{X}$, consider the application of a  Monte Carlo method to approximate $\bm{\zeta}$, then
	$||\bm{\zeta} - \widehat{\bm{\zeta}}||_2 \to 0$
	as $M \to \infty$, where $M$ is the number of samples.
	Hence,  the approximation $\widehat{J} = \sum_{i=1}^N\beta_i \widehat{J}_i$  has an error 
	\begin{align*}
	|J - \widehat{J}| = \mathcal{O}(\lambda(r)),
	\end{align*}
	where $\lambda(r) \to 0$ as the fill distance $r \to 0$ and $M \to \infty$.
	%	{ tengo dudas sobre este teorema... segun Wendland2005 (pag 185) escalar la basis function con un $\delta$ de forma proporcional a la fill distance $r$ hace que el error de interpolacion no vaya a cero, sino a la constante de proporcionalidad entre ambos...}
	%\vspace{-0.2cm}
	\begin{proof}
		The term $||\bm{\zeta} - \widetilde{\bm{\zeta}}||_2 \to 0$ as the number of Monte Carlo samples $M \to \infty$ \cite{Robert04}. 
	\end{proof}
\end{thm}

\begin{conj}\label{Thm:ErrorBoundotro} {\bf (for a decreasing bandwidth $h$ and fixed $M$)} \label{Teo8}
	%Let us consider a bounded support $\mathcal{X}$.
	Given a bounded support $\mathcal{X}$, consider  a noisy approximation $\widehat{\bm{\zeta}}$ of ${\bm{\zeta}}$.
	Assume that we decrease $h$ as the number of nodes $N$ grows (in order to control the instability term, i.e., the magnitude of $\norm{{\bf K}^{-1}}_2$). Hence,   the approximation $\widehat{J} = \sum_{i=1}^N\beta_i \widehat{J}_i$  has an error 
	\begin{align*}
	|J - \widehat{J}| = \mathcal{O}(\lambda(r)) + b,
	\end{align*}
	where  $b$ is some constant bias, $\lambda(r) \to 0$ as $r \to 0$, and making $h\rightarrow 0$ when $N\rightarrow \infty$.
\end{conj}
\noindent
Note that, as $h$ approaches $0$, the interpolation matrix  ${\bf K}$ becomes a diagonal matrix, with the maximum values of the kernels in the diagonal. Thus, controlling the maximum values of the kernel functions, we can control the minimum value of the eigenvalues, such that the interpolation matrix ${\bf K}$ be well-conditioned.
Moreover, recall that we are using an interpolative approach and the probabilistic interpretation in Section \ref{Sec:FromIntToReg} is not strictly required. Therefore, we have more flexibility in the choice and/or tuning of the kernel functions. Indeed,  one could consider different bandwidths (one for each kernel function), bigger in regions with lower density of points, while smaller bandwidths in regions with a higher density of nodes. This would improve the numerical stability. 
%We stated that we can avoid this issue if we adjust the bandwidth of the Gaussian parameter as we add more number of nodes, so this separation distance virtually increases.
\newline
\begin{Rem}
	The interplant based on NN kernels does not suffer the uncertainty problem, since they have compact non-overlapping supports. Namely, we can interpret that the bandwidths are automatically tuned.
\end{Rem}

%%%%%%%%%%%%%%%%%%%%%%%%%%%%%%%%
\subsection{Results for local interpolators}\label{NNteo}
%%%%%%%%%%%%%%%%%%%%%%%%%%%%%%%%
In a local interpolation method, the addition and/or a change of one node, only affects the solution in a subset of the support domain.  This scenario corresponds to the use of the constant NN kernels. Recall that the interpolant based on constant kernels, $$\widehat{\pi}(\x) = \sum_{i=1}^N \pi(\x_i) \mathbb{I}_{\mathcal{R}_i}(\x),$$
where $\mathcal{R}_i$ denotes the Voronoi region associated with node $\x_i$. 
Let us first state a result for sufficiently smooth $\pi(\x)$.
If $\pi(\x)$ is Lipschitz continuous, i.e., for all $\x,\z \in\mathcal
X$ we have $|\pi(\z) - \pi(\x)| \leq L_0 || \z - \x ||$ for some constant $L_0$, then we have the following result.
\begin{thm}\label{Thm:LipsTarget}
	Given the NN interpolant  $\widehat{\pi}(\x)$, 
	if $\pi(\x)$ is Lipschitz continuous we have that  $\norm{\pi - \widehat{\pi}}_\infty \leq L_0 r$,
	where $L_0$ is the Lipschitz constant and $r$ is the fill distance introduced in Eq. \eqref{eq:fillDist}. Then, from Theorem \ref{Thm:errorBoundInftyNorm}, we have
	\begin{align*}
	|J - \widetilde{J}| \leq |\mathcal{X}|\norm{f}_\infty L_0 r.
	\end{align*}
	Moreover, given a sequence of nodes $\{\x_i\}_{i=1}^N$ generated as in Proposition \ref{PropNN}, and since $r \leq C_{d_x,\mathcal{X}} {N^{-1/d_x}}{\log N}$, we have the following (pessimistic) bound
	\begin{align*}
	|J - \widetilde{J}| = \mathcal{O}\left(N^{1/d_x}\log N\right).
	\end{align*}
	\begin{proof}
		See Appendix \ref{LipsProof}.
	\end{proof} 
\end{thm}
\noindent
Now, recall the approximation of $\widetilde{J}$  given by
\begin{align*}
\widetilde{J}  &= \int_\mathcal{X} f(\x)\widehat{\pi}(\x)d\x \approx S_N = \sum_{i=1}^N\pi(\x_i)f(\x_i)C_i ,
\end{align*}
where $S_N$ is the Riemann approximation, which has been also discussed  in Sect. \ref{Sec:ApproxVoroRegions}, and $C_i = \int_{\mathcal{R}_i}d\x$, i.e., the measure of $\mathcal{R}_i$. Here, we used the approximation $\int_{\mathcal{R}_i}f(\x)d\x \approx f(\x_i)C_i$. We will show that $S_N$ converges to $J = \int_\mathcal{X}f(\x)\pi(\x)d\x$ as we add more nodes according to one of the proposed acquisition functions, that is, as $t\to \infty$. As with Gaussian kernels, the convergence is related with how well the nodes fill space. Here, the role of fill distance is played by the maximum of the measures $C_i$. The theorem below states that, as we fill the space, the measures $C_i$ converges to zero. Recall that the Voronoi partition $\{\mathcal{R}_i\}_{i=1}^N$ generated from the set of nodes $\{\x_i\}_{i=1}^N$ corresponds to the subdivision of $\mathcal{X}$ in $N$ non-overlapping pieces. 

\begin{propo}
	Consider a sequence of points $\x_1,\dots,\x_N$ covering the space $\mathcal{X}$, then for the associated Voronoi regions $\mathcal{R}_i$, we have that $\max_iC_i \to 0$ as $N \to \infty$. 
%	\vspace{-0.3cm}
	\begin{proof}
		See the proofs of Theorems 1 and 4 in \cite{devroye2017measure}. 
	\end{proof}
\end{propo}

\begin{thm}\label{TeoRiemann}
	Let $\pi(\x)$ be a continuous and bounded target pdf (up to a normalizing constant) defined on a bounded support $\mathcal{X}\subset\mathbb{R}^{d_x}$. 
	Let $f(\x): \mathcal{X} \to \mathbb{R}$ bounded on $\mathcal{X}$. Consider the integral 	
	$J = \int_{\mathcal{X}} f(\x)\pi(\x)d\x$.
	Let us consider a Voronoi partition of $\mathcal{X}$, generated by the nodes $\{\x_i\}_{i=1}^N$,  defined as $\mathcal{R}_1,\dots,\mathcal{R}_N$ (recall that $C_i=|\mathcal{R}_i|$). 
	Given the Riemann sum  $S_N = \sum_{i=1}^N f(\x_i)\pi(\x_i)C_i$,
	the convergence of $S_N \to J$ is guaranteed as $\max_i C_i \rightarrow 0$ when $N\to \infty$. 
	%\vspace{-0.15cm}
	\begin{proof}
		See Sect 8.3 in \cite{protter2012first}.
	\end{proof}
\end{thm}
\noindent Above, we have assumed that $C_i$ are known. However, we can have very accurate Monte Carlo estimates without requiring additional evaluations of the target $\pi(\x)$ (but just of the interpolant  $\widehat{\pi}(\x)$), i.e., only with a slight increase in the overall computation cost.

%%%%%%%%%%%%%%%%%%%%%%%%%%%%%%%%%%%%%%%%%%
%%%%%%%%%%%%%%%%%%%%%%%%%%%%%%%%%%%%%%%%%%
\section{Numerical experiments} \label{NumEx}
%%%%%%%%%%%%%%%%%%%%%%%%%%%%%%%%%%%%%%%%%%
%%%%%%%%%%%%%%%%%%%%%%%%%%%%%%%%%%%%%%%%%%
In this section, we provide several numerical tests in order to show the performance of the proposed adaptive quadrature schemes and compare them with benchmark approaches in the literature. The first example corresponds to a nonlinear banana-shaped density in dimension $d_x=$2, 3, 4 and 5. The second test is a multimodal scenario with dimension $d_x$=10. Finally, we test our schemes in a challenging astronomic inference problem of detecting the number of exoplanets orbiting a star.

%\vspace{-0.3cm}
%%%%%%%%%%%%%%%%%%%%%%%%%%%%%%%%%%%%%%%%%%
\subsection{Banana target}\label{sec:bananaTar}
%%%%%%%%%%%%%%%%%%%%%%%%%%%%%%%%%%%%%%%%%%
As {a} first example, we consider a banana-shaped target pdf,
\begin{align}\label{eq:BananaTarget}
\bar{\pi}(\x) \propto \exp \left\{ -\frac{(\eta_1-Bx_1-x_2^2)^2}{2\eta_0^2}  - \sum_{i=1}^{d_x} \frac{x_i^2}{2\eta_{i}^2}\right\},
\end{align}
with $\x \in \mathcal{X} = [-10,10]^{d_x}$, $B=4$, $\eta_0=4$ and $\eta_i= 3.5$ for $i=1,...,d_x$ . We consider $d_x=\{2,3,4,5\}$ (i.e., different dimensions) and compute in advance the {\it true} moments of the target (i.e., the groundtruth) by using a costly grid, in order to check the performance of the different techniques.

%%%%%%%%%%%%%%%%%%%%%%%%%%%%%%
%%%%%%%%%%%%%%%%%%%%%%%%%%%%%%
\begin{figure*}[!t]
	\centering
	%	\centerline{
	\subfigure[]{\includegraphics[width=0.3\textwidth]{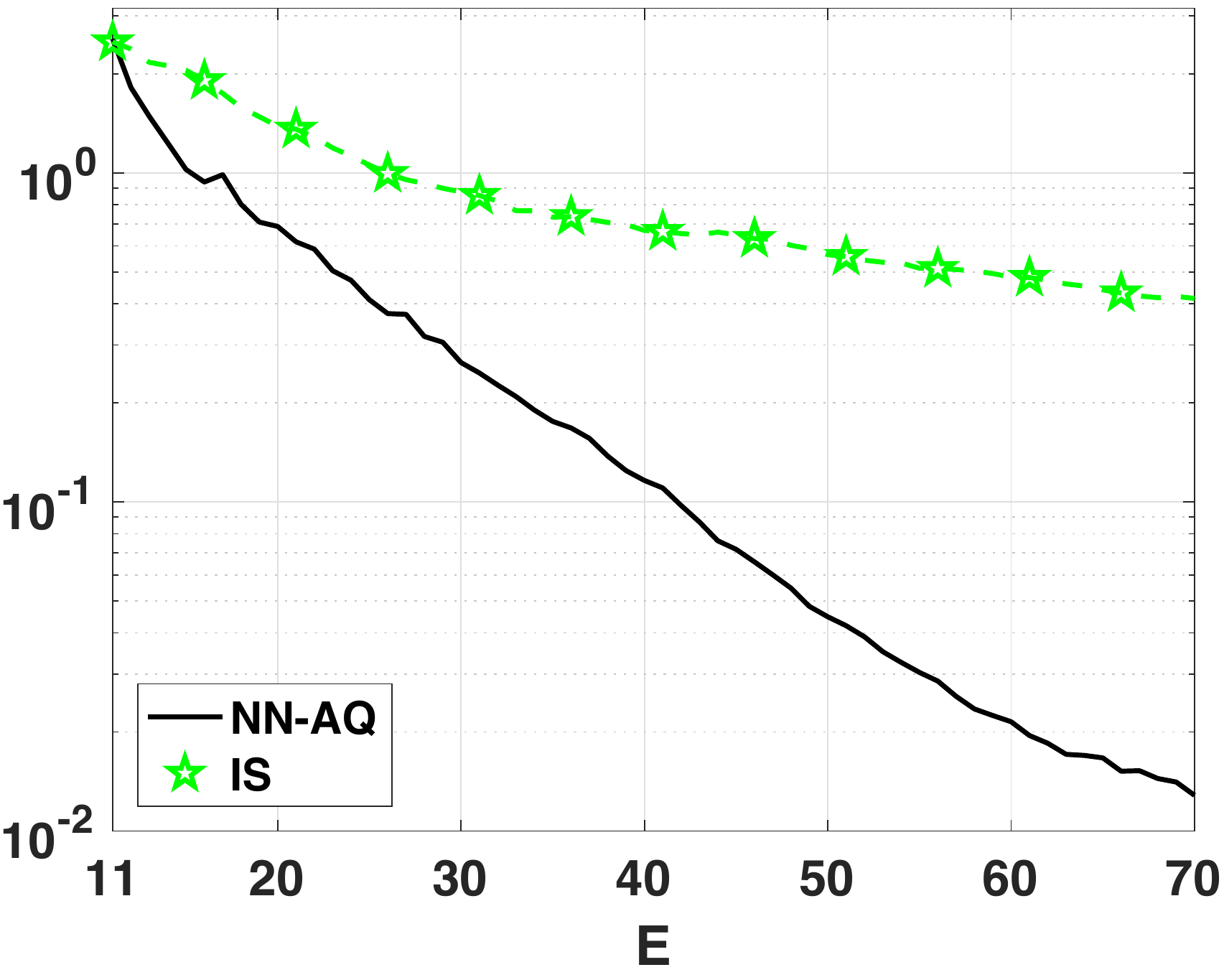}}
	\subfigure[]{\includegraphics[width=0.3\textwidth]{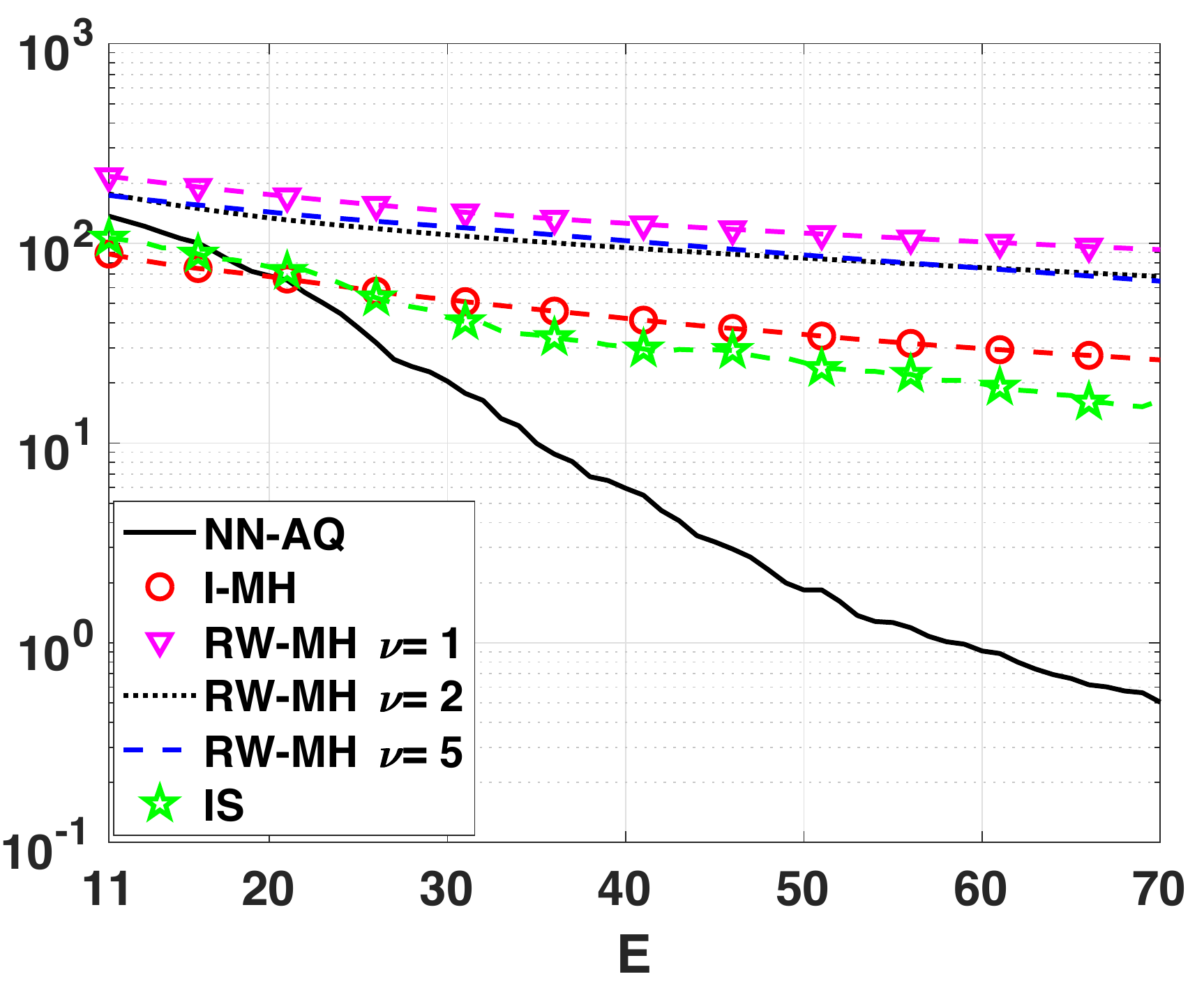}}
	\subfigure[]{\includegraphics[width=0.3\textwidth]{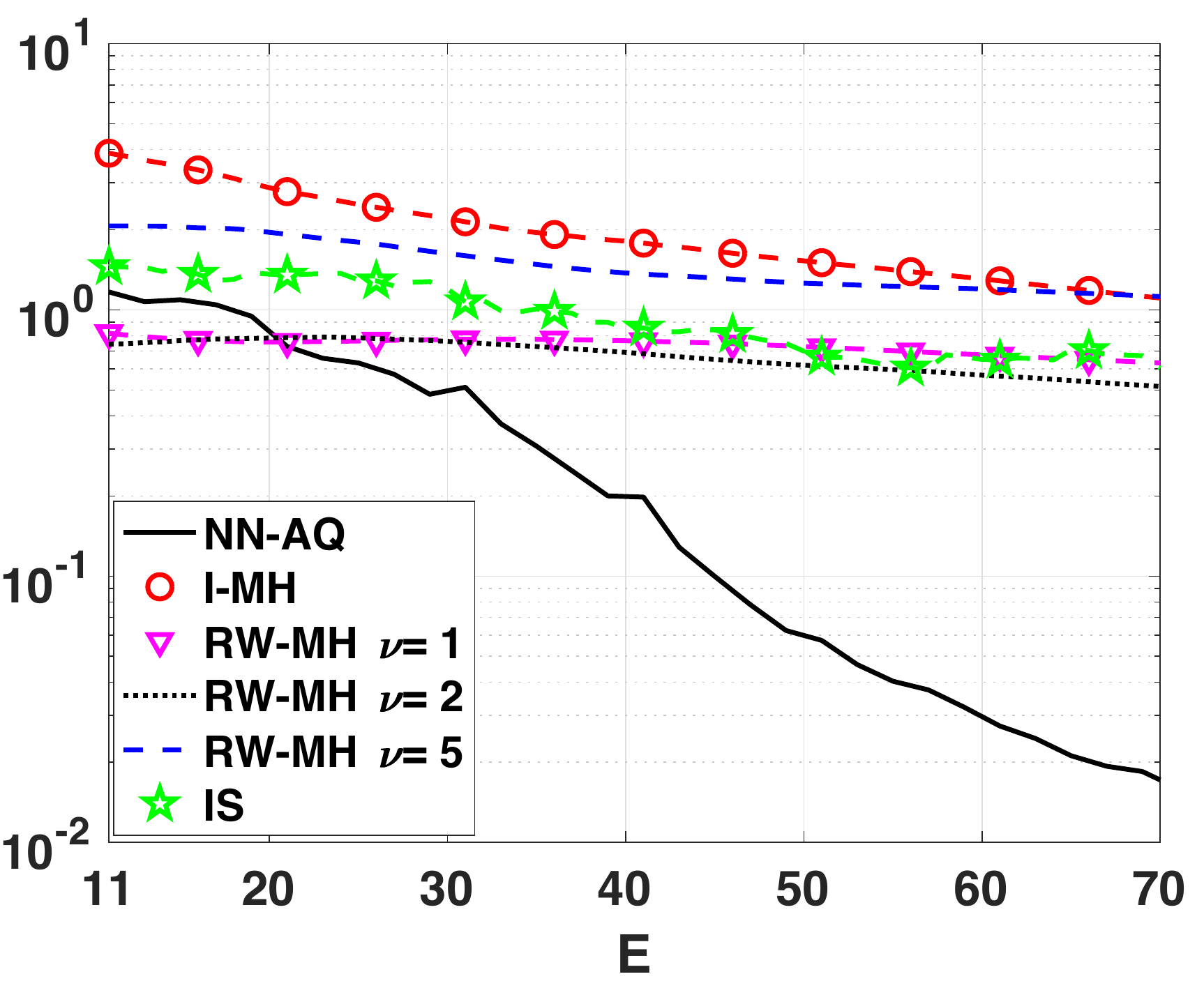}}
	%		\subfigure[]{\includegraphics[width=0.3\textwidth]{rel_MSE_versus_DIM_ssp}}
	%	}
	%\vspace{-0.3cm}
	\caption{\footnotesize{\bf (a)} Rel-MSE in log-scale for $Z$ as function of number of target evaluations $E$.   {\bf (b)} Rel-MSE in log-scale for $\bm{\mu}$ as function of number of target evaluations $E$.  {\bf (c)} Rel-MSE in log-scale for estimating $[\sigma_1^2, \sigma_2^2]$ as function of number of target evaluations $E$.
	}
	\label{fig_SIMU1}
\end{figure*}
%%%%%%%%%%%%%%%%%%%%%%%%%%%%%%
%%%%%%%%%%%%%%%%%%%%%%%%%%%%%%

%%%----------------------------%%%%
\subsubsection{Experiment 1}
%%%----------------------------%%%%
We set $d_x=2$ and test the different algorithms in order to compute the vector mean $\bm{\mu} = [-0.4, 0]$ and the diagonal of the covariance matrix  $[\sigma_1^2, \sigma_2^2] =[1.3813, 8.9081]$.
Moreover, our schemes are also able to estimate $Z$, whose ground-truth is $Z = 7.9979$, thus we also measure the error in this estimation.  We compare the performance in terms of Relative Mean Square Error (Rel-MSE), averaged over 500 independent runs,  using different methodologies: {(\bf a)} NN-AQ starting with $N_0=10$ nodes randomly chosen in $[-10,10]\times[-10,10]$ and $M=10^5$; {\bf (b)} an independent MH algorithm (I-MH) with random initialization in $[-10,10]\times[-10,10]$; {\bf (c)} random-walk MH algorithms (RW-MH) with different proposal variance, and random initialization in $[-10,10]\times[-10,10]$; {\bf (d)} an IS algorithm.  The proposal density for both I-MH and IS is a uniform in $[-10,10]\times[-10,10]$, whereas for the RW-MHs is a Gaussian density centered at the current state of the chain with covariance matrix $v^2{\bf I}$ where $v\in\{1,2,5\}$ (so we consider $3$ different RW-MHs).
\newline
For a fair comparison, we need that all methods have the same number $E$ of target evaluations (fixing $E=70$).
Since NN-AQ, I-MH and RW-MH require one new target evaluation per iteration,  we run $T=70$ iterations for I-MH and RW-MH ($E=T$), and $T-N_0=60$ iterations for NN-AQ. In this regard, the IS algorithm use 70 samples drawn from the uniform proposal. 
Hence, all methods need $T=70$ target evaluations. The results are given in Figures  
\ref{fig_SIMU1}(a)-(b). Note that the estimation of $Z$ via MCMC techniques is not straightforward (e.g., see \cite{ourRev}). 
\newline
{\bf Discussion 1.} We can observe that NN-AQ outperforms the other methods in terms of Rel-MSE in estimation. Moreover, in Fig. \ref{fig_SIMU1}(a)-(b) we can see that the decrease is much greater, as $E$ grows, than the other methods. Namely, NN-AQ has more benefits with new evaluations of $\pi(\x)$.

%%%----------------------------%%%%
\subsubsection{Experiment 2}
%%%----------------------------%%%%
%{\bf Experiment 2.} 
In this case, we fix the number of target evaluations $E$,  and vary $d_x=\{2,3,4,5\}$. The Rel-MSE in the estimation of $Z$ is given in Table \ref{table2} (with $E\in\{100,1000\}$). 
\newline
{\bf Discussion 2.} In this experiment, $E$ is fixed along different dimensions. The results given in Table \ref{table2}, with fixed $E$, does not show all the potential of NN-AQ. However, NN-AQ outperforms IS in all the dimensions $d_x$ considered when $E=1000$.  
\begin{table}[!h]
	\centering
	%\small
	\caption{\footnotesize \textbf{Relative MSE of $Z$ with $E\in\{100,1000\}$ for different $d_x$}}
%	\vspace{0.2cm}
	{\footnotesize
		\begin{tabular}{|c|c|c|c|c|c|}
			%\hline
			\hline	
			methods& $E$ & $d_x=2$ & $d_x=3$ & $d_x=4$ &  $d_x=5$  \\
			\hline
			%		$Z$ & 70.1672 & 615.5905 & 5.4007e+03 & 4.7381e+04 \\
			%		\hline
			\multirow{2}{*}{NN-AQ}	& 100  & 0.0027 & 0.1127  & 0.3798 & 1.9730 \\
			&  1000	& 	4$\cdot10^{-4}$	& 0.0023  & 0.0140 &  0.0374  \\
			\hline 
			\multirow{2}{*}{IS} & 100 & 0.2645  &  0.4427  & 0.7627  &  1.1115 \\
			& 1000 &  0.0226 & 0.0378  &  0.0641 &   0.1094 \\
			\hline
		\end{tabular}
	}	
	\label{table2}
\end{table}

%%%%%%%%%%%%%%%%%%%%%%%%%%%%%%%%%%%%%%%%%%%
%%%%%%%%% FIGURA NN-AQ VS VARIANTES %%%%%%%
%%%%%%%%%%%%%%%%%%%%%%%%%%%%%%%%%%%%%%%%%%%
\begin{figure*}[!t]
	\centering
	%	\centerline{
	\subfigure[]{\includegraphics[width=0.3\textwidth]{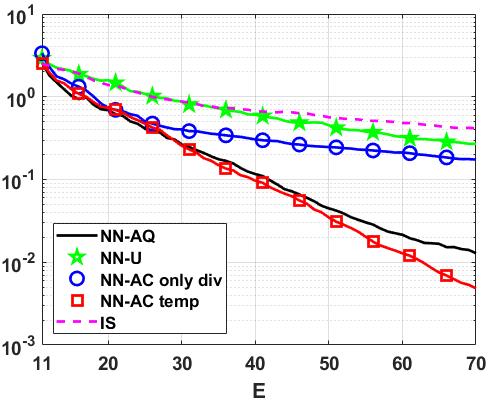}}
	\subfigure[]{\includegraphics[width=0.3\textwidth]{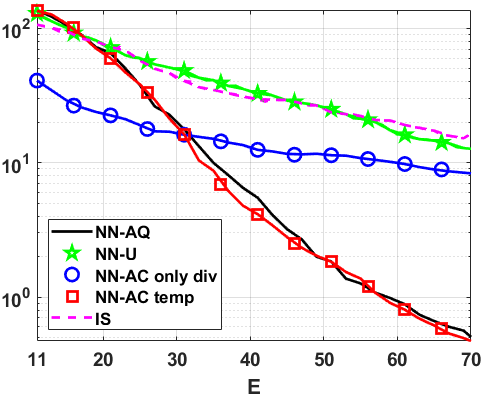}}
	\subfigure[]{\includegraphics[width=0.3\textwidth]{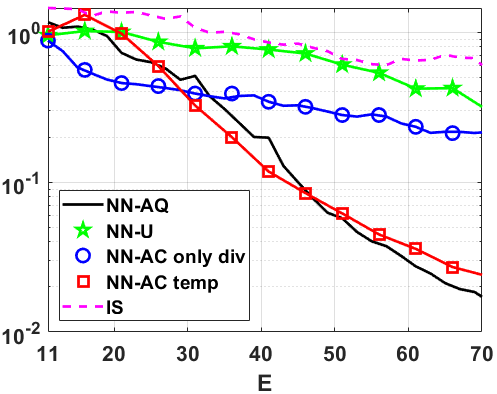}}
	%	}
	\vspace{-0.3cm}
	\caption{\footnotesize {\bf (a)} Rel-MSE in log-scale for $Z$ as function of number of target evaluations $E$. {\bf (b)} Rel-MSE in log-scale for $\bm{\mu}$ as function of number of target evaluations $E$.  {\bf (b)} Rel-MSE in log-scale for estimating $[\sigma_1^2, \sigma_2^2]$ as function of number of target evaluations $E$. }
	\label{fig_SIMU2}
\end{figure*}
%%%%%%%%%%%%%%%%%%%%%%%%%%%%%%%%%%%%%%%%%%%
%%%%%%%%%%%%%%%%%%%%%%%%%%%%%%%%%%%%%%%%%%%

%%%%%%%%%%%%%%%%%%%%%%%%%%%%%%%%%%%%%%%%%%%%%%%%%%%%%
%%%%%%%%%%%% FIGURE NN-AQ VS GK-AQ %%%%%%%%%%%%%%%%%%
%%%%%%%%%%%%%%%%%%%%%%%%%%%%%%%%%%%%%%%%%%%%%%%%%%%%%
\begin{figure*}[!t]
	\centering
	%	\centerline{
	%%%\vspace{0.5cm}
	\subfigure[]{\includegraphics[width=0.23\textwidth]{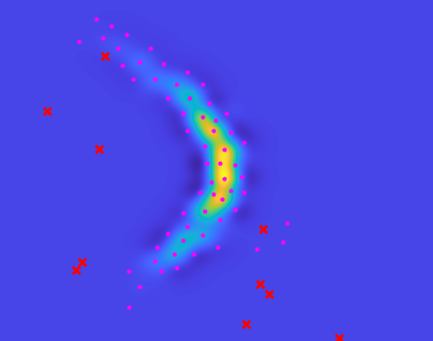}}
	\subfigure[]{\includegraphics[width=0.23\textwidth]{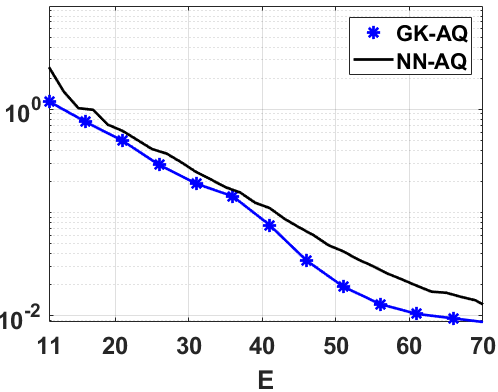}}
	\subfigure[]{\includegraphics[width=0.23\textwidth]{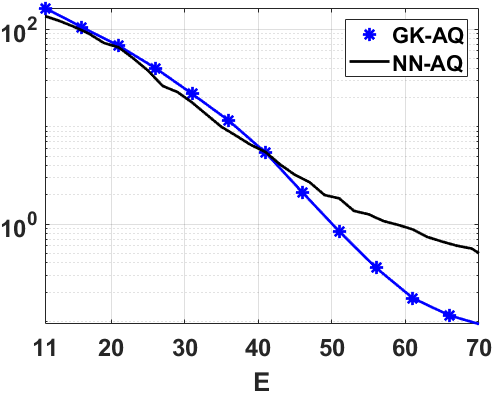}}
	\subfigure[]{\includegraphics[width=0.23\textwidth]{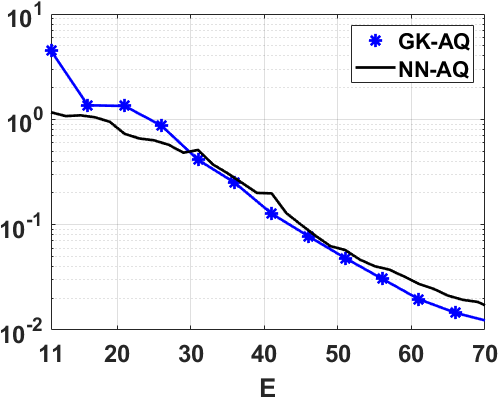}}
	%	}
	\vspace{-0.3cm}
	\caption{\footnotesize{\bf (a)} Example of application of GK-AQ with 10 starting points (red cross-marks) and T=60 iterations (red dots), i.e., $E=70$ target evaluations. {\bf (b)} Rel-MSE in log-scale for $Z$ as function of number of target evaluations $E$. {\bf (c)} Rel-MSE in log-scale for $\bm{\mu}$ as function of number of target evaluations $E$.  {\bf (d)} Rel-MSE in log-scale for estimating $[\sigma_1^2, \sigma_2^2]$ as function of number of target evaluations $E$. 
	}
	\label{fig_SIMU3}
\end{figure*}
%%%%%%%%%%%%%%%%%%%%%%%%%%%%%%%%%%%%%%%%%%%%%%%%%%%%%
%%%%%%%%%%%%%%%%%%%%%%%%%%%%%%%%%%%%%%%%%%%%%%%%%%%%%

\vspace{-0.3cm}

%%%----------------------------%%%%
\subsubsection{Experiment 3}
%%%----------------------------%%%%
For $d_x=2$, we compare now IS, NN-AQ, and three variants of NN-AQ: (i) NN-U,  where the optimization step in \eqref{eq:MaximizationOfAc} is substituted with sampling uniformly the new node in $[-10,10]\times[-10,10]$ (i.e., without using an acquisition function), (ii) NN-AQ {\it only diversity}, which uses the acquisition in \eqref{temperedA} with $\alpha=0$, $\beta=1$, i.e., with only the diversity term $D_t(\x)$, and (iii) NN-AQ {\it tempered}, which uses the acquisition in \eqref{temperedA}  with $\alpha=0$, $\beta_t = \frac{200}{t}$, i.e., $A_t(\x)=[D_t(\x)]^{\beta_t}$. % where $D_t(\x) = \min_{i=1,\dots,N_t} \norm{\x-\x_i}_p$. 
Note that the adaptation in NN-AQ {\it only diversity} can be viewed as filling the space in a deterministic way. Note also that the adaptation in NN-AQ {\it tempered} will encourage more exploration than NN-AQ in the early iterations. 
Again, we compare the error in estimating $Z$, $\bm{\mu}$ and $[\sigma_1^2, \sigma_2^2]$ as a function of target evaluations $E$ (up to $E=70$). The results are given in Figures \ref{fig_SIMU2}(a)-(b).  
\newline
{\bf Discussion 3.} We can observe that NN-AQ and NN-AQ {\it tempered} outperform the others in terms of Rel-MSE in estimation. Moreover, in Fig. \ref{fig_SIMU2}(a)-(b) we can see that {for} NN-AQ and NN-AQ {\it tempered}, {the RMSE decreases at a faster rate} as $E$ grows, than the NN-U and NN-AQ {\it only diversity}, highlighting the importance of taking into account the current interpolant  to locate the new nodes. It can be seen that NN-AQ {\it only diversity} works much better than NN-U in the early iterations. We explain these results by the fact that NN-AQ {\it only diversity} tends to cover the space more efficiently in these early iterations since it avoids placing new nodes near the existing ones. However, as $E$ grows, the performance of NN-U and NN-AQ {\it only diversity} is similar since both end up filling uniformly the space. 
Interestingly, NN-U performs better than IS as $E$ increases, which demonstrate the power of the interpolative approach even when the new nodes are randomly chosen.

%%%----------------------------%%%%
\subsubsection{Experiment 4}
%%%----------------------------%%%%
For $d_x=2$, we investigate the performance of GK-AQ in the estimation of $Z$, $\bm{\mu}$ and $[\sigma_1^2, \sigma_2^2]$ as function of $E$. NN-GK employs the acquisition in \eqref{eq:AcforGauss}.
The kernel bandwidth $h$ is fitted using the procedure in Appendix \ref{App:MagicalHeuristicLuca}. As commented in Sect. \ref{sec:hyperparTuning}, we consider a small noise of $\sigma = 10^{-2}$ for numerical stability. We will compare the performance against NN-AQ. The results are given in Figures \ref{fig_SIMU3}(a)-(d), along with an example of  GK-AQ interpolant , with $E=70$, obtained in a specific run.

\noindent{\bf Discussion 4.} %Using the procedure in Appendix \ref{App:MagicalHeuristicLuca} for fitting the bandwidth seems to give 
The  results are shown in Figures \ref{fig_SIMU3}(b)-(d).  GK-AQ  outperforms NN-AQ in this particular experiment. However, it is important to remark that the results of GK-AQ may worsen considerably if $h$ is not selected adequately (we have used the procedure in  App. \ref{App:MagicalHeuristicLuca}), in contrast to NN-AQ which is free of hyperparameter tuning and hence more robust.

%%%%%%%%%%%%%%%%%%%%%%%%%%%%%%%%%%%%%%%%%%
\subsection{Multimodal target}
%%%%%%%%%%%%%%%%%%%%%%%%%%%%%%%%%%%%%%%%%%
In this experiment, we consider a multimodal Gaussian target in $d_x=10$,
\begin{align*}
\bar{\pi}(\x) = \frac{1}{3}\mathcal{N}(\x|\bm{\mu}_1, {\bf \Sigma}_1) + \frac{1}{3}\mathcal{N}(\x|\bm{\mu}_2, {\bf \Sigma}_2) + \frac{1}{3}\mathcal{N}(\x|\bm{\mu}_3, {\bf \Sigma}_3),
\end{align*}
with $\bm{\mu}_1 = [5,0,\dots,0]$, $\bm{\mu}_2=[-7,0,\dots,0]$, $\bm{\mu}_3=[1,\dots,1]$ and ${\bf \Sigma}_1={\bf \Sigma}_2={\bf \Sigma}_3 = 4^2{\bf I}_{10}$. 
We want to test the performance of the different methods in estimating the normalizing constant $Z=1$.
We consider an application of GK-AQ with $N_0=500$ initial nodes, random in $[-15,15]^{10}$, and $T=1000-N_0$, hence fixing the number of evaluations to $E=1000$. We compare it against three sophisticated AIS schemes, namely PMC, LAIS and AMIS \cite{bugallo2017adaptive}. For PMC, we choose Gaussian proposal pdfs and test different number of proposals $L \in \{10,100,200,500\}$, whose means are also initialized at random in $[-15,15]^{10}$. At each iteration one sample is drawn from each proposal, hence the algorithm is run for $T_\text{PMC} = \frac{1000}{L}$ iterations for a fair comparison. 
As a second alternative, we consider the deterministic mixture weighting approach for PMC, which is shown to have better overall performance, denoted DM-PMC. 
For LAIS, we also consider different number of proposals $L \in \{10,100,200,500\}$. More specifically, we consider two versions of LAIS: the {\it one-chain} version and an {\it ideal} version. In ideal LAIS, the means of the $L$ Gaussian proposals are drawn exactly from $\post(\x)$. The one-chain application of LAIS (OC-LAIS) requires to run a MCMC algorithm targeting $\post(\x)$ to obtain the $L$ proposal means, hence it requires $L$ evaluations of the target. At each iteration one sample is drawn from the mixture of the $L$ Gaussian proposals, hence we run the algorithm for $T_\text{LAIS} = 1000 - L$ iterations for a fair comparison.  We used a Gaussian random walk Metropolis to obtain the $L$ means in the one-chain scenario.
Finally, we consider AMIS with several combinations of number of iterations $T_\text{AMIS}$ and number of samples per iteration $R$. At each iteration, $R$ samples are drawn from a single Gaussian proposal, hence the total number of evaluations is $E = RT_\text{AMIS}$. In this case, we test $E \in \{1000, 2000, 3000, 5000\}$, so the comparison is not fair except for $E=1000$. For PMC, LAIS and AMIS, as well as for the random walk proposal within the Metropolis algorithm, the covariance of the Gaussian proposals was fixed to $h^2{\bf I}_{10}$ (for $h=1,...,6$), where $h$ is the initial bandwidth parameter used in GK-AQ.\footnote{Recall that, for GK-AQ, the final bandwidth is tuned as described in App. \ref{App:MagicalHeuristicLuca}.}
All the methods are compared through the mean absolute error (MAE) in estimating $Z$, and the results are averaged over 500 independent simulations.
The results are shown in Table \ref{table3} and Table \ref{table4}. For each method, the best and worst MAE are boldfaced.
\newline
{\bf Discussion.} We can observe that GK-AQ obtains the best range of MAE values $[0.078,0.4782]$ and the best results for $h=1$.  For  $h>1$, we can see in Tables \ref{table3}-\ref{table4} that the lowest MAE values are obtained by ideal LAIS with $L=500$ and $h=3$. We stress that  ideal LAIS is not available in practice, since we usually cannot sample directly from $\post(\x)$. Regardless of the ideal LAIS scheme (not applicable in practice),  GK-AQ provides the best results. Moreover, we see that GK-AQ with $h=3$ is the best performing method in this experiment, since it achieves a lower MAE than PMC, DM-PMC and OC-LAIS for every combination of $L$ and $h$. Table \ref{table4} shows that AMIS performs worse than GK-AQ for $E=1000$ (fair comparison), but even with much more AMIS evaluations $E \in \{2000,3000\}$ (unfair comparison in favor of AMIS). AMIS needs to reach a big enough value of $E$ ($E=5000$),  to beat GK-AQ in terms of MAE.

\begin{table*}[!h]
	\centering
	%\small
	\caption{ \textbf{MAE of $Z$ with $E=1000$} (best and worst MAE of each method are boldfaced)}
	\vspace{0.2cm}
	%	{\footnotesize
	\begin{tabular}{|c c|c|c|c|c|c|c|}
		\hline	
		% & & & & & & &  \\
		\multicolumn{2}{|c|}{{\bf Methods}}  & ${\bf h=1}$ & ${\bf h=2}$ & ${\bf h=3}$ & ${\bf h=4}$ &  ${\bf h=5}$ & ${\bf h=6}$  \\		  
		\hline
		\hline
		\multicolumn{2}{|c|}{GK-AQ}     	& {\bf 0.4782} & 0.1741 & {\bf 0.0780}  & 0.1362 &  0.1497 &  0.2322\\
		\hline
		\hline
		\multirow{ 4}{*}{PMC} & $N=10$& 0.9993 & 0.9526	& 0.8603 & 0.6743 & 0.6024 & 0.6155 \\
		%\hline 
		& $N=100$& 0.9998 & 0.9896 & 0.8853 & 0.6761 & 0.5192 & {\bf 0.4544}\\ 
		& $N=200$& {\bf 1.0002} & 0.9893 & 0.8816 & 0.7099 & 0.6389  &  0.5384\\
		%\hline
		&$N=500$& 0.9995 & 0.9916 & 0.9741 & 0.8700 & 0.7421 & 0.6544 \\
		\hline
		\hline
		\multirow{ 4}{*}{DM-PMC}  & $N=10$&  0.9991 & 0.9478 &	0.8505 & 0.6009 & 0.5352 & 0.5814 \\
		%\hline 
		&$N=100$& 0.9997 & 0.8719 & 0.4490 & 0.2425 &  {\bf 0.1901} & 0.2193 \\ %\hline
		&$N=200$&  0.9999	& 0.9321 & 0.5708 & 0.3257 & 0.2374 & 0.2524 \\
		%\hline
		& $N=500$& {\bf 1.0000} & 0.9888 & 0.7969 & 0.5009 & 0.3684 & 0.3800 \\
		\hline
		\hline
		\multirow{ 4}{*}{Ideal LAIS} &$N=10$& {\bf 0.9992}	& 0.8114 & 0.2579 & 0.0863 & 0.0819 & 0.1091 \\
		%\hline 
		&$N=100$& 0.9918 & 0.3638 & 0.0547 &  0.0407 & 0.0598  & 0.1053 \\ %\hline
		&$N=200$&  0.9846 & 0.2486 & 0.0352 & 0.0411 & 0.0680 & 0.1093 \\
		%\hline
		& $N=500$& 0.9687 & 0.1852 & {\bf 0.0335} & 0.0473 & 0.0891 & 0.1353 \\
		\hline
		\hline
		\multirow{ 4}{*}{OC-LAIS} &$N=10$& {\bf 1.0000}	& 1.0000 & 0.9992 & 0.9883 & 0.9468 & 0.9079 \\
		%\hline 
		&$N=100$& 0.9999 & 0.8731 & 0.4434 & 0.2785 & 0.2392 & 0.2870  \\ %\hline
		&$N=200$& 0.9982 & 0.7028 & 0.2418 & 0.1243 & 0.1406 & 0.2070 \\
		%\hline
		& $N=500$& 0.9937 & 0.4949 & 0.1221 & {\bf 0.0857} & 0.1195 & 0.1786 \\
		\hline
	\end{tabular}
	%	}	
	\label{table3}
\end{table*}

\begin{table*}[!h]
	\centering
	%\small
	\caption{ \textbf{MAE of $Z$ of AMIS with $E\in \{1000, 2000, 3000, 5000\}$}}
	\vspace{0.2cm}
	%	{\footnotesize
	\begin{tabular}{|c c|c|c|c|c|c|c|}
		\hline	
		% & & & & & & &  \\
		\multicolumn{2}{|c|}{{\bf Methods}}  & ${\bf h=1}$ & ${\bf h=2}$ & ${\bf h=3}$ & ${\bf h=4}$ &  ${\bf h=5}$ & ${\bf h=6}$  \\		  
		\hline
		\hline
		\multicolumn{2}{|c|}{GK-AQ (E=1000)}     	& {\bf 0.4782} & 0.1741 & {\bf 0.0780}  & 0.1362 &  0.1497 &  0.2322\\
		\hline
		\hline
		\multirow{3}{*}{AMIS}  &$M=10$&  0.9998 & 0.9997 & 0.9997 & 0.9996 & 0.9996 & 0.9995 \\
		%\hline 
		\multirow{3}{*}{$E=1000$} &$M=100$& 1.0000 & 1.0000 & 1.0000 & 0.9999 & 0.9997 &  0.9990 \\ %\hline
		&$M=200$& 1.0000 & 1.0000 & 1.0000 & 1.0000 & 0.9998 & 0.9994 \\
		%\hline
		& $M=500$& {\bf 1.0000} & 1.0000 & 1.0000 & 1.0000 & 0.9998 & {\bf 0.9989} \\
		\hline
		\hline
		\multirow{3}{*}{AMIS}  &$M=10$& 0.9155 & 0.9117 & 0.8981 & 0.8987 & 0.8891 & {\bf 0.8878} \\
		%\hline 
		\multirow{ 3}{*}{$E=2000$}&$M=100$& 0.9998 & 0.9986 &  0.9934 & 0.9784 & 0.9559 & 0.9072
		\\ %\hline
		&$M=200$& 1.0000 & 1.0000 & 0.9998 & 0.9981 & 0.9888 & 0.9712 \\
		%\hline
		& $M=500$& {\bf 1.0000} & 1.0000 & 1.0000 & 0.9998 & 0.9984 & 0.9953 \\
		\hline
		\hline
		\multirow{3}{*}{AMIS}  &$M=10$& 0.3293 & 0.3402 & {\bf 0.3051} & 0.3381 & 0.3540 & 0.3443 \\
		%\hline 
		\multirow{ 3}{*}{$E=3000$}	&$M=100$& 0.9725 & 0.9040 & 0.7963  & 0.6384 & 0.4964 & 0.3816 \\
		&$M=200$& 0.9998 & 0.9977 & 0.9884 & 0.9527 & 0.8308 & 0.7119 \\
		%\hline
		& $M=500$& {\bf 1.0000} & 1.0000 & 0.9998 & 0.9988  & 0.9859 & 0.9566
		\\ 
		\hline
		\hline
		\multirow{3}{*}{AMIS}&$M=10$& 0.0766 & 0.0768 & {\bf 0.0695} & 0.0722 & 0.0699 & 0.0725 \\
		%\hline 
		\multirow{ 3}{*}{$E=5000$} &$M=100$& 0.1626 & 0.1176 & 0.0957 & 0.0810 & 0.0737 &  0.0656 \\
		&$M=200$& 0.8771 & 0.6040 & 0.2824 & 0.1473 & 0.1163 & 0.0899 \\
		%\hline
		& $M=500$& {\bf 1.0000} & 0.9982 & 0.9904 & 0.9449 & 0.7944 &  0.4532 \\ 
		\hline
	\end{tabular}
	%	}	
	\label{table4}
\end{table*}

%%%%%%%%%%%%%%%%%%%%%%%%%%%%%%%%%%%%%%%%%%
\subsection{Applications to exoplanet detection}
%%%%%%%%%%%%%%%%%%%%%%%%%%%%%%%%%%%%%%%%%%
In recent years, the problem of revealing objects orbiting other stars has acquired large attention. Different techniques have been proposed to discover exo-objects but, nowadays, the radial velocity technique is still the most used \cite{Gregory2011,Barros2016,Affer2019,Trifonov2019}. The problem consists in fitting a dynamical model to data acquired at different moments spanning during long time periods (up to years). The model is highly non-linear and,  for certain sets of parameters, its evaluation is quite costly in terms of computation time.  This is due to the fact that its evaluation involves numerically integrating a differential equation, or using an iterative procedure for solving a non-linear equation (until a certain condition is satisfied). This loop can be very long for some sets of parameters. 

\begin{table}[t]
	\centering
	\caption{Description of parameters in Eq.~\eqref{eq:rv}.}
	\small
	\begin{tabular}{lll} % Column formatting, @{} suppresses leading/trailing space
		\hline
		Parameter & Description & Units \\
		\hline
		\multicolumn{3}{l}{For each planet}\\
		\hline
		$K_i$        & amplitude of the curve & m\,s$^{-1}$ \\
		${u}_{i,t}$      & true anomaly     & rad \\
		$\omega_{i}$      & longitude of periastron & rad \\ 
		$e_i$        & orbit's eccentricity    & \ldots \\
		$P_i$        & orbital period        & s \\
		$\tau_i$       & time of periastron passage & s \\
		\hline
		\multicolumn{3}{l}{\footnotesize Below: not depending on the number of objects/satellite }\\
		\hline
		$V_0$      & mean radial velocity   & m\,s$^{-1}$ \\
		%$\mathbf{s}$         & jitter               & m\,s$^{-1}$ \\
		\hline
	\end{tabular}
	\label{tab:rvpar}
\end{table}

%%%----------------------------%%%%
\subsubsection{Likelihood function}
%%%----------------------------%%%%
When analyzing radial velocity data of an exoplanetary system, it is commonly accepted that the \emph{wobbling} of the star around the center of mass is caused by the sum of the gravitational force of each planet independently and that they do not interact with each other. Each planet follows a Keplerian orbit and the radial velocity of the host star is given by
\begin{equation}
{y}_{t} = V_0 + \sum\limits_{i = 1}^{S} K_i \left[ \cos \left( {u}_{i,t} + \omega_{i} \right) + e_i \cos \left( \omega_{i} \right) \right] +\xi_t,
\label{eq:rv}
\end{equation}
with $t=1,\ldots,T$.\footnote{More generally, we can have $y_{t_j}$ with $j=1,...,T$.} The number of objects in the system is $S$. Both ${y}_{t}$, ${u}_{i,t}$ depend on time $t$, and $\xi_t$ is a Gaussian noise perturbation with variance $\sigma_e^2$. We consider the noise variance $\sigma_e^2$ an unknown parameter as well.
% For the sake of simplicity, we consider this value known, $\sigma_e^2=1$. 
The meaning of each parameter in Eq.~\eqref{eq:rv} is given in Table~\ref{tab:rvpar}. The likelihood
function is jointly defined by \eqref{eq:rv} and some indicator variables described below. 
The angle ${u}_{i,t}$ is 
the true anomaly of the planet $i$ and it can be determined from
\begin{equation*}
\frac{d{u}_{i,t}}{dt} = \frac{2\pi}{P_i} \frac{\left( 1 + e_i \cos {u_{i,t}} \right)^2}{\left( 1 - e_i \right)^\frac{3}{2}}
\label{eq:trueanomaly}
\end{equation*}
This equation has {an} analytical solution. As a result, the true anomaly $u_{i,t}$ can be determined from the mean anomaly  $M_{i,t}$. However, the analytical solution contains a non-linear term that needs to be determined by iterating. First, we define the mean anomaly $M_{i,t}$ as
\begin{equation*}
M_{i,t} = \frac{2\pi}{P_i} \left( t - \tau_i \right),
\label{eq:meananomaly}
\end{equation*}
where $\tau_i$ is the time of periastron passage of the planet $i$ and $P_i$ is the period
of its orbit (see Table~\ref{tab:rvpar}). Then, through the Kepler's equation, 
\begin{equation}
M_{i,t} = E_{i,t} - e_i \sin E_{i,t},
\label{eq:kepler}
\end{equation}
where $E_{i,t}$ is the eccentric anomaly. Equation~\eqref{eq:kepler} has no analytic solution and it must be solved by an iterative procedure. A Newton-Raphson method is typically used to find the roots of this equation \cite{Press2002}. For certain sets of parameters, this iterative procedure can be particularly slow and the computation of the likelihood becomes quite costly.  We also have
\begin{equation}
\tan \frac{u_{i,t}}{2} = \sqrt{ \frac{1 + e_i}{1 - e_i}} \, \tan \frac{E_{i,t}}{2}, 
\label{eq:eccentricanomaly}
\end{equation}
%
%The result of the model for a given time $t$ is compared to the received data $y_{r,t}$.
%%%%The variables $\omega_{i,t}$'s, for $i=1,\ldots,N$, can vary with time.
Therefore, the variable of interest $\mathbf{x}$ is the vector of dimension $d_X=1+5S$ (where $S$ is the number of planets),
\begin{align*}
\mathbf{x}= [V_0, K_1, \omega_{1}, e_1, P_1, \tau_1, \ldots, K_S, \omega_{S}, e_S, P_S, \tau_S],
\end{align*}
For a single object (e.g., a planet or a natural satellite), the dimension of $\mathbf{x}$ is $d_X = 5+1=6$, with two objects the dimension of $\mathbf{x}$  is $d_X = 11$, etc. All the Eqs. from \eqref{eq:rv} to \eqref{eq:eccentricanomaly} induce a likelihood function
$\ell({\bf y}|\mathbf{x},\sigma_e)=\prod_{t=1}^T\ell(y_{t}|\mathbf{x},\sigma_e)$,
%\end{align*}
where ${\bf y}=\{y_{1},\ldots,y_{T}\}$.

%%%----------------------------%%%%
\subsubsection{Prior and posterior densities}
%%%----------------------------%%%%
The prior $g({\bf x})$ is defined as multiplication of indicator variables  $V_0\in [-20,20]$, $K_{i} \in [0,\max y_{i,t}- \min y_{i,t}]$, $e_{i} \in [0,1]$, $P_i \in [0,365]$, $\omega_{i,t} \in [0,2\pi]$,  $\tau_{i} \in [0,30]$,
(i.e., the prior is zero outside these intervals), for all $i=1,\ldots,S$. This means that the prior density is zero when the particles fall out of these intervals. Note that the interval of $\tau_{i}$ is conditioned to the value $P_i$. This parameter is the time of periastron passage, i.e. the time passed since the object crossed the closest point in its orbit. It has the same units of $P_i$ and can take values from 0 to $P_i$. The complete posterior is  $$p(\mathbf{x}|{\bf y},\sigma_e)=\frac{1}{p({\bf y}|\sigma_e)} \ell({\bf y}|\mathbf{x},\sigma_e) g({\bf x}).$$ 
We are interested in inferring the parameters ${\bf x}$ and, more specifically,  computing the marginal likelihood
$$
Z=p({\bf y}|\sigma_e)=\int_{\mathcal{X}} \ell({\bf y}|\mathbf{x},\sigma_e) g({\bf x}) d{\bf x},
$$ 
obtained integrating out ${\bf x}$, in order to infer the number of planets. The noise variance $\sigma_e^2$ is also inferred after the sampling, by maximizing $Z=p({\bf y}|\sigma_e)$, i.e., $\widehat{\sigma}_e^2=\arg\max\limits_{\sigma_e} p({\bf y}|\sigma_e)$.

%%%----------------------------%%%%
\subsubsection{Experiments}
%%%----------------------------%%%%
Given a set of data $\y$ generated according to the model (see the initial parameter values below), our goal is to infer the number $S$ of planets in the system. For this purpose, we have to approximate the model evidence $Z=p({\bf y}|\sigma_e)$ of each model. 
In all experiments, we consider $60$ total number of observations.
We consider three different experiments: {\bf (E1)} $S=0$, i.e., no object, {\bf (E2)} $S=1$ (one object) and {\bf (E2)} the case of two objects $S=2$. We set $V=2$, in all cases. For the first object in {\bf E1} and {\bf E2}, we set $K_{1}=25$, $\omega_{1}=0.61$, $e_{1}=0.1$, $P_{1}=15$, $\tau_{1}=3$. For {\bf E2}, we also consider a second object with $K_{2}=5$, $\omega_{2}=0.17$, $e_{2}=0.3$, $P_{2}=115$, $\tau_{2}=25$ (in that case $S=2$).  All the data are generated with $\sigma_e^2=2$.
%% Note that the signal-to-noise ratio (SNR) associate to the second object is quite low (so that the detection of this planet is a very difficult task)
The rest of trajectories are generated according to the transition model (and the corresponding measurements $y_{t}$ according to the observation model). 

%%%%%Note that ABC cannot be applied since we are not able to generate arti 

%%%----------------------------%%%%
\subsubsection{Methods}
\begin{figure*}[!t]
	\centering
	%	\centerline{
	\subfigure[Zero planets]{\includegraphics[width=0.30\textwidth]{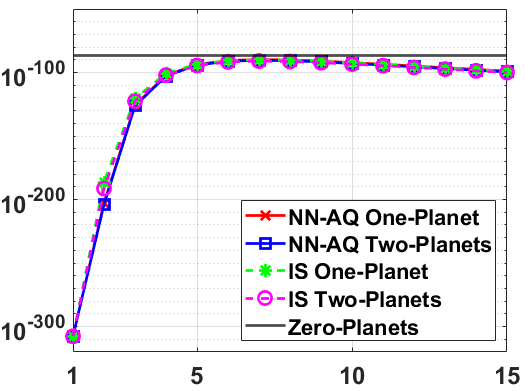}}
	\subfigure[One planet]{\includegraphics[width=0.30\textwidth]{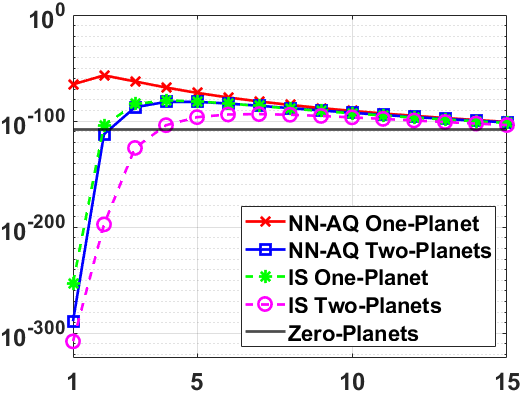}}		\subfigure[Two planets]{\includegraphics[width=0.30\textwidth]{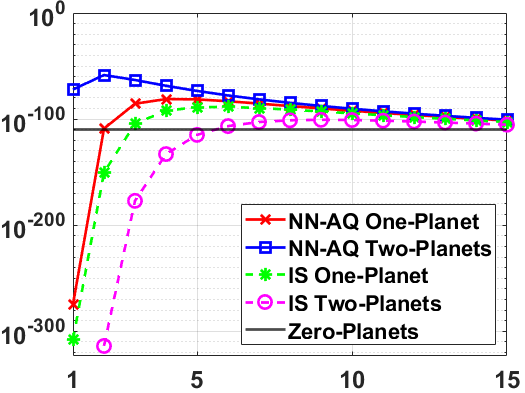}}
	%	}
	%\vspace{-0.3cm}
	\caption{\footnotesize Plot of marginal likelihood estimates of Model 1 (one-planet) and Model 2 (two-planets) versus $\sigma$ for the three data sets. The straight lines represent the known marginal likelihoods of Model 0 (zero planets) for each data set. {\bf (a)} data set with zero planets, {\bf (b)} data set with one planet, {\bf (c)} data set with two planets.   }
	\label{fig_Exoplanet}
\end{figure*}

%%%----------------------------%%%%
For each experiment, three models (i.e. three different target pdfs) are considered: a model with $S=0$ (Zero-Planets), a model with $S=1$ (One-Planet) and a model with $S=2$ (Two-Planets). The goal is to estimate the marginal likelihood of these models and then correctly detect the number of planets, i.e., $S=0$ for {\bf (E1)}, $S=1$ for {\bf (E2)} and $S=2$ for {\bf (E3)}. The marginal likelihoods corresponding to the Zero-planets models are available in closed form and need not be estimated (the model is simply Gaussian in that case).  %Hence, we only need to consider the estimation of One-Planet and Two-Planets models in each experiment.
For this purpose, we apply NN-AQ (with $M=10^7$) and  and an IS procedure. We allocate a budget of $4\cdot 10^6$ evaluations of the target. In IS, this budget is used to draw $4\cdot 10^6$ samples from the priors. While NN-AQ uses first $4\cdot 10^6 - 5000$ of these samples to look for a good initialization, more specifically, the sample with the highest target evaluation is kept, along with 9 more samples taken at random, to use them as initial nodes. Then, NN-AQ is run for $5000$ iterations. Both One-Planet and Two-Planets models are estimated for different values of $\sigma_e=1,2,\dots,15$. Note that we do not need to evaluate the target again when considering different $\sigma_e$, i.e., a single target evaluation can be reused for all values of $\sigma_e$. The results are shown in Figure \ref{fig_Exoplanet}.

%%%%%%%%%%%%%%%%%%%%%%%%%%%%%%%%%%%%%%%%%%%%%%%%%%%%%
%%%%%%%%%%%% FIGURE EXOPLANETAS NN-AQ VS NAIVE %%%%%%%%%%%%%%%%%%
%%%%%%%%%%%%%%%%%%%%%%%%%%%%%%%%%%%%%%%%%%%%%%%%%%%%%
%%%%%%%%%%%%%%%%%%%%%%%%%%%%%%%%%%%%%%%%%%%%%%%%%%%%%
%%%%%%%%%%%%%%%%%%%%%%%%%%%%%%%%%%%%%%%%%%%%%%%%%%%%%

%%%----------------------------%%%%
\subsubsection{Results}
%%%----------------------------%%%%

For each experiment {\bf (E1)}-{\bf (E3)}, Figure \ref{fig_Exoplanet}(a)-(c) depicts the estimations of $Z$ of the different models provided by NN-AQ and IS, versus $\sigma_e$. The horizontal lines correspond to the known marginal likelihoods of Zero-Planets models. Overall, NN-AQ outperforms IS and predicts correctly the number of planets as well as the true value of $\sigma_e$ (indeed, the curves corresponding to NN-AQ reach a maximum at $\sigma_e=2$).
Figure \ref{fig_Exoplanet}(a) shows that the estimations provided by NN-AQ and IS correctly rank the Zero-planets model ($S=0$) as the most probable one. Figure \ref{fig_Exoplanet}(b) shows both NN-AQ and IS predict correctly the One-Planet model ($S=1$) to be the correct one. However, for $\sigma_e=2$, IS barely differentiates between the Zero-Planet and One-Planets models. Further, for $\sigma_e=1$, it wrongly predicts Zero-Planets as the best one. Conversely, NN-AQ is able to predict 
the correct model for every value of $\sigma_e$, and besides, also predicts the true value $\sigma_e=2$. In Figure \ref{fig_Exoplanet}(c), the difference in performance of NN-AQ and IS is more acute. While NN-AQ is able to correctly predict the Two-Planets model ($S=2$) as the most probable for all values of $\sigma_e$, IS is unable to detect that second planet and, therefore, considers the One-Planet model more probable. As in the previous case, IS fails at detecting any planet for small values of $\sigma_e$. Again, NN-AQ predict the correct value of $\sigma_e$.

\section{Conclusions}

In this work, we have described a general framework for adaptive interpolative quadrature schemes, 
leveraging {an in-}depth study of different fields and related techniques in the literature, such as  Bayesian quadrature algorithms, scattered data approximations, emulation, experimental design and active learning schemes. 
%%%%%%
The nodes of the quadrature are adaptively chosen by maximizing a suitable acquisition function, which depends on the current interpolant and the positions of the nodes. This maximization does not require extra evaluations of the true posterior. The proposed methods supply also a surrogate model  (emulator) which approximates the true posterior density, that can be also employed in further statistical analyses. 
Two specific schemes, based on Gaussian and NN bases,  have been described. In both cases, a non-negative estimation $\widehat{Z}$ of the marginal likelihood $Z$ is ensured.
\newline
In the proposed framework, we also relax the assumptions regarding the kernel-basis functions with respect to other approaches in the literature, e.g., the bases could be non-symmetric. For instance, the NN bases are non-symmetric functions and their use has different important benefits: {\bf (a)} they  ensure obtaining non-negative interpolation coefficients and estimators $\widehat{Z}$, {\bf (b)} the linear system is directly solved without the need of inverting any matrix (the interpolation matrix is always diagonal), and  {\bf (c)} the bandwidth of the bases are automatically selected. 
Our scheme also allows selecting different kernel functions for each node point.
Therefore, the quadrature rules in Bayesian quadrature are a special case of our proposed scheme. Indeed, Bayesian quadrature considers a single symmetric and semi positive definite kernel function.
%In this sense, the proposed framework extends the applicability of Bayesian quadrature, 
%which requires using a single symmetric and semi positive definite kernel function. Obviously, our schemes do not provide 
%The combinations of our schemes with MC and deterministic quadrature rules has been considered. 
An  importance sampling interpretation has been also provided.  It is important to remark that the true posterior is only evaluated at the nodes selected sequentially by the algorithm, and the rest of other computations does not query the true model. 
The convergence of the proposed quadrature rules has been discussed, jointly with other theoretical results.  The new algorithms are powerful techniques as also shown by several numerical experiments.

\bibliographystyle{IEEEtranN}
\bibliography{bibliografia}

\begin{appendix}

	%%%%%%%%%%%%%%%%%%%%%%%%%%%%%%%%%%%%%%%%%%%%%%%%%%%%%
	%%%%%%%%%%%% FIGURE %%%%%%%%%%%%%%%%%%
	%%%%%%%%%%%%%%%%%%%%%%%%%%%%%%%%%%%%%%%%%%%%%%%%%%%%%
	\begin{figure*}[!t]
		\centering
		\centerline{
			\subfigure[]{\includegraphics[width=0.3\textwidth]{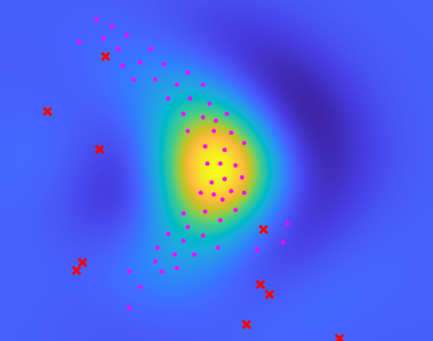}}
			\subfigure[]{\includegraphics[width=0.3\textwidth]{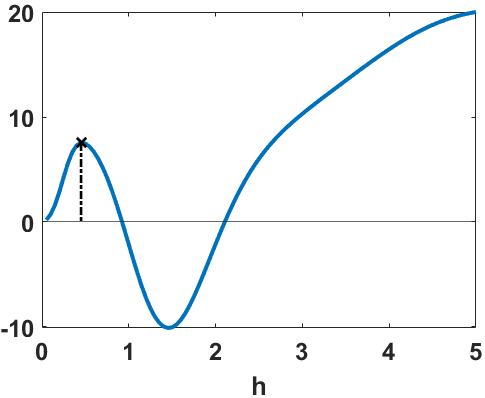}}
			%		\subfigure[]{\includegraphics[width=0.27\textwidth]{banana_GK_hFixed}}
			\subfigure[]{\includegraphics[width=0.3\textwidth]{banana_GK_png}}
		}
		\vspace{-0.3cm}
		\caption{\footnotesize {\bf (a)} GK based interpolant with $N_T=70$ nodes and $h=2.7$, fitted by maximizing the marginal likelihood. {\bf (b)} Plot of $\widehat{Z}$ as function of $h$. The value of $h$ at which $\widehat{Z}$ attains the local maximum is used to build the interpolant in our procedure.  {\bf (d)} GK based interpolant with $N_T=70$ nodes and $h=0.51$ fitted with the heuristic. }
		\label{fig_heuristic}
	\end{figure*}
	%%%%%%%%%%%%%%%%%%%%%%%%%%%%%%%%%%%%%%%%%%%%%%%%%%%%%
	%%%%%%%%%%%%%%%%%%%%%%%%%%%%%%%%%%%%%%%%%%%%%%%%%%%%%
	%%%%%%%%%%%%%%%%%%%%%%%%%%%%%%%%%%%%%%%%%%%%%%%%%%%%%

	\section{Procedure for tuning the Gaussian kernel bandwidth}\label{App:MagicalHeuristicLuca}
	%-- sencillo: utilizar h fijo en el proceso adaptativo y aplicar esta heuristica al final
	%
	%-- pero tambien se podria hacer durante la adaptacion en alguna iteracion 
	
	In this Appendix, we propose a procedure for fitting the bandwidth parameter $h$ of the Gaussian kernel (GK),
	\begin{align}
	k_G(\x,\x_i) = \frac{1}{(2\pi)^{\frac{d_x}{2}}h^{d_x}}
	\exp\left(-\frac{1}{2h^2}(\x-\x_i)^{\top}(\x-\x_i) \right ),
	\end{align}
	when building the GK based interpolant of Sect. \ref{sec:GaussKernels} for a given number of nodes. Assume we have run the GK-AQ algorithm (with some fixed $h_0$), so we have a total of $N_T$ nodes. Now, for any $h$, we may solve the linear system (Eq. \eqref{eq:InterpCoeffs}), obtain the coefficients $\{\beta_i\}_{i=1}^{N_T}$ and calculate
	\begin{align}
	\widehat{Z} = \sum_{i=1}^{N_T}\beta_i.
	\end{align}
	Note that, although not explicit, the $\beta_i$'s, and hence $\widehat{Z}$, depend on $h$. The proposed procedure consists of taking $h$ as the value where $\widehat{Z}$ attains its first local  maximum.
	Starting from a small value $h$ close to zero and increasing it, the estimation $\widehat{Z}$ is growing reaching a maximum. Then, $h$ is starting to become ``too big'', producing too much overlapping among the kernel areas. The values of the elements out the diagonal of ${\bf K}$ grow, and some of the coefficients $\beta_i$ are negative, and the estimation $\widehat{Z}$ decreases. As $h$ becomes greater and greater, the matrix ${\bf K}$ tends to become ill-conditioned, and the absolute values of  $\beta_i$'s grows.
	Figure \ref{fig_heuristic} compares the GK based interpolant of the target from Sect \ref{sec:bananaTar} with two different choices of $h$ and $N_T=70$ nodes. Figure \ref{fig_heuristic}(a) plots the interpolant taking $h$ as the value which minimizes the marginal likelihood (see Sect. \ref{sec:hyperparTuning}). Note that this value of $h$ is too big given the dispersion of the nodes. While Figure \ref{fig_heuristic}(c) plots the interpolant taking $h$ as the value where the curve of $\widehat{Z}$ (Figure \ref{fig_heuristic}(b)) attains its local maximum. This choice of $h$ seems to fit better the existing nodes. 
	Note also that, for some values of $h$, $\widehat{Z}$ may be negative.

	%{\color{red} recordar que aqui hemos empleado un $\sigma_\text{nois}=0.01$... si se uso uno demasiado pequeño entonces no hay maximo local }
	%\newline
	%{\color{red}In the above, we have considered running the GK-AQ algorithm with some initial and fixed $h_0$, and then using the heuristic to fit $h$ at the end of the run, once we have obtained the $N_T$ nodes. However, intuitively, the magnitude of $h_0$ should decrease as $N_t$ grows. For this reason, we may consider applying the heuristic to fit $h$ as the algorithm runs at a set of given times....}

	\section{Probabilistic interpretation of $J$}
	\label{App:BayesQuadview}
	
	Let us consider $J = \int_\mathcal{X}f(\x)\pi(\x)d\x$, which is the numerator of \eqref{eq:IntegralOfInter}, our integral of interest $I$. In section \ref{Sec:FromIntToReg}, we have seen that, when $k(\x,\x_i)=k(\x_i,\x)$ (i.e., a symmetric basis function), the interpolant  $\widehat{\pi}(\x) = \sum_{i=1}^N\beta_ik(\x,\x_i)$ has the probabilistic interpretation of being the mean of the posterior distribution of (the ``unknown'') $\pi(\x)$ after observing ${\bf d} = [\pi(\x_1),\dots,\pi(\x_N)]^\top$, i.e., $\mathbb{E}[\pi(\x)|{\bf d}] = \widehat{\pi}(\x)$.  The distribution on $\pi(\x)$ induces a posterior distribution on $J$, which is a Gaussian with mean
	\begin{align}\label{eq:meanOfBQ}
	\mathbb{E}[J|{\bf d}] =\widetilde{J}= \int_\mathcal{X}f(\x)\widehat{\pi}(\x)d\x,
	\end{align}
	and variance given by
	\begin{align}
	\text{var}[J|{\bf d}] = \int\int k(\x,\x')f(\x)f(\x')d\x d\x' - \bm{\zeta}^\top{\bf K}^{-1}\bm{\zeta},
	\end{align}
	where $\bm{\zeta} = [J_1,\dots,J_N]$ and $J_i = \int_\mathcal{X}f(\x)k(\x,\x_i)d\x$. This interpretation corresponds to the so-called Bayesian quadrature, which uses Eq. \eqref{eq:meanOfBQ} as approximation of $J$. Note that Eq. \eqref{eq:meanOfBQ} is the quadrature obtained by substituting the true $\pi(\x)$ with its interpolant  $\widehat{\pi}(\x)$, which coincides with the numerator of $\widehat{I}$ in Eq. \eqref{I approx by substituting pi with interpolator}.
	
	%%%%%%%%%%%%%%%%%%%%%%%%%%%%%%%%%%%%%%%
	%%%%%%%%%%%%%%%%%%%%%%%%%%%%%%%%%%%%%%%
	\section{Recursive inversion of a bordered matrix}\label{Inverse of bordered matrix}
	%%%%%%%%%%%%%%%%%%%%%%%%%%%%%%%%%%%%%%%
	%%%%%%%%%%%%%%%%%%%%%%%%%%%%%%%%%%%%%%%
	
	The most costly step when calculating $\bm{\beta}$ in \eqref{eq:InterpCoeffs} consists in inverting the $N\times N$ matrix $({\bf K})_{i,j} = k(\x_i,\x_j)$ ($i,j \in \{1,\dots,N\}$). Moreover, every time a new node is added, the $\beta_i$ must be recomputed, so the step of computing the inverse has to be done again. This time the matrix is bigger due to adding a new node, that is, it has an additional row and column. We show that knowing ${\bf K}^{-1}$ help us to compute the inverse of augmented matrices (called ``bordered matrix'', i.e., adding a ``border'' of new row and column to an existing matrix). 
	\newline
	Let us denote with ${\bf K}_{N}$ the matrix built using $N$ nodes, and let ${\bf K}_{N+1}$ be the matrix with $N+1$ nodes. Of course we have
	\begin{align}
	{\bf K}_{N+1} = \begin{pmatrix}
	{\bf K}_{N} & {\bf k}_N \\
	{\bf k}^T_N  & k
	\end{pmatrix}
	\end{align} 
	where ${\bf k}_N = (k(\x_1,\x_{N+1}), k(\x_2,\x_{N+1}),\dots,k(\x_N,\x_{N+1}))^T$ and $k=k(\x_{N+1},\x_{N+1})$. The $(N+1)\times (N+1)$ inverse of ${\bf K}_{N+1}$ can be expressed in terms of ${\bf K}^{-1}_N$ as follows
	%\begin{align}
	%	{\bf K}_{N+1}^{-1} = \begin{pmatrix}
	%	\left( {\bf K}_N - \frac{1}{ k}{\bf k}_N {\bf k}^T_N \right)^{-1} & 
	%	-{\bf K}^{-1}_N{\bf k}_N\left( k - {\bf k}^T_N {\bf K}^{-1}_N{\bf k}_N \right)^{-1} \\
	%	-\frac{1}{k}{\bf k}^T_N\left( {\bf K}_N - \frac{1}{k}{\bf k}_N{\bf k}^T_N \right)^{-1} &
	%	\left( k - {\bf k}_N^T {\bf K}_N^{-1}{\bf k}_N \right)^{-1}
	%	\end{pmatrix}
	%\end{align}
	%Lo malo de esta formula es que se necesita saber tambien la inversa de ${\bf K}_N - \frac{1}{ k}{\bf k}_N {\bf k}^T_N$ ....
	%\newline
	%\newline
	%Por lo visto, si que es posible expresar ${\bf K}^{-1}_{N+1}$ solo conociendo ${\bf K}^{-1}_N$ (visto en stackexchange: https://math.stackexchange.com/questions/1391227/is-there-a-formula-for-the-inverse-of-this-bordered-matrix):
	\begin{align}\label{eq:iterKInverse}
	{\bf K}_{N+1}^{-1} = \begin{pmatrix}
	{\bf A} & {\bf b} \\
	{\bf c} & s
	\end{pmatrix},
	\end{align}
	where 
	\begin{align*}
	{\bf A}&={\bf K}^{-1}_N + {\bf K}_N^{-1}{\bf k}_N\left( k - {\bf k}_N^T {\bf K}_N^{-1}{\bf k}_N \right)^{-1} {\bf k}_N^T{\bf K}_N^{-1} \in \mathbb{R}^{N\times N}, \\
	{\bf b}&=-{\bf K}^{-1}_N{\bf k}_N\left( k - {\bf k}^T_N {\bf K}^{-1}_N{\bf k}_N \right)^{-1}  \in \mathbb{R}^{N\times 1}, \\
	{\bf c}	&=-\left( k - {\bf k}^T_N {\bf K}^{-1}_N{\bf k}_N \right)^{-1}{\bf k}^T_N{\bf K}^{-1}_N   \in \mathbb{R}^{1\times N}, \\
	s	&=\left( k - {\bf k}_N^T {\bf K}_N^{-1}{\bf k}_N \right)^{-1}  \in \mathbb{R}.
	\end{align*}
	Note that computing $s=\left( k - {\bf k}_N^T {\bf K}_N^{-1}{\bf k}_N \right)^{-1}$ is not costly since it is an scalar value. 
	% If we have a noise term $\sigma>0$, we are interested in the inverse of ${\bf K}_{N+1} + \sigma^2{\bf I}_{N+1}$ given $\left( {\bf K}_{N} + \sigma^2{\bf I}_{N} \right)^{-1}$. We can keep using Eq. \eqref{eq:iterKInverse} where we need to substitute ${\bf K}_N^{-1}$ with $\left( {\bf K}_{N} + \sigma^2{\bf I}_{N} \right)^{-1}$, and $k$ with $k + \sigma^2$.

	\section{Proofs}\label{App:Proofs}
	
	\subsection{Proof to theorem 1}  \label{Teo1} %\ref{Thm:errorBoundInftyNorm}
	%	\begin{proof}
	We have that
	\begin{align*}
	|J - \widehat{J}| &= \left|\int_\mathcal{X}f(\x)\pi(\x)d\x - 
	\int_\mathcal{X}f(\x)\widehat{\pi}(\x)d\x  \right| \\
	&=\left|\int_\mathcal{X}f(\x)\left(\pi(\x)-\widehat{\pi}(\x)\right)d\x\right|.
	\end{align*}
	It is easy to see that, for any $g(\x)$ we have $-|g(\x)| \leq g(\x) \leq |g(\x)|$ for all $\x$, and that $-\int |g(\x)|d\x \leq \int g(\x)d\x \leq \int |g(\x)|d\x$, so we have $\left|\int g(\x)d\x \right| \leq \int |g(\x)|d\x$.
	Using this result we can state the first inequality
	\begin{align*}
	|J - \widehat{J}| &=\left|\int_\mathcal{X}f(\x)\left(\pi(\x)-\widehat{\pi}(\x)\right)d\x\right| \\
	&\leq \int_\mathcal{X}\left|f(\x)\right|\left|\pi(\x)-\widehat{\pi}(\x)\right|d\x \\
	&= \norm{f(\pi - \widehat{\pi})}_1.
	\end{align*}
	The second inequality of the theorem follows from Holder's inequality
	\begin{align*}
	\norm{f(\pi - \widehat{\pi})}_1 \leq \norm{f}_2\norm{\pi - \widehat{\pi}}_2.
	\end{align*}
	Finally, the last inequality of the theorem is obtained after manipulating the $\norm{f}_2$ and $\norm{\pi - \widehat{\pi}}_2$,
	{\footnotesize
	\begin{align*}
	\norm{f}_2\norm{\pi - \widehat{\pi}}_2&= \left(\int_\mathcal{X}|f(\x)|^2d\x\right)^{\frac{1}{2}}\left(\int_\mathcal{X}|\pi(\x)-\widehat{\pi}(\x)|^2d\x\right)^{\frac{1}{2}} \\
	&\leq \left(|\mathcal{X}|\max |f(\x)|^2   \right)^{\frac{1}{2}}\left(|\mathcal{X}|\max |\pi(\x)-\widehat{\pi}(\x)|^2   \right)^{\frac{1}{2}} \\
	&= |\mathcal{X}| \max |f(\x)| \max |\pi(\x)-\widehat{\pi}(\x)| \\
	&= |\mathcal{X}|\norm{f}_\infty\norm{\pi-\widehat{\pi}}_\infty.
	\end{align*}
}
	%\end{proof}
	
	%%%%%%%%%%%%%%%%%%%%%%%%%%%%%%%
	\subsection{Proof to theorem 2} \label{Teo2}
	%%%%%%%%%%%%%%%%%%%%%%%%%%%%%%%
	%\ref{Thm:OptimalWeights}
	We provide the main concepts and elements of the proof. For more details, see \cite{sommariva2006numerical,briol2019probabilistic}.
	Let $J = \int_\mathcal{X}f(\x)\pi(\x)d\x$
	and $\widetilde{J} = \sum_{i=1}^N\nu_i \pi(\x_i)$
	be the integral of interest and the quadrature using points $\{\x_i\}_{i=1}^N$, respectively. Recall that we also denote  $\bm{\nu} = [\nu_1,\dots, \nu_N]^\top$.
	%At this point, the vector $\bm{\nu} = [\nu_1,\dots, \nu_N]^\top$ is not specified and we seek to find an expression for the optimal weights by assuming that $\pi$ belongs to a function space.
	\newline
	Consider that $\pi$ is a function belonging to the reproducing kernel Hilbert space of functions $\mathcal{H}$ originated from the symmetric and positive definite kernel function $k(\x,\x')$. Hence, $J$ and $\widetilde{J}$ are functionals over that RKHS
	\begin{align*}
	J[\pi] &= \int_\mathcal{X}f(\x)\pi(\x)d\x, \\
	\widetilde{J}[\pi] &= \sum_{i=1}^N\nu_i \pi(\x_i), \enskip \pi \in \mathcal{H}. 
	\end{align*}
	where we write explicitly $J[\cdot]$ is the functional that integrates w.r.t. $f(\x)$, while $\widetilde{J}[\cdot]$ is the functional that integrates w.r.t. the weighted sum $\sum_{i=1}^N\nu_i\delta_{\x_i}$, where $\delta_{\x_i}$ denotes the point evaluation in $\x_i$.
	The integration error associated with $\widetilde{J}$ is characterized by the norm, in the dual space $\mathcal{H}^*$, of the error functional
	\begin{align}\label{eq:WorseCaseError}
	\norm{J - \widetilde{J}}_{\mathcal{H}^*} = \sup_{\norm{\pi}_{\mathcal{H}}\leq 1}\left|\widetilde{J}[\pi] - J[\pi]\right|,
	\end{align}
	where $\norm{\cdot}_{\mathcal{H}}$ and $\norm{\cdot}_{\mathcal{H}^*}$ denote the norm in $\mathcal{H}$ and $\mathcal{H}^*$ respectively. Eq. \eqref{eq:WorseCaseError} is also called worst-case error (WCE). Define the functions
	\begin{align}
	k_f(\x) = \int_\mathcal{X} f(\x')k(\x,\x')d\x',
	\end{align}
	and 
	\begin{align}
	k_{\widetilde{f}}(\x) = \sum_{i=1}^N \nu_ik(\x,\x_i),
	\end{align}
	where $k_f$, $k_{\widetilde{f}} \in \mathcal{H}$. These functions exist as consequence of $\int_\mathcal{X}k(\x,\x)f(\x)d\x < \infty$. It can be shown that 
	$\norm{J - \widetilde{J}}_{\mathcal{H}^*} = \norm{k_f - k_{\widetilde{f}}}_\mathcal{H}$, 
	and 
	{\footnotesize
	\begin{align}
	\norm{J - \widetilde{J}}^2_{\mathcal{H}^*} &= \bm{\nu}^\top {\bf K}\bm{\nu} - 2\bm{\nu}^\top\bm{\zeta} + \int_\mathcal{X}\int_\mathcal{X}f(\x)f(\x')k(\x,\x')d\x d\x',
	\end{align}
}
	for a vector of weights $\bm{\nu} \in \mathbb{R}^N$, the matrix $({\bf K})_{1\leq i,j\leq N} = k(\x_i,\x_j)$, and the vector of integrals $\bm{\zeta} = [k_f(\x_1),\dots,k_f(\x_N)]^\top$. Conditional on the fixed states $\{\x_i\}_{i=1}^N$, the weights $\bm{\nu}$ that minimizes the above expression are given by $\bm{\nu} = {\bf K}^{-1}\bm{\zeta}$. These are the weights that arises if we build the interpolant  $\widehat{\pi}$ of $\pi$ at points $\{\x_i\}_{i=1}^N$, using $k(\x,\x')$ as the basis function, and substitute it in $J$ to obtain the quadrature.

	\subsection{Proof to theorem 6}\label{Sect_ErrorBoundFillAndSeparDists}
	%\ref{Thm:ErrorBoundFillAndSeparDists}
	Let $J$ be the integral of interest, and let $\tilde{J} = \sum_{i=1}^{N}\beta_iJ_i$ and $\widehat{J} = \sum_{i=1}^{N}\beta_i\widehat{J}_i$ be the approximations using, respectively, the exact $J_i$ and the noisy estimation $\widehat{J}_i$ . Recall that the coefficients $\beta_i$ are written in matrix form as $\bm{\beta} = {\bf K}^{-1}{\bf d}$ where ${\bf K}$ is the interpolation matrix and ${\bf d}$ is the vector of evaluations of $\pi$. Let us denote $\bm{\zeta} = [J_1,\dots,J_N]^\top$ and $\widehat{\bm{\zeta}} = [\widehat{J}_1,\dots,\widehat{J}_N]^\top$. Denoting the dot product in $\mathbb{R}^N$ as $\langle\cdot,\cdot\rangle$, we can express $\widetilde{J} = \langle \bm{\zeta}, \bm{\beta} \rangle$ and $\widehat{J} = \langle \widehat{\bm{\zeta}}, \bm{\beta}\rangle$. Thus
	\begin{align*}
	|J - \widehat{J}| &= |J -\langle \widehat{\bm{\zeta}}, \bm{\beta}\rangle| \\
	&= |J - \langle\bm{\zeta} - \bm{\zeta} + \widehat{\bm{\zeta}}, \bm{\beta}\rangle| \\
	&= |J - \langle\bm{\zeta}, \bm{\beta} \rangle +
	\langle  \bm{\zeta}, \bm{\beta}\rangle -  \langle \widehat{\bm{\zeta}}, \bm{\beta} \rangle| \\
	&\leq |J - \widetilde{J}| + |\langle \bm{\zeta} - \widehat{\bm{\zeta}}, \bm{\beta}\rangle| \\
	&= |J - \widetilde{J}| + |\langle{\bf K}^{-1}( \bm{\zeta} - \widehat{\bm{\zeta}}), \bm{d}\rangle| \\
	&\leq ||f(\pi - \widehat{\pi})||_1 + ||{\bf K}^{-1}(\bm{\zeta} - \widetilde{\bm{\zeta}})||_2 \norm{{\bf d}}_2 \\
	&\leq |\mathcal{X}| \left||f|\right|_\infty ||\pi - \widehat{\pi}||_\infty + ||{\bf K}^{-1}||_2 ||\bm{\zeta} - \widehat{\bm{\zeta}}||_2\left||{\bf d}|\right|_2 
	\end{align*}
	where the norm $||{\bf K}^{-1}||_2$ represents the largest singular value of ${\bf K}^{-1}$. The bounds $\norm{\pi - \widehat{\pi}}_\infty = \lambda(r)$ and $|| {\bf K}^{-1} ||_2  =\mathcal{O}( \upsilon(s,h))$ for different RBF can be found respectively in Chapters 11.3 and 12.2 of \cite{wendland2004scattered}. For further details, see  Proposition 1 in \cite{sommariva2006numerical}.

	\subsection{Proof to theorem 8} \label{LipsProof}
	%\ref{Thm:LipsTarget}
	Let us consider the target $\pi(\x)$ and the interpolant  $\widehat{\pi}(\x)$ based on NN constant kernels. Note that for all $\x \in \mathcal{X}$ we have $\widehat{\pi}(\x) = \pi(\x^*)$, where $\x^* = \arg \min_i \norm{\x - \x_i}$, i.e., the node that is closest to $\x$. Lipschitz continuity implies that  $|\pi(\z) - \pi(\x)| \leq L_0\norm{\z - \x}$ for all $\z,\x \in \mathcal{X}$. Hence, 
	\begin{align*}
	\norm{\pi - \widehat{\pi}}_\infty &= \max_{\x \in \mathcal{X}} |\pi(\x) - \widehat{\pi}(\x)| \\
	&= \max_{\x \in \mathcal{X}} |\pi(\x) - \pi(\x^*)| \\
	&\leq L_0 \max_{\x \in \mathcal{X}}\norm{\x - \x^*} \\	
	&= L_0 \max_{\x \in \mathcal{X}}\min_i \norm{\x - \x_i} \\
	&=L_0 r,
	\end{align*}
	where we used the definition of fill distance $r$, i.e.,  
	$$
	r = \max_{\x \in \mathcal{X}} \min_i \norm{\x - \x_i}.
	$$
	For further details, see \cite{butler2017measure,Pronzato17}.

\end{appendix}

\end{document}